\documentclass{aa}  
\usepackage{graphicx}
\usepackage{natbib}
\bibpunct{(}{)}{:}{a}{}{,}
\usepackage{txfonts}
\begin{document}
   \title{Pulsations detected in the line profile variations of red giants}

   \subtitle{Modelling of line moments, line bisector and line shape}

   \author{S. Hekker \inst{1} \fnmsep \inst{2}
    \and C. Aerts \inst{2} \fnmsep \inst{3}
     \and J. De Ridder \inst{2}
      \and F. Carrier \inst{2} \fnmsep \inst{4}}

   \offprints{S. Hekker, \\
                    email: saskia@strw.leidenuniv.nl}
   
   \institute{Leiden Observatory, Leiden University, P.O. Box 9513, 2300 RA
Leiden, The Netherlands
                  \and
              Instituut voor Sterrenkunde, Katholieke Universiteit Leuven,
Celestijnenlaan 200 B, B--3001, Belgium     
                    \and
              Department of Astrophysics, University of Nijmegen, PO Box 9010,
6500 GL Nijmegen, The Netherlands 
                     \and
              Observatoire de Gen\`{e}ve, 51 chemin de Maillettes, 1290
Sauverny, Switzerland
}

   \date{Received <date>; accepted <date>}

% \abstract{}{}{}{}{} 
% 5 {} token are mandatory
 
  \abstract
{So far, red giant oscillations have been studied from radial velocity and/or
light curve variations, which reveal frequencies of the oscillation modes. To characterise radial and non-radial oscillations, line profile variations are a valuable diagnostic. Here we present for the first time a line profile analysis of pulsating red giants, taking into account the small line profile variations and the predicted short damping and re-excitation times. We do so by modelling the time variations in the cross correlation profiles in terms of oscillation theory.}
{The performance of existing diagnostics for mode identification is investigated for known oscillating giants which have very small line profile variations. We modify these diagnostics, perform simulations, and characterise the radial and non-radial modes detected in the cross correlation profiles. }
{Moments and line bisectors are computed and analysed for four giants. The robustness of the discriminant of the moments against small oscillations with finite lifetimes is investigated. In addition, line profiles are simulated with short damping and re-excitation times and their line shapes are compared with the observations.}
{For three stars, we find evidence for the presence of non-radial pulsation modes, while for \object{$\xi$ Hydrae} perhaps only radial modes are present. Furthermore the line bisectors are not able to distinguish between different pulsation modes and are an insufficient diagnostic to discriminate small line profile variations due to oscillations from exoplanet motion.}
   {}

 \authorrunning{S. Hekker et al.}
  \maketitle
%
%________________________________________________________________

\section{Introduction}
Techniques to perform very accurate radial velocity observations are refined
during the last decade. Observations with an accuracy of a few m\,s$^{-1}$ are obtained
regularly (see e.g. \citet{marbut}, \citet{queloz2001a}) and detections of amplitudes of
1~m\,s$^{-1}$ are  nowadays possible with e.g. HARPS \citep{pepe2003}.  This
refinement not only forced a breakthrough in the detection of extra solar
planets
but also in the observation of solar like oscillations in distant
stars. Solar like oscillations are excited by turbulent convection near the
surface of cool stars of spectral type F, G, K or M and show radial velocity variations with amplitudes of typically a few
cm\,s$^{-1}$
to a few m\,s$^{-1}$, and with periods ranging from a few minutes for main sequence
stars to about half an hour for subgiants and a couple of hours for giants.

For several stars on or close to the main sequence, solar like oscillations were
detected for a decade (see e.g. \citet{Kjeldsen1995}, \citet{bouchy2001b}, \citet{bouchy2003}, \citet{bedding2003}). More
recently, such type of oscillations has also been firmly established in several
red (sub)giant stars.  \citet{frandsen2002}, \citet{deridder2006b},
\citet{carrier2006a}, (see also \citet{barban2004}) and \citet{carrier2003} used the CORALIE and ELODIE
spectrographs to obtain long term high resolution time series of the three red
giants \object{$\xi$ Hydrae}, \object{$\epsilon$ Ophiuchi} and \object{$\eta$
Serpentis} and of the subgiant \object{$\delta$ Eridani}, respectively. They
unravelled a large frequency separation in the radial velocity Fourier
transform, with a typical value expected for solar like oscillations in the respective
type of star (a few $\mu$Hz for giants and a few tens $\mu$Hz for subgiants), according to theoretical predictions (e.g.
\citet{dziembowski2001}). It was already predicted by \citet{dziembowski1971} that
non-radial oscillations are highly damped in evolved stars, and that, most
likely, any detectable oscillations will be radial modes.

So far, red (sub)giant
oscillations have only been studied from radial velocity or light
variations. Line profile variations are a very valuable diagnostic to detect
both radial and non-radial heat driven coherent oscillations (e.g. \citet{aerts2003}
and references therein), and to characterise the wavenumbers ($\ell,m$) of such self-excited oscillations. It is therefore worthwhile to try and detect them for red
(sub)giants with confirmed oscillations and, if successful, to use them for empirical
mode identification. This would provide an independent test for the theoretical
modelling of the frequency spectrum. 

It is also interesting to compare the
line profile diagnostics used for stellar oscillation analysis with those
usually adopted
to discriminate oscillations from exoplanet signatures, such as the line bisector and
its derived quantities. Recently, \citet{gray2005} pointed out that the wide range of bisector shapes he found must contain information about the velocity fields in the atmospheres of cool stars, but that the extraction of information about the velocity variations requires detailed modelling. Here we perform such modelling in terms of non-radial oscillation theory. We do so by considering different types of line characteristics derived from cross correlation profiles. \citet{dall2006} already concluded that bisectors are not suitable to analyse solar like oscillations. We confirm this finding and propose much more suitable diagnostics. We nevertheless investigate how line bisector quantities behave
for confirmed oscillators, in order to help future planet hunters in
discriminating the cause of small line profile variations in their data.

The main problems in characterising the oscillation modes of red (sub)giants are,
first, the low amplitudes of the velocity variations, which results in
very small changes in the line profile. Secondly, the damping and re-excitation
times are predicted to be very short. Indeed, \citet{stello2004} derived an
oscillation mode lifetime in \object{$\xi$ Hydrae} of only approximately two days. In general, mode lifetimes are difficult to compute for different stellar parameters from current available theory, cf. \citet{houdek1999} versus \citet{stello2004}. \citet{stello2006} derived the observed mode lifetime of \object{$\xi$ Hydrae} from extensive simulations and find a large difference with theoretical predictions.
The
known mode identification methods from line profile variations all use an
infinite mode lifetime so far. Here, we provide line profile variations
simulated for stochastically excited modes, and investigate how the damping
affects the behaviour of the diagnostics in this case. We apply our methodology
to the four case studies of \object{$\xi$ Hydrae} (HD100407, G7III),
\object{$\epsilon$ Ophiuchi} (HD146791, G9.5III), \object{$\eta$ Serpentis}
(HD168723, K0III) and \object{$\delta$ Eridani} (HD23249, K0IV). Some basic properties of these stars are listed in Table~\ref{propstar}.

\begin{table*}
\begin{minipage}{17cm}
\caption{Basic stellar parameters of the four stars: Effective temperature ($T_{\mathrm{eff}}$) in Kelvin, rotational velocity ($\upsilon \mathrm{sin}i$) in km\,s$^{-1}$, parallax ($\pi$) in mas, distance ($d$) in pc, the apparent magnitude ($m_{v}$) and absolute magnitude ($M_{V}$) in the V band.}
\label{propstar}
\centering
\renewcommand{\footnoterule}{}
\begin{tabular}{lcccc}
\hline\hline
parameter & \object{$\epsilon$ Ophiuchi}\footnote{\citet{deridder2006b}, $^{b}$ \citet{taylor1999}, $^{c}$ \citet{carrier2003}, $^{d}$ \citet{glebocki2000}, $^{e}$ \citet{esa1997}} & \object{$\eta$ Serpentis} & \object{$\xi$ Hydrae} & \object{$\delta$ Eridani}\\
\hline
$T_{\mathrm{eff}}$ [K] & $4887 \pm 100$ & $4855 \pm 19^{b}$ & $5010 \pm 15^{b}$ & $5050\pm100^{c}$\\
$\upsilon \mathrm{sin}i$ [km\,s$^{-1}$] & $3.4 \pm 0.5$ & $2.6\pm 0.8^{d}$ & $2.4^{d}$ & $2.2\pm0.9^{d}$\\
$\pi$ [mas] & $30.34\pm0.79$ & $52.81\pm0.75^{e}$ & $25.23\pm0.83^{e}$ & $110.58\pm0.88^{e}$\\
$d$ [pc] & $33.0\pm 0.9$ & $18.9 \pm 0.3$ & $39.6\pm1.5$ & $9.0\pm0.1$\\
$m_{v}$ [mag] & $3.24\pm0.02$ & $3.23\pm0.02^{e}$ & $3.54\pm0.06^{e}$ & $3.52\pm0.02^{e}$\\
$M_{V}$ [mag] & $0.65\pm0.06$ & $1.85\pm0.05$ & $ 0.55\pm0.04$ & $3.75\pm0.2$\\
\hline
\end{tabular}
\end{minipage}
\end{table*}

The paper is organised as follows. In Sect.~2 the data set at our disposal and
different observational diagnostics are described. In Sect.~3 the diagnostics
for mode identification are described, while in Sect.~4 the damping and
re-excitation effects in theoretically generated spectral lines are investigated.
In Sect.~5 the results obtained from the simulations and observations are compared and in Sect.~6 some conclusions are drawn.

\section{Observational diagnostics}
\subsection{Spectra}
For all four stars we have spectra at our disposal obtained with the fibre fed
\'{e}chelle spectrograph CORALIE mounted on the Swiss 1.2~m Euler telescope at La
Silla (ESO, Chile). The observations for \object{$\xi$ Hydrae} were made during
one full month (2002 February 18 - March 18). This is the same dataset as used
by \citet{frandsen2002} for the detection of the solar like oscillations. For
\object{$\epsilon$ Ophiuchi} and \object{$\eta$ Serpentis} the solar like
observations described by \citet{deridder2006b} and \citet{carrier2006a} are
obtained from a bi-site campaign, using CORALIE and ELODIE (the fibre fed
\'{e}chelle spectrograph mounted on the French 1.93~m telescope at the Observatoire
the Haute-Provence) during the summer of 2003. The observations of \object{$\delta$ Eridani} are obtained with CORALIE during a twelve day campaign in November 2001. For the line profile analysis,
described in this paper, only the data obtained with CORALIE are used. These
spectra range in wavelength from 387.5 nm to 682 nm and the observation times
were adjusted to reach a signal to noise ratio of at least 100 at 550 nm (60-120 for \object{$\delta$ Eridani}) without averaging out a too large fraction of the pulsation phase. More details about the observation strategy is available in the publications describing the first detections of the solar like oscillations of the respective stars.

\begin{figure}
\centering
\resizebox{\hsize}{!}{\includegraphics{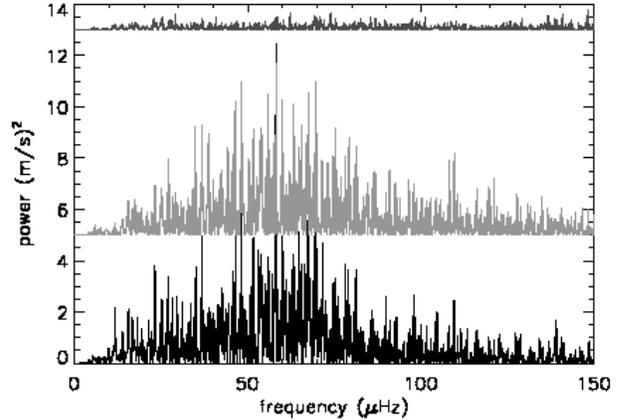}}
\caption{Power spectrum of \object{$\epsilon$ Ophiuchi}. The lower black one is
obtained by \citet{deridder2006b}, the grey one in the middle is obtained from
$\langle \mathrm{v} \rangle$ and the top one is obtained from the bisector velocity span. For clarity
the latter two are shifted.}
\label{periodogramepsoph}
\end{figure}

\begin{figure}
\centering
\resizebox{\hsize}{!}{\includegraphics{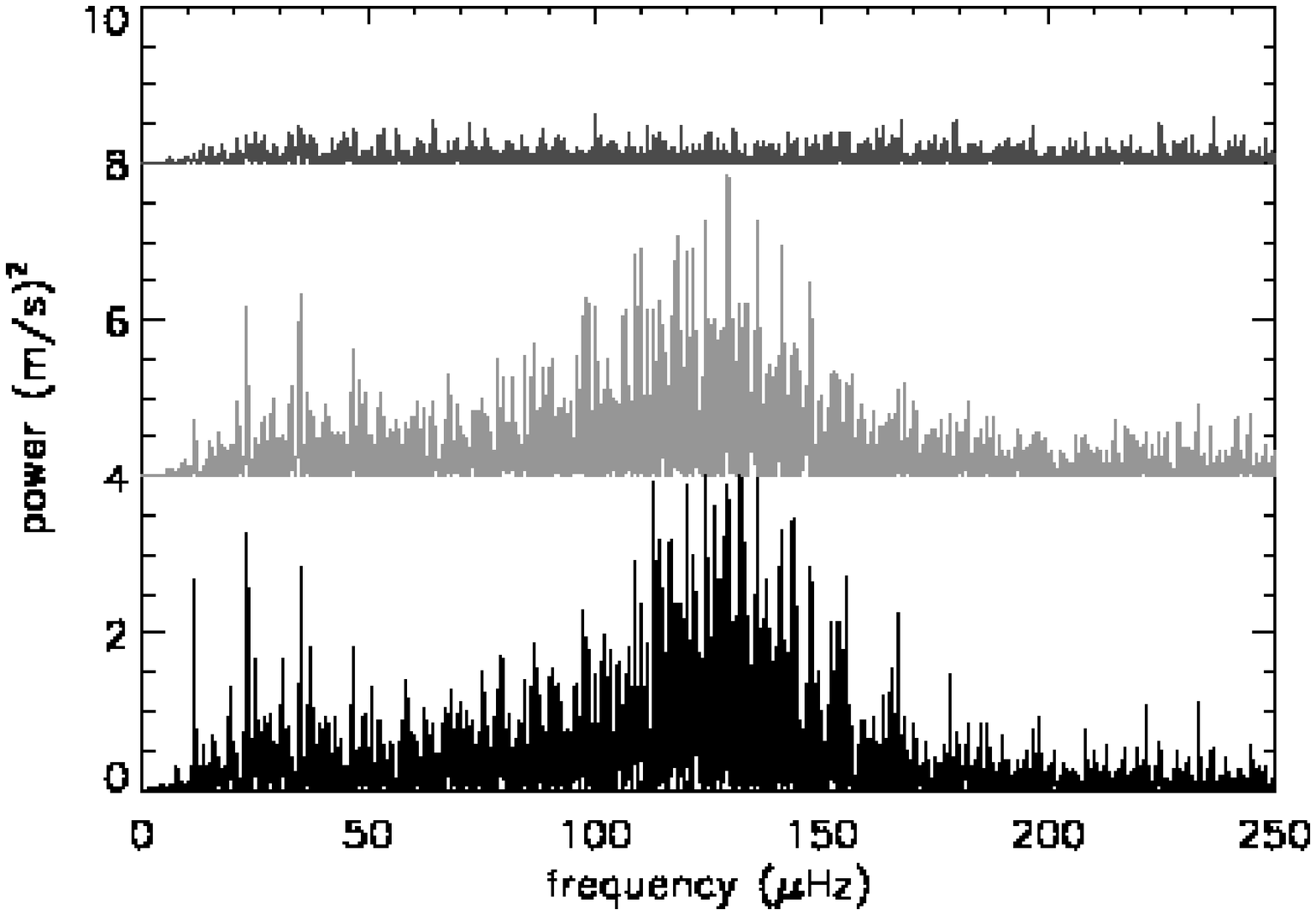}}
\caption{Power spectrum of \object{$\eta$ Serpentis}. The lower black one is
obtained by \citet{carrier2006a}, the grey one in the middle is obtained from
$\langle \mathrm{v} \rangle$ and the top one is obtained from the bisector velocity span. For clarity
the latter two are shifted.}
\label{periodogrametaser}
\end{figure}

\begin{figure}
\centering
\resizebox{\hsize}{!}{\includegraphics{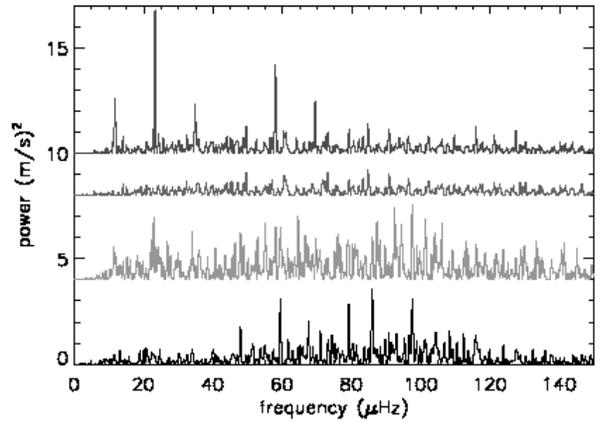}}
\caption{Power spectrum of \object{$\xi$ Hydrae}. The lower black one is
obtained by \citet{frandsen2002}, the grey one is obtained from $\langle \mathrm{v} \rangle$. The two top
power spectra are obtained from the bisector velocity span. The top one is the
original, while the lower one is corrected for the 1 c\,d$^{-1}$ ($11.57\mu$Hz)
frequency due to the diurnal cycle. For clarity the latter three power spectra
are shifted.}
\label{periodogramxihy}
\end{figure}

\begin{figure}
\centering
\resizebox{\hsize}{!}{\includegraphics{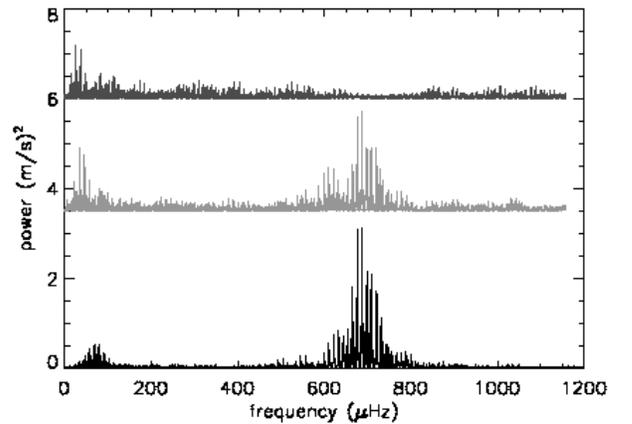}}
\caption{Power spectrum of \object{$\delta$ Eridani}. The lower black one is
obtained by \citet{carrier2003}, the grey one in the middle is obtained from $\langle \mathrm{v} \rangle$. The top
power spectrum is obtained from the bisector velocity span. For clarity the latter two power spectra
are shifted.}
\label{periodogramderi}
\end{figure}

\subsection{Cross correlation profiles}
The line profile variations of the four pulsating red (sub)giants are analysed with moments \citep{aerts1992} and with the line bisector \citep{gray2005}. The moments are
often used to analyse stellar oscillations excited by a heat mechanism, while the
line bisector is usually adopted to discover planet
signatures. Both methods can be applied to one single spectral line, but also to
a cross correlation profile, which has an increased signal to noise ratio:
\begin{equation}
SNR_{cross-cor}=\sqrt{M<(SNR)^{2}>}
\label{SNRCCF}
\end{equation} 
with $M$ the number of spectral lines used for the cross correlation and $<(SNR)^{2}>$ the average of the squared signal to noise ratios of each spectral line used. The cross correlation profiles are constructed from a mask in such a way that the weak, strong and blended lines are excluded. They are a good representation of an average spectral line of the star whenever the lines are formed at not too different line formation regions in the atmosphere.

\citet{mathias1996} and \citet{chadid2001} already performed a line profile analysis on cross
correlation profiles for a $\delta$ Scuti star using moments \citep{aerts1992}. Compared to the analysis of a single spectral line this only
introduces some additional terms in the moments for which can be corrected. Furthermore, the
line bisector analysis is widely used on cross correlation profiles (see
e.g. \citet{queloz2001b}, \citet{setiawan2003}, \citet{martinez2005}, \citet{dall2006}). \citet{dall2006} showed that the cross correlation bisector contains the same information as single line bisectors.

The red (sub)giants analysed in the present work show low amplitude radial
velocity variations, which result in very small line profile variations. The
INTER-TACOS (INTERpreter for the Treatment, the Analysis and the COrrelation of
Spectra) software package developed at Geneva Observatory \citep{baranne1996},
was used to cross correlate the observed spectrum with a mask containing
atmospheric lines of red (sub)giants with a resolution of 100 m\,s$^{-1}$. On top of that we
interpolated to a resolution of 10 m\,s$^{-1}$ to reach the highest possible precision in the computation of the moments.

\subsection{Frequency analysis}
In order to see whether the line moments and line bisector are sensitive enough
to determine the very small line profile variations present in pulsating red
(sub)giants, a frequency analysis is performed and compared with the frequencies
obtained previously. \citet{frandsen2002}, \citet{deridder2006b}, \citet{carrier2006a} and \citet{carrier2003} used radial velocity variations, obtained with the optimum weight method
described by \citet{bouchy2001a}, for their frequency analysis.

\subsubsection{Moments}
A line profile can be described by its moments \citep{aerts1992}. The first
moment $\langle \mathrm{v} \rangle$ represents the centroid velocity of the line profile, the second
moment $\langle \mathrm{v}^{2} \rangle$ the width of the line profile and the third moment
$\langle \mathrm{v}^{3} \rangle$ the skewness of the line profile. The quantity $\langle \mathrm{v} \rangle$ is thus a particular
measure of the radial velocity and should therefore show similar frequency
behaviour. We expect it to perform less well than the radial velocity measure derived from the optimum weight method \citep{bouchy2001a} (see below), but $\langle \mathrm{v} \rangle$ is a mode identification diagnostic \citep{aerts1992}, while other radial velocity measures are not. This is why we re-compute and re-analyse the radial velocity from $\langle \mathrm{v} \rangle$ here. 

Frequencies are determined with the conventional method of iterative sinewave
fitting (`prewhitening'). A Scargle periodogram \citep{scargle1982} is used with
frequencies between 0 and 15 cycles per day (c\,d$^{-1}$) (173.6 $\mu$Hz) and with a
frequency step of 0.0001 c\,d$^{-1}$ (0.001 $\mu$Hz) for the red giants. For the subgiant frequencies between 0 and 150 c\,d$^{-1}$ and a frequency step of 0.001 c\,d$^{-1}$ are used. The significance of the
frequencies is calculated with respect to the average amplitude of the
periodogram after prewhitening as described in \citet{kuschnig1997}. The errors in the frequency are obtained with the method described by \citet{breger1999} and have typical values of $10^{-4}$ c\,d$^{-1}$ ($10^{-9} \mu$Hz) for the giants and $10^{-3}$ c\,d$^{-1}$ ($10^{-8} \mu$Hz) for the subgiant.

The frequencies of $\langle \mathrm{v} \rangle$ are compared to the frequencies obtained from radial
velocities derived from the optimum weight method, mentioned in earlier publications. In
Figs~\ref{periodogramepsoph},~\ref{periodogrametaser}, \ref{periodogramxihy} and \ref{periodogramderi} the power spectra of \object{$\epsilon$ Ophiuchi},
\object{$\eta$ Serpentis}, \object{$\xi$ Hydrae} and \object{$\delta$ Eridani} are shown, respectively. The
lower black power spectra are the ones obtained by \citet{deridder2006b},
\citet{carrier2006a}, \citet{frandsen2002} and \citet{carrier2003}, while the grey ones in the middle
are obtained from $\langle \mathrm{v} \rangle$ in the present work. 

Although the same data is used, a comparison is made between power spectra of two different representations of the radial velocity variations, computed with different methods. The radial velocity derived from the optimum weight method is intrinsically more precise than the one from $\langle \mathrm{v} \rangle$ of the cross correlation. It not only uses the full spectrum instead of a box shaped mask, but also relies on weights of the individual spectra. These differences influence the noise and the amplitudes in the power spectra. On top of that the reference point of each night is determined by the observation with the highest signal to noise ratio in case of the optimum weight method and by the average of the night in case $\langle \mathrm{v} \rangle$ is used. As nightly reference points are effectively a high pass filter, different filters induce differences at low frequencies in the power spectra.

Despite these differences the comparison between the power spectra in the literature and obtained from
$\langle \mathrm{v} \rangle$ is very good, especially for \object{$\epsilon$ Ophiuchi} and \object{$\delta$ Eridani}. For
\object{$\eta$ Serpentis}, the dominant frequencies obtained by
\citet{carrier2006a} are slightly different. This is mainly due to the additional
ELODIE data, which contains a substantial number of observations and therefore alters the time sampling of the data. In case only the CORALIE data are taken into account, the dominant
frequencies match very well. For $\xi$ Hydrae the dominant frequencies obtained
by \citet{frandsen2002} are more distinct than those obtained
from $\langle \mathrm{v} \rangle$. This is due to the low amplitude of this star, which implies a larger relative difference between the different computations of the radial velocity than for the other two giants.
Nevertheless, the overall shape of the power spectra obtained from the radial velocity and from $\langle \mathrm{v} \rangle$ is comparable. The
fact that $\langle \mathrm{v} \rangle$ shows the same behaviour as the radial velocity obtained with
the method described by \citet{bouchy2001a}, indicates that this moment
diagnostic is able to detect low amplitude oscillations of
red (sub)giants.

The behaviour of $\langle \mathrm{v}^{2} \rangle$ obtained from a cross correlation is not exactly the
same as for a single spectral line. \citet{chadid2001} show that the constant of
a sinusoidal fit through $\langle \mathrm{v}^{2} \rangle$ is different in case of a cross correlation
compared to a single spectral line. Furthermore, a fit through $\langle \mathrm{v}^{3} \rangle$
obtained from a cross correlation contains a non zero constant, which is not the
case for a single spectral line. This indicates that we should be very cautious
in interpreting the constants of the moments. \citet{chadid2001} attribute the
different behaviour of $\langle \mathrm{v}^{2} \rangle$ and $\langle \mathrm{v}^{3} \rangle$ obtained from cross
correlation to the influence of possible blending with very faint lines which alters the absolute values of $\langle \mathrm{v}^{2} \rangle $ and $\langle \mathrm{v}^{3} \rangle$, but not their variation.

For the observed red giants, $\langle \mathrm{v}^{2} \rangle$ shows variations in the average value
per night, which is probably caused by changing instrumental conditions.
Although a correction for this behaviour is applied by shifting the values of
each night to the average value of all observations, $\langle \mathrm{v}^{2} \rangle$ does not behave
as predicted from theory. Frequencies for $\langle \mathrm{v}^{2} \rangle $ are expected to be: $\nu_{i}$,
$\nu_{i}+\nu_{j}$ or $\nu_{i}-\nu_{j}$, with $\nu_{i,j}$ the frequencies
obtained in $\langle \mathrm{v} \rangle$ \citep{mathias1994}, but are not found in the observations. For all four stars, the frequencies obtained for $\langle \mathrm{v}^{3} \rangle$ are the same as for $\langle \mathrm{v} \rangle$, which is as expected. 

\subsubsection{Line Bisector}
The line bisector is a measure of the displacement of the centre of the red and
blue wing from the core of the spectral line at each residual flux. The line
bisector is often characterised by the bisector velocity span, which is defined
as the horizontal distance between the bisector positions at fractional flux
levels in the top and bottom part of the spectral line, see for instance
\citet{brown1998}. The bisector velocity span is supposed to remain constant
over time in case of substellar companions.

The bisector velocity span is calculated in this work as the difference between
the bisector at a fractional flux level of 0.80 and 0.20. The frequencies of the
bisector velocity span are determined in the same way as the frequencies of the moments.

Bisector velocity spans are calculated for simulated spectral lines of stars pulsating with different modes, using \object{$\epsilon$ Ophiuchi}'s amplitude. In case of noiseless line profiles and infinite mode lifetimes, the dominant frequency of modes with $m=0$ is $2\nu$, with $\nu$ the dominant frequency obtained for $\langle \mathrm{v} \rangle$. In case $m\neq0$, $\nu$ is obtained as the dominant frequency. For infinite mode lifetimes the dominant frequency obtained for modes with $m=0$ is low, but at $2\nu$ the power spectrum also shows a clear excess. For modes with $m\neq0$ we obtained $\nu$ as the dominant frequency. This is completely in line with the behaviour expected for $\langle \mathrm{v}^{2} \rangle$ \citep{aerts1992}.

To give an estimate of the minimum signal to noise ratio needed to detect a dominant frequency, with the amplitude detected for \object{$\epsilon$ Ophiuchi} in the bisector velocity span, we added white noise to the simulated spectral lines. These simulations reveal that for pulsations with infinite as well as finite mode lifetimes with $m=0$, a signal to noise ratio of 100\,000 is not enough to reveal a frequency in the bisector velocity span, while a signal to noise ratio of order a few 10\,000 would suffice for modes with $m\neq0$. The signal to noise ratio of the cross correlation profiles of our targets does not exceed a few 1\,000 (see Equation~\ref{SNRCCF}). This signal to noise level is much less than the minimum value needed to detect the oscillations.

The top graphs in Figs~\ref{periodogramepsoph},~\ref{periodogrametaser},~\ref{periodogramxihy} and \ref{periodogramderi} show the power spectra of the bisector velocity span for
\object{$\epsilon$ Ophiuchi}, \object{$\eta$ Serpentis}, \object{$\xi$
Hydrae} and \object{$\delta$ Eridani}, respectively. The bisector velocity spans for \object{$\epsilon$
Ophiuchi} and \object{$\eta$ Serpentis} do not show any dominant frequency. The
power spectrum of the bisector velocity span of \object{$\xi$ Hydrae} shows
peaks at 1 c\,d$^{-1}$ ($11.57\mu$Hz) and at its 1 day aliases
(Fig.~\ref{periodogramxihy} top). The aliases are due to the diurnal cycle and
in case these are removed no dominant frequencies in the bisector velocity span
can be seen (Fig.~\ref{periodogramxihy} second power spectrum from the top). For \object{$\delta$ Eridani} the bisector velocity span shows some peaks at low frequencies, but from asteroseismological considerations solar like pulsations can not occur at these low frequencies in subgiants. These peaks are probably caused by instrumental effects.
From the present test case, one can thus conclude that, at least for low amplitude oscillations, discrimination between exoplanet companions and pulsations is not possible with the bisector velocity
span on data with realistic signal to noise ratio. This result is consistent with \citet{dall2006}.

\section{Theoretical mode diagnostics}
In order to characterise wavenumbers $(\ell,m)$ for the modes present in the red (sub)giants, the mathematical description of the line bisector \citep{brown1998} and of the moments \citep{aerts1992} are investigated.

\citet{brown1998} introduced a mathematical description for the line bisector in
terms of orthogonal Hermite functions. The Hermite coefficients $h_{i}$ describe
the line shapes.  Information on the oscillation mode, inclination and velocity
parameters of the star was obtained by \citet{brown1998} from a comparison with theoretically
generated line profile variations during a pulsation cycle, assuming infinite
mode lifetimes. To obtain the wavenumbers of the oscillations, the Hermite
coefficients or bisector velocity span obtained from observations are
compared with the ones calculated for the simulated line profiles. The main
drawback of this description in Hermite functions is the lack of a pulsation
theory directly coupled to the Hermite coefficients or bisector velocity span.
For this reason we do not use it here.

The variations in the observed moments
of the line profiles are compared with their theoretical expectations derived
from oscillation theory \citep{aerts1992}. The moments are a function of wavenumbers $\ell,m$,
inclination $i$, pulsation velocity amplitude $\upsilon_{p}$, projected rotational
velocity $\upsilon\sin i$ and intrinsic width of the line profile $\sigma$.
The comparison is performed objectively with a discriminant
(\citet{aerts1992}, \citet{aerts1996}, \citet{briquet2003}) which selects the most likely set of
parameters ($\ell,m,i,\upsilon_{p},\upsilon\sin i,\sigma$). Due to the fact
that several combinations of the wavenumbers and velocity parameters result in
almost the same line profile variation, the ``few'' best solutions resulting
from the discriminant should be investigated carefully before drawing conclusions about the mode identity.

The fact that $\langle \mathrm{v} \rangle$ and $\langle \mathrm{v}^{3} \rangle$ are sensitive to the very small line
profile variations and the direct connection between the moments and pulsation
theory makes moments, in principle, suitable for the
analysis of pulsating red (sub)giants, provided that we test its robustness against small amplitudes and 
the finite lifetimes of the modes. This is what we have done in the present work.

\subsection{Discriminant}
In the present analysis the discriminant is determined for $\ell=0,1,2$,
inclination angles ranging from $5\degr$ until $85\degr$ with steps of $1\degr$,
projected rotational velocity $\upsilon\sin i$ and intrinsic width of the
line profile $\sigma$ between 0 and 5 km\,s$^{-1}$ with steps of 0.5 km\,s$^{-1}$. Furthermore a
limbdarkening coefficient $u(\lambda)$ of 0.5 is used \citep{vanhamme1993}.

Due to the fact that the red (sub)giants show low amplitude variations, the
discriminant might give inclination angles close to an inclination angle of
complete cancellation (IACC) \citep{chadid2001}. Complete cancellation occurs when the star is looked upon in such way that the contribution to $\langle \mathrm{v} \rangle$ of each point on the stellar disk exactly cancels out the contribution of another point on the stellar disk. Therefore the angles close to $0\degr$ and $90\degr$ are not taken into account, as well as solutions close to
the IACC at $i=54.7\degr$ for $\ell=2, m=0$.

As the discriminant is designed for oscillations inducing moment variations larger than the equivalent width of the line (moment of order zero), a check needs to be performed whether this technique is applicable in the present work.  Therefore three series of spectral line profiles with infinite lifetimes and different pulsation velocity amplitudes are generated using the 628 observation times of \object{$\epsilon$ Ophiuchi}.  The dominant
frequency of \object{$\epsilon$ Ophiuchi}, 5.03 c\,d$^{-1}$ (58.2 $\mu$Hz), is used as
an input parameter at an inclination of $i=35\degr$ for wavenumbers
$\ell=0,1,2$ with positive $m$ values, a projected rotational velocity $\upsilon\sin i=3.5$ km\,s$^{-1}$ and an intrinsic width $\sigma=4$ km\,s$^{-1}$.
Furthermore, the equivalent width of the spectral lines is taken to be 10 km\,s$^{-1}$, while the minimum amplitude of the pulsation velocity ($\upsilon_{p}$) is 0.04 km\,s$^{-1}$ and increases with a factor of 10 for the different series. For these series the moments are calculated as well as the discriminant. For a pulsation velocity amplitude of 0.04 km\,s$^{-1}$, which resembles the data best, $\langle \mathrm{v}^{2} \rangle$ of modes with $m=0$ all show the expected harmonic of the frequency \citep{mathias1994}, but with a different constant. The discriminant values only differ by an order of 0.01 km\,s$^{-1}$. This is also the case for a pulsation velocity amplitude of 0.4 km\,s$^{-1}$, which indicates that the discriminant is not suitable for the analysis of small line profile variations with an amplitude below 1 km\,s$^{-1}$. For a pulsation velocity amplitude of 4.0 km\,s$^{-1}$ the difference between the discriminant values increases and the correct mode is present among the best options. This indicates that the discriminant is a useful analysis tool for spectral line variations with amplitude larger than about half the equivalent width of the spectral line, but not for red (sub)giants.

\subsection{Amplitude and phase distribution}

\begin{figure*}
\begin{minipage}{4.25cm}
\centering
\includegraphics[width=4.25cm]{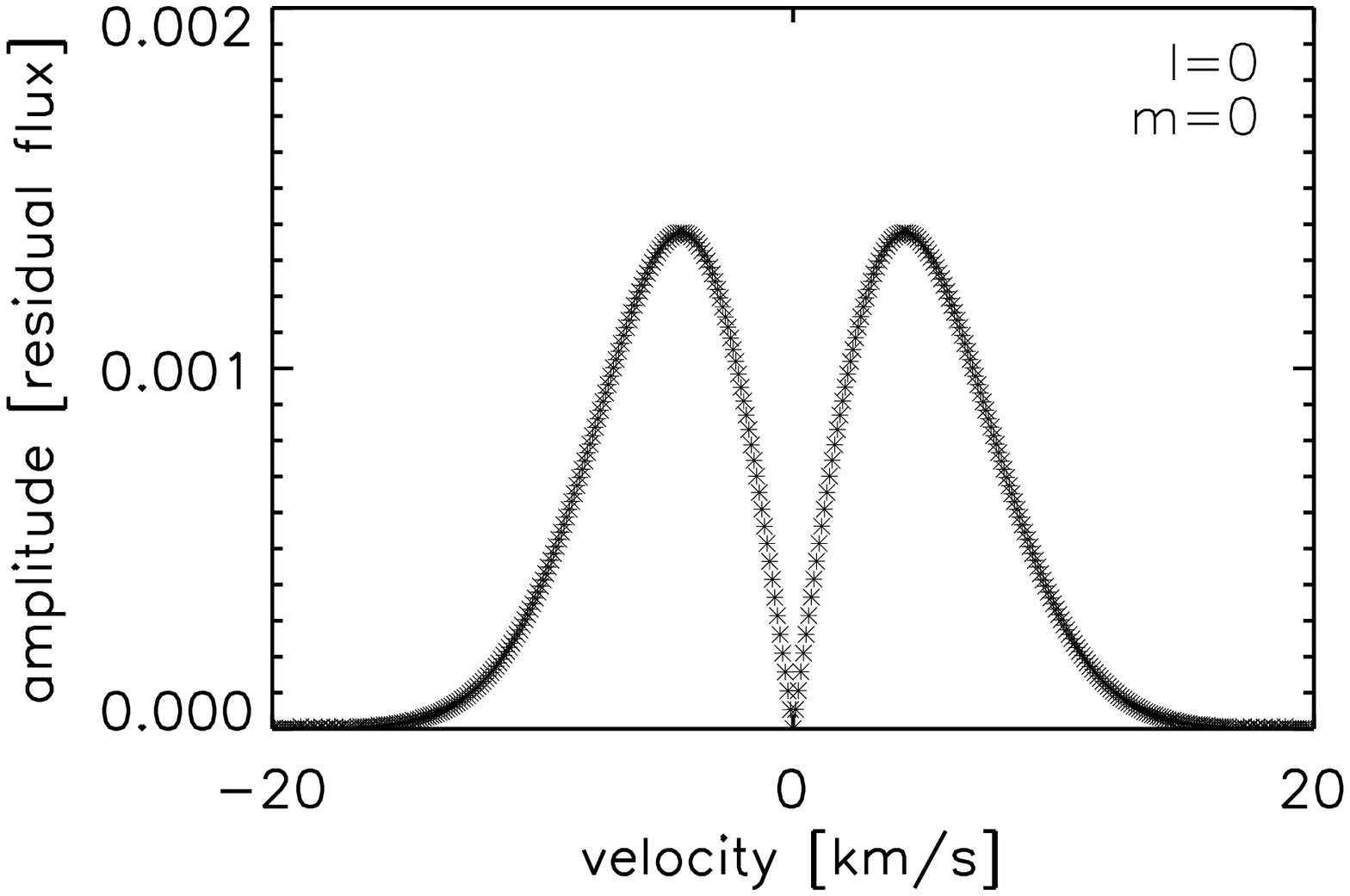}
\end{minipage}
\hfill
\begin{minipage}{4.25cm}
\centering
\includegraphics[width=4.25cm]{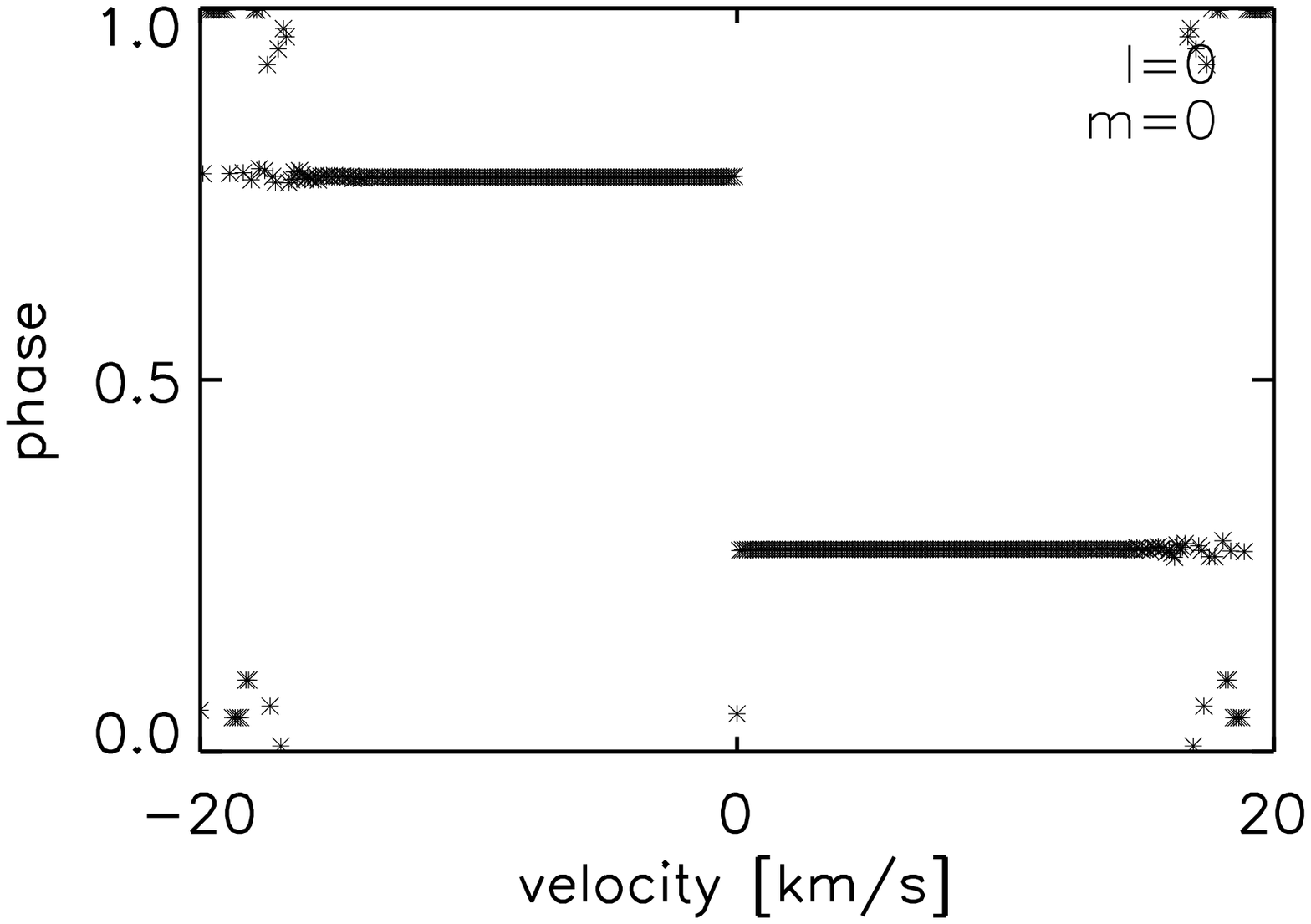}
\end{minipage}
\hfill
\begin{minipage}{4.25cm}
\centering
\includegraphics[width=4.25cm]{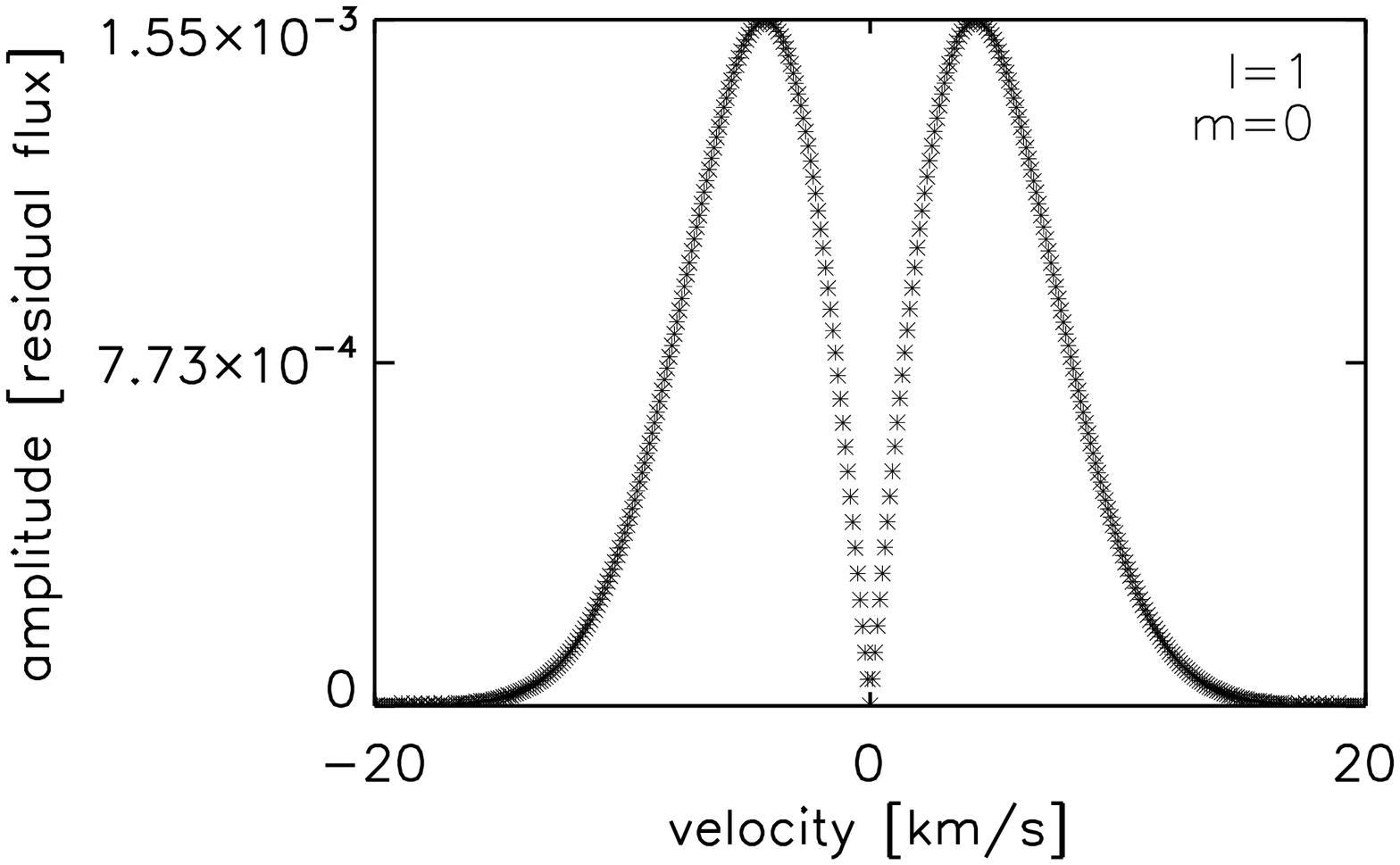}
\end{minipage}
\hfill
\begin{minipage}{4.25cm}
\centering
\includegraphics[width=4.25cm]{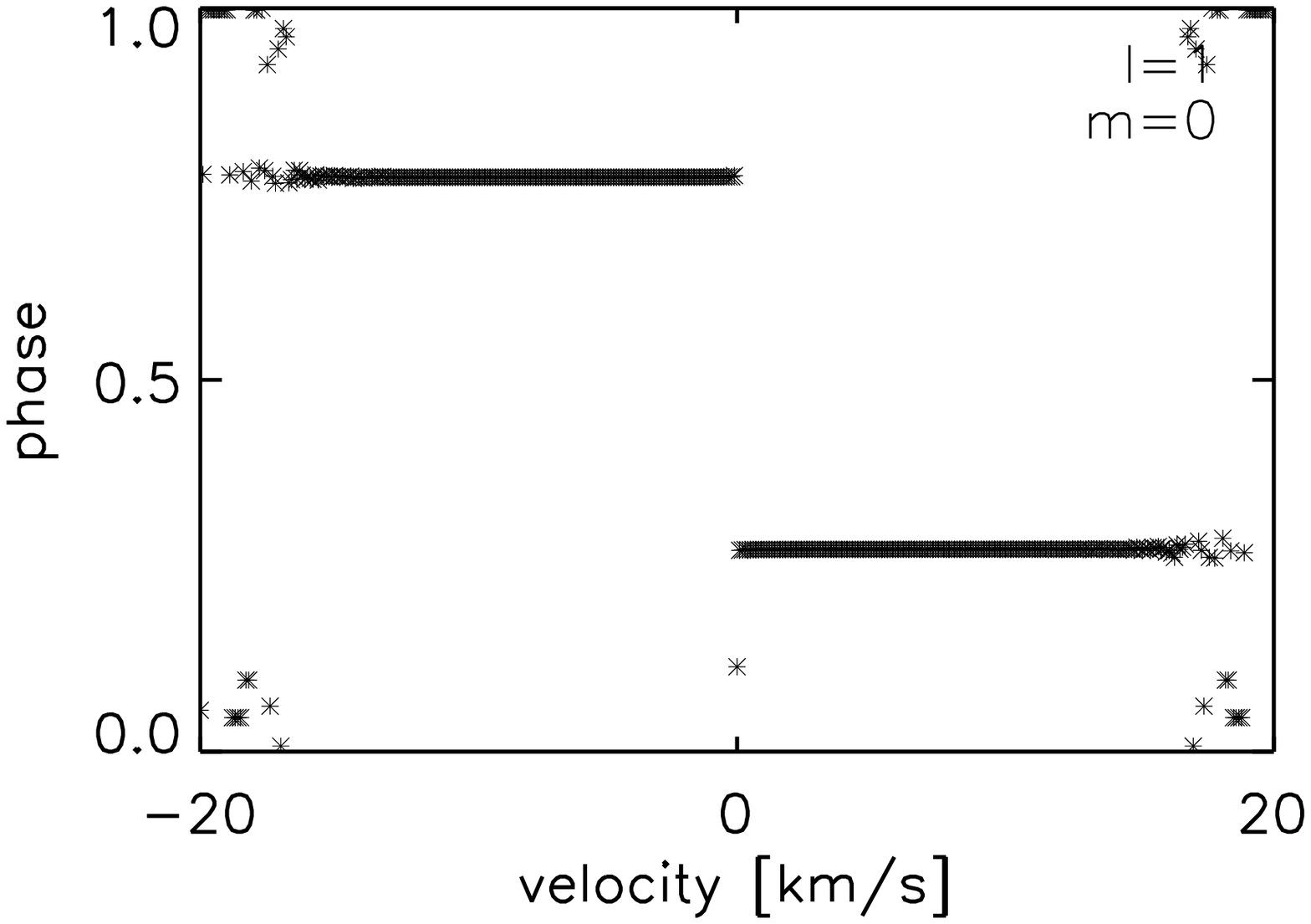}
\end{minipage}
\hfill
\begin{minipage}{4.25cm}
\centering
\includegraphics[width=4.25cm]{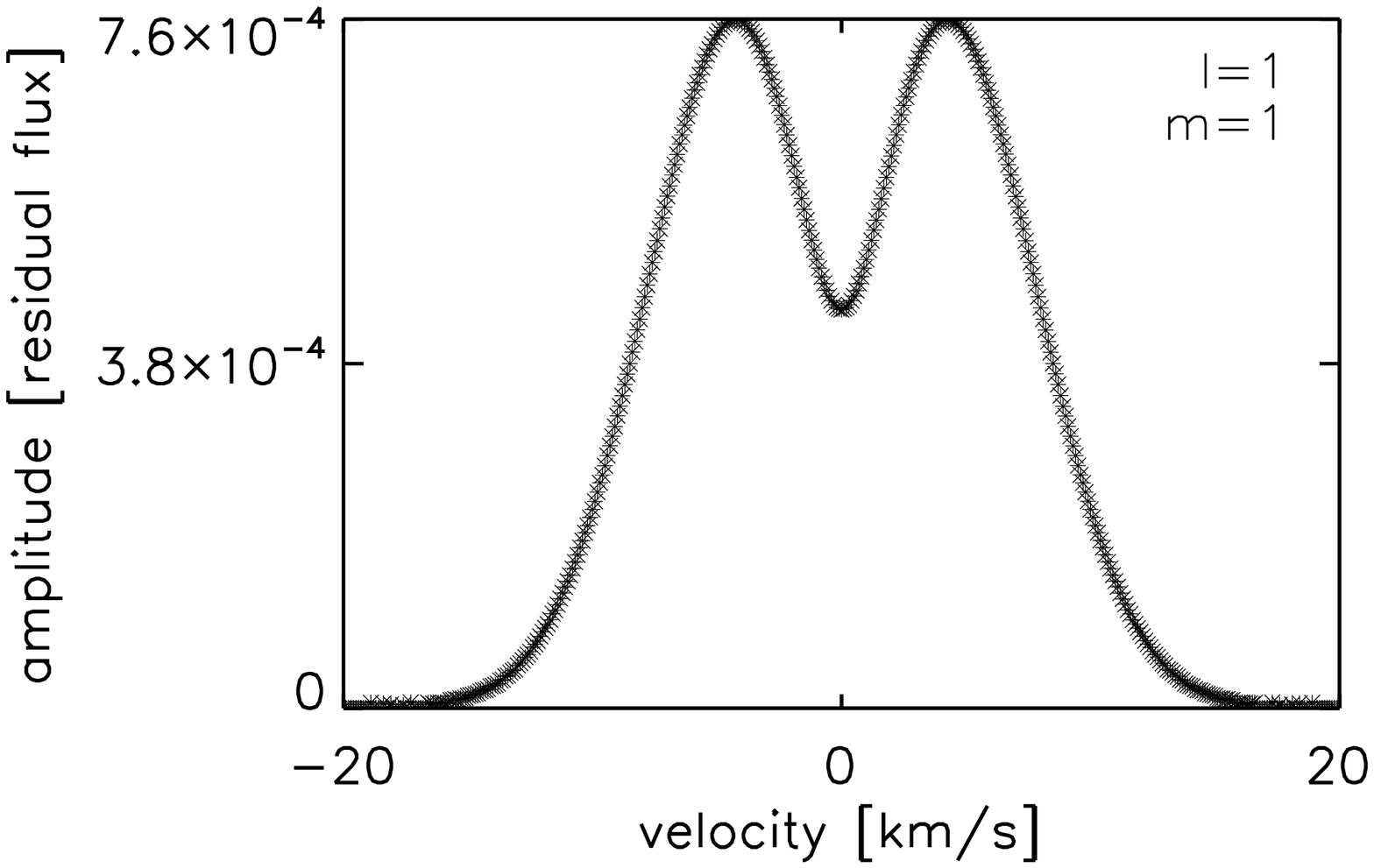}
\end{minipage}
\hfill
\begin{minipage}{4.25cm}
\centering
\includegraphics[width=4.25cm]{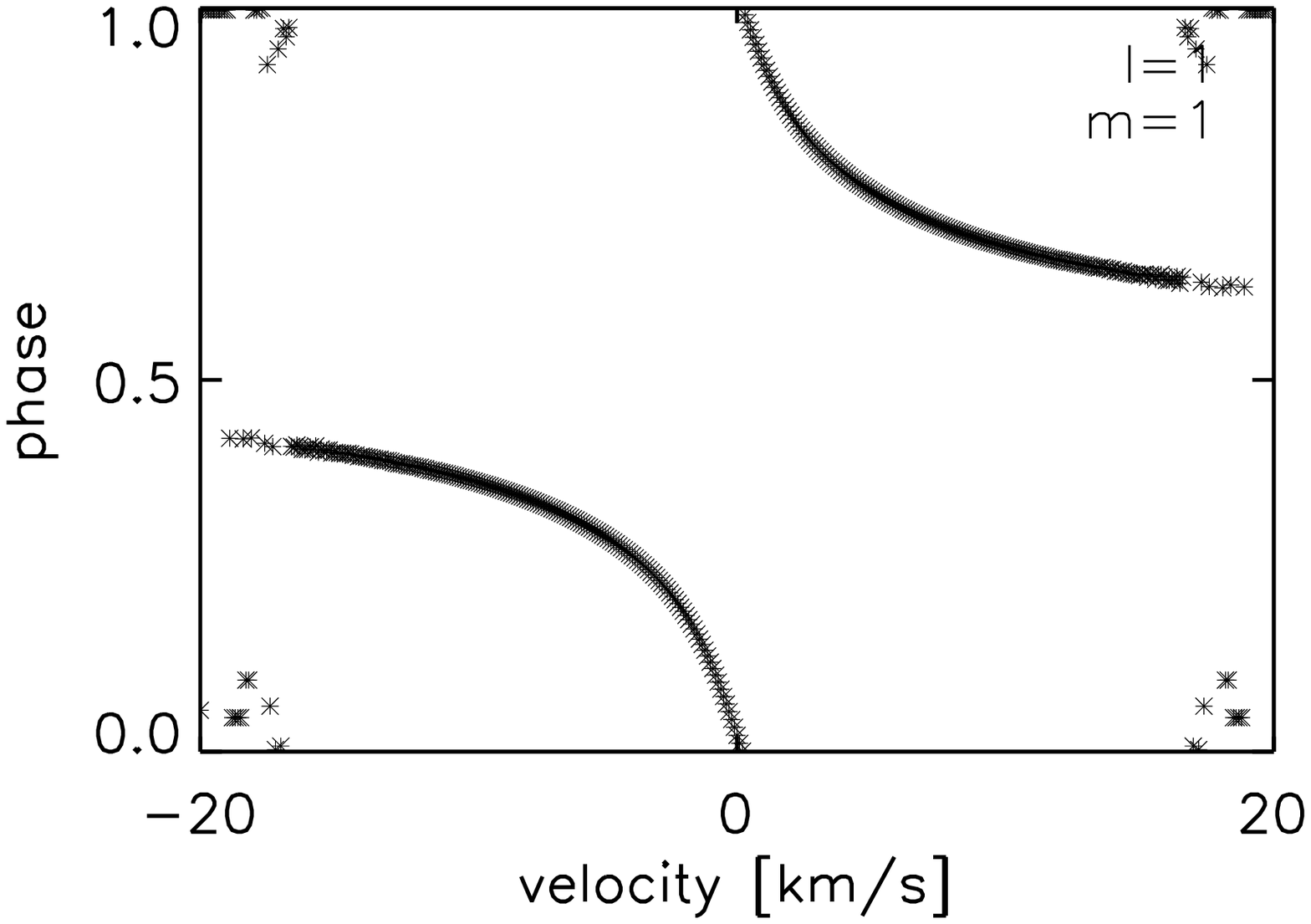}
\end{minipage}
\hfill
\begin{minipage}{4.25cm}
\centering
\includegraphics[width=4.25cm]{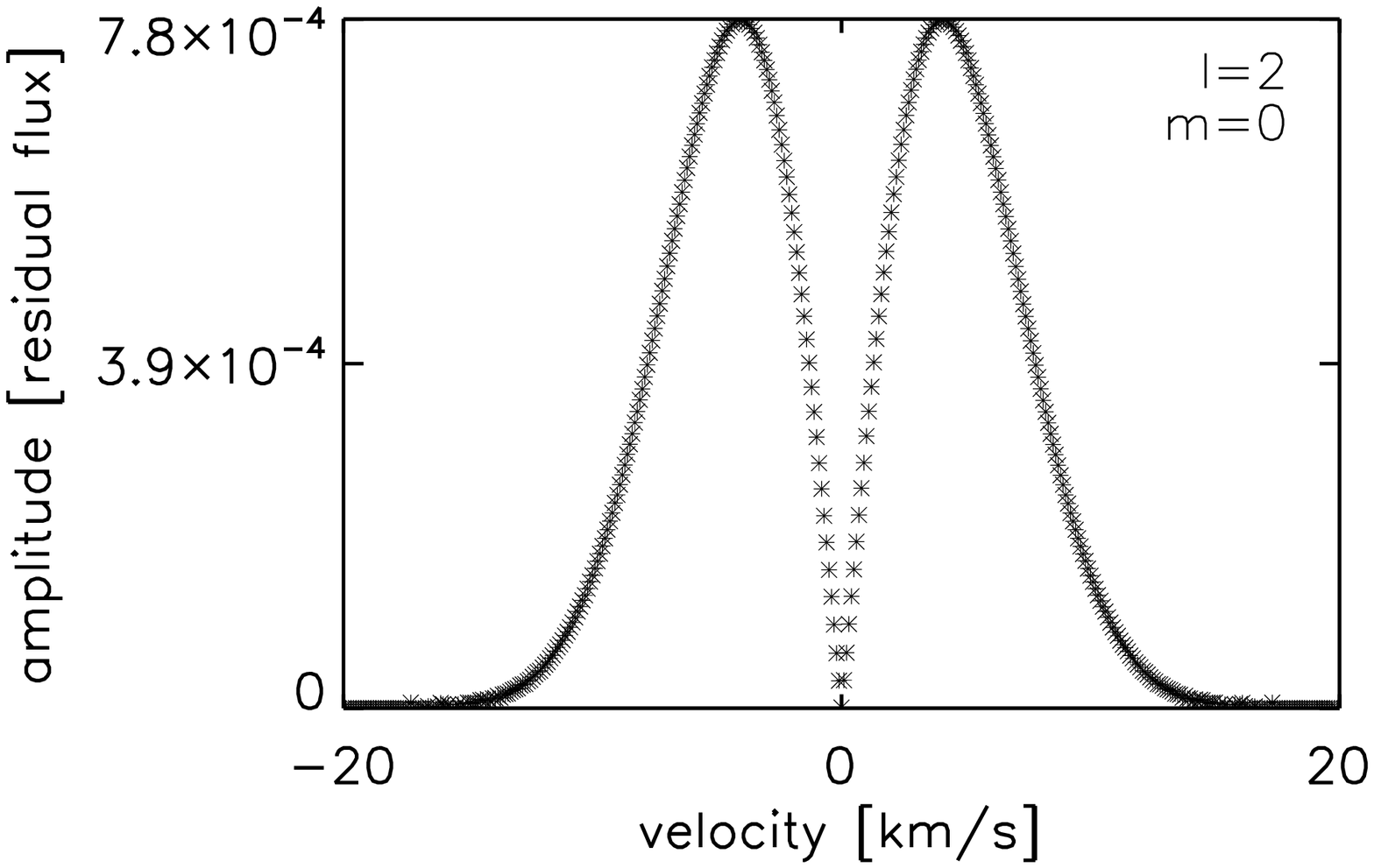}
\end{minipage}
\hfill
\begin{minipage}{4.25cm}
\centering
\includegraphics[width=4.25cm]{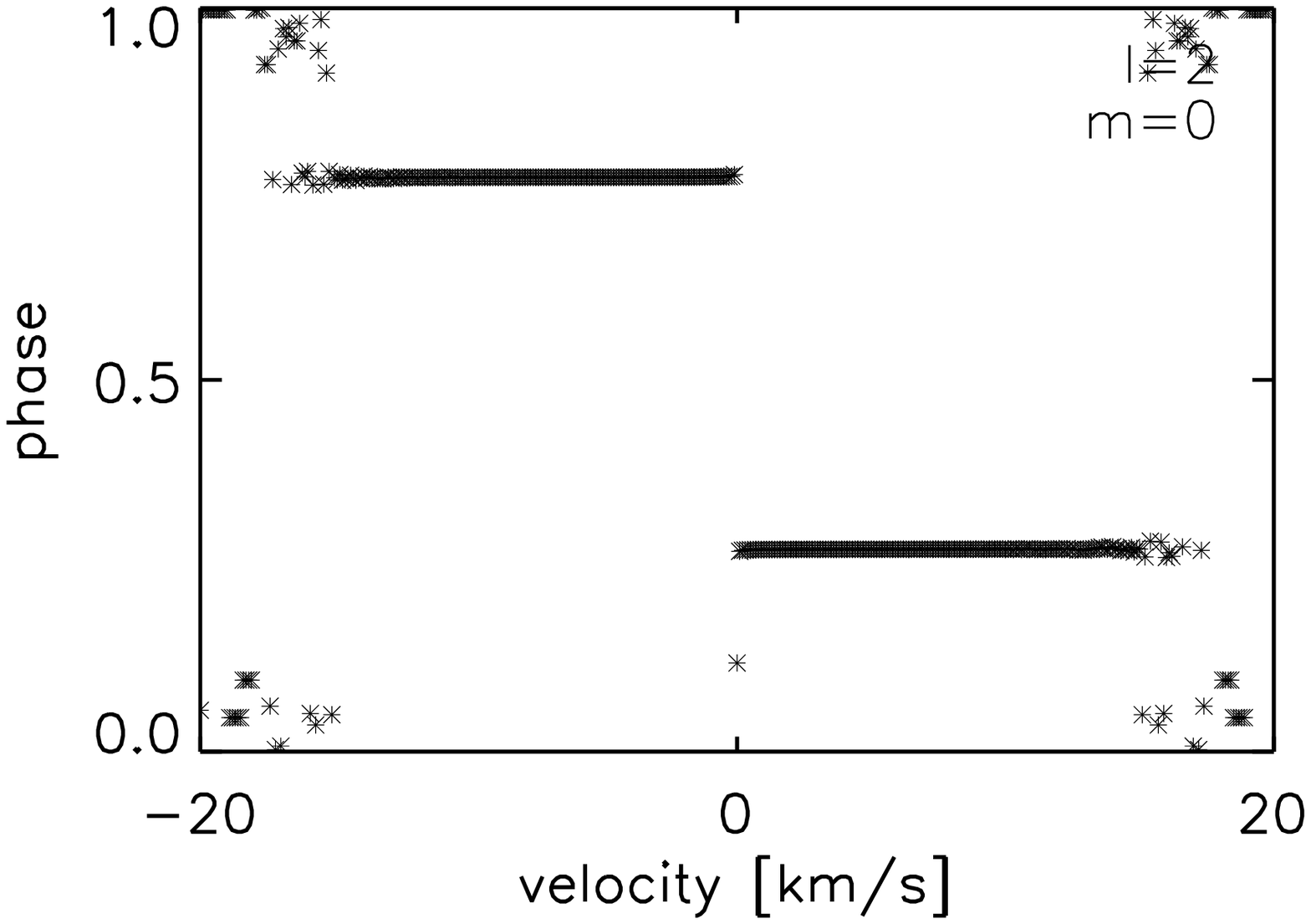}
\end{minipage}
\hfill
\begin{minipage}{4.25cm}
\centering
\includegraphics[width=4.25cm]{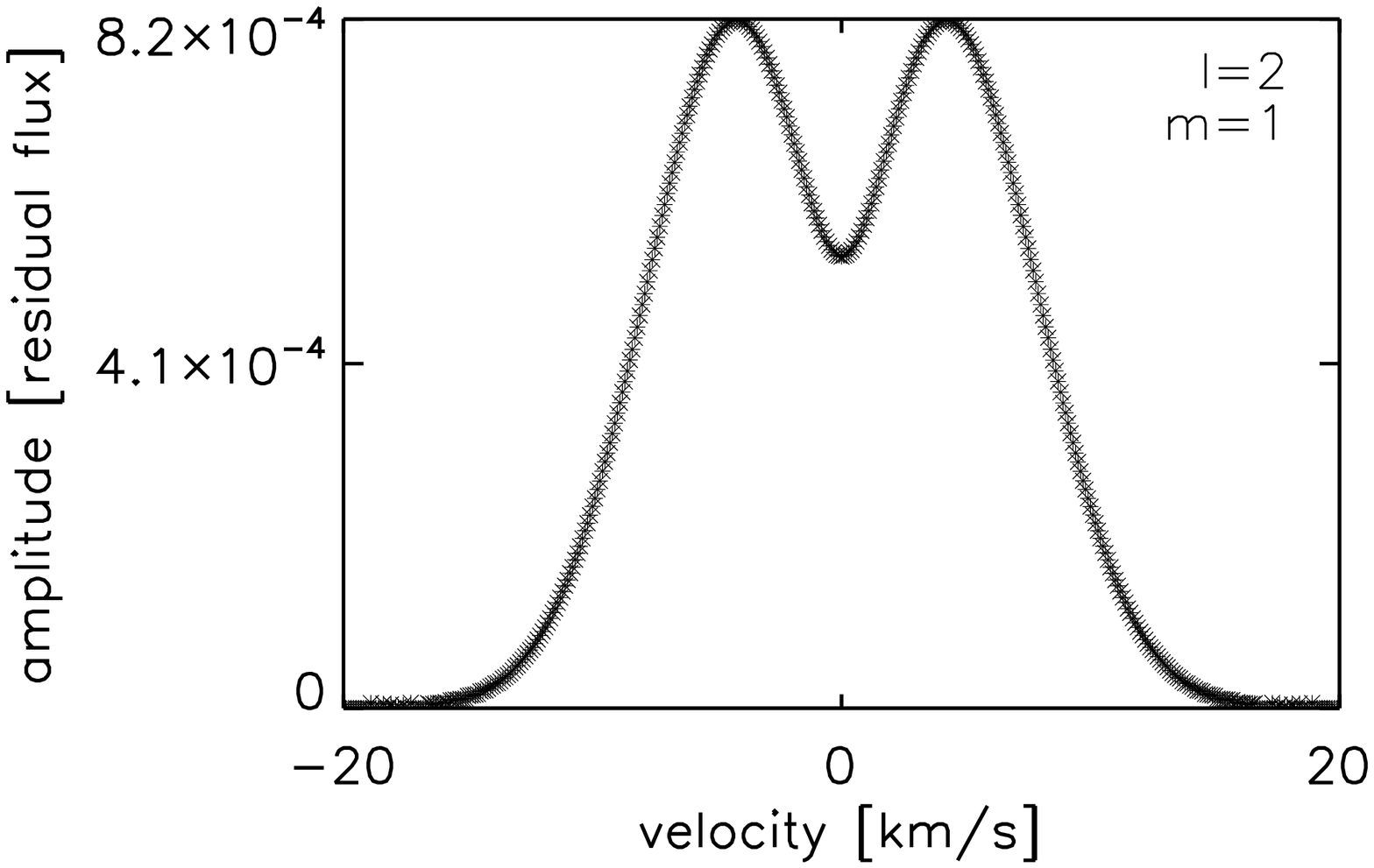}
\end{minipage}
\hfill
\begin{minipage}{4.25cm}
\centering
\includegraphics[width=4.25cm]{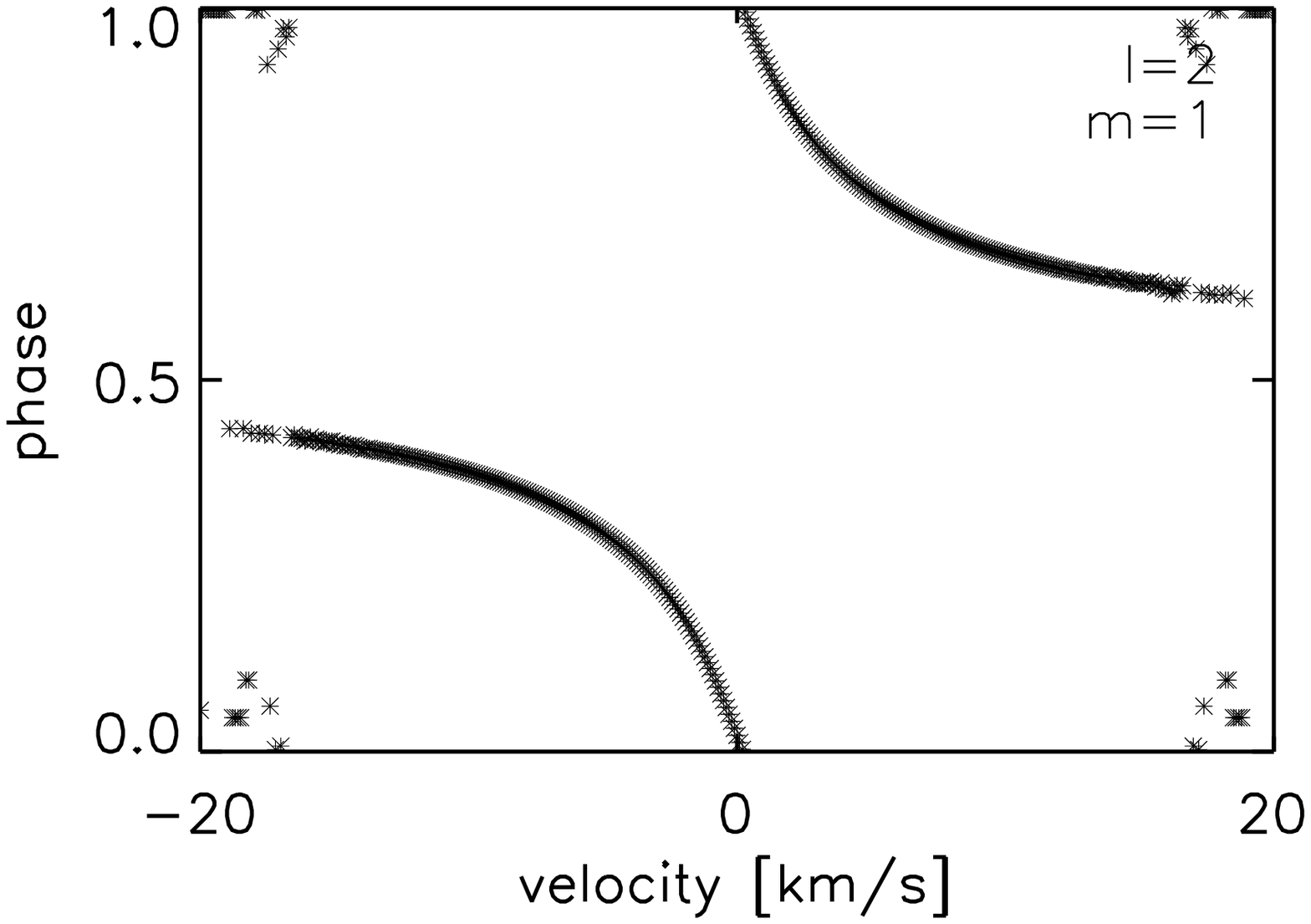}
\end{minipage}
\hfill
\begin{minipage}{4.25cm}
\centering
\includegraphics[width=4.25cm]{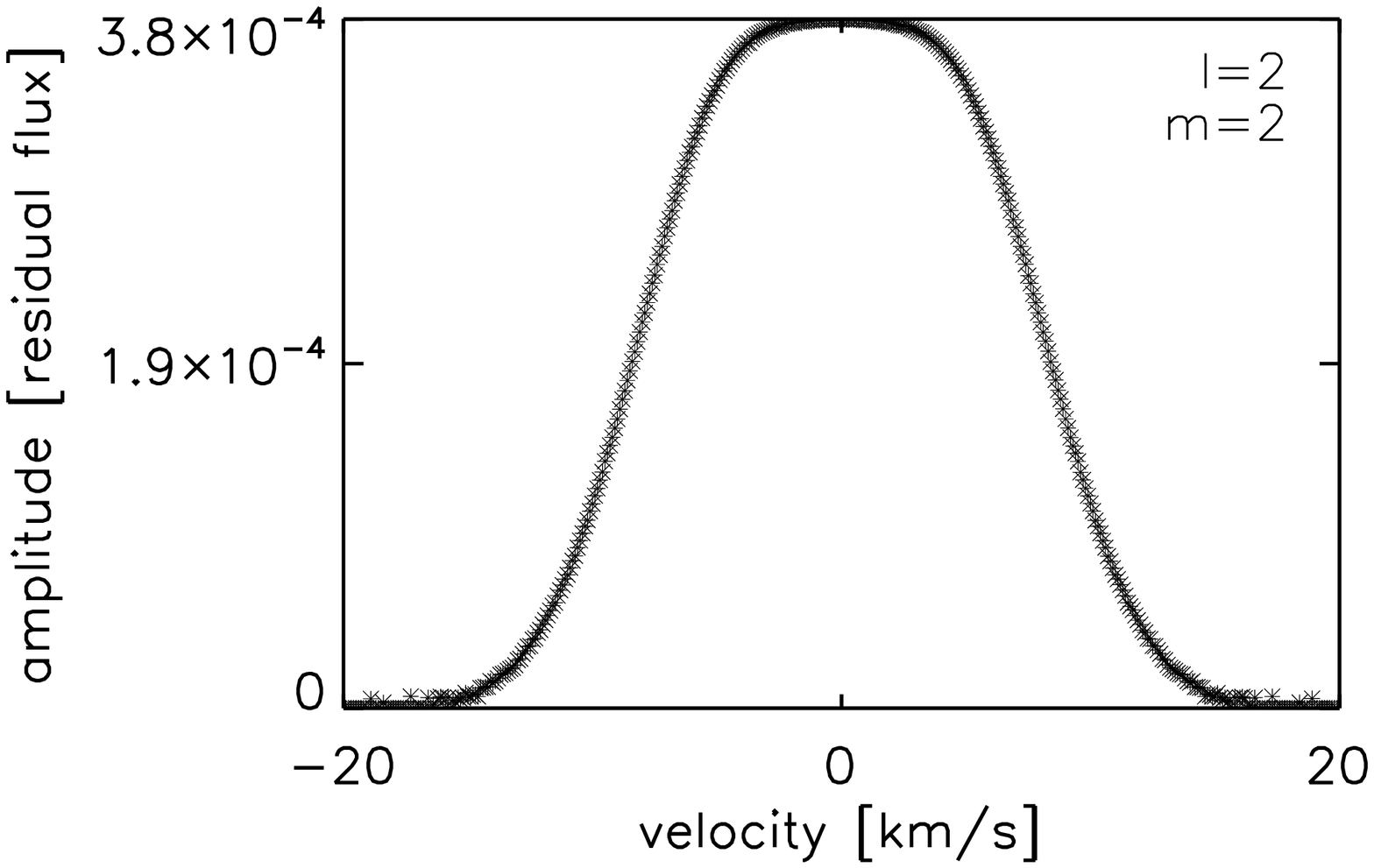}
\end{minipage}
\hfill
\begin{minipage}{4.25cm}
\centering
\includegraphics[width=4.25cm]{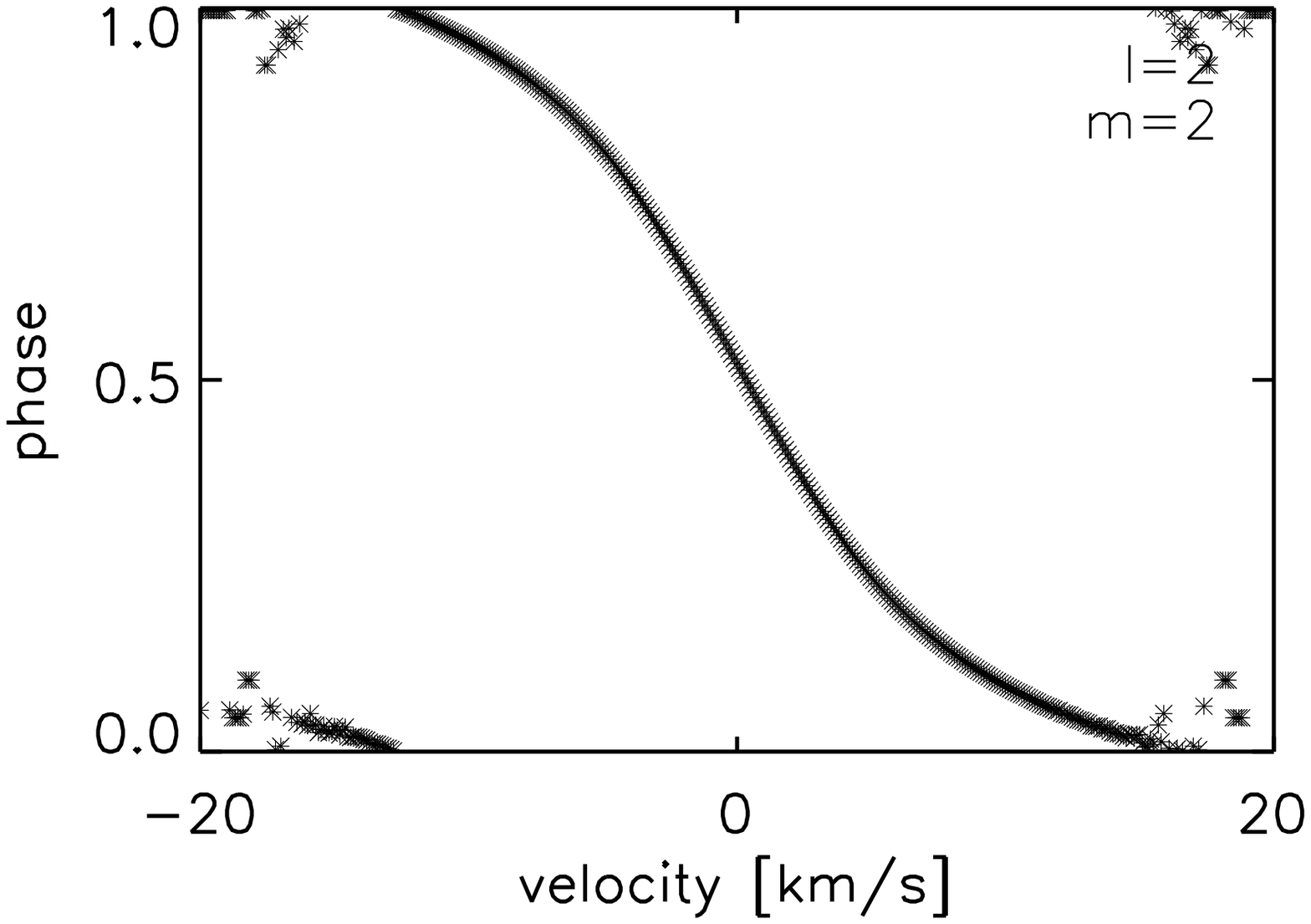}
\end{minipage}
\hfill
\caption{The amplitude (left) and phase (right) distributions for simulated line profiles with infinite damping time at an inclination angle of $i=35\degr$. Top left: $\ell=0$, $m=0$, top right: $\ell=1$, $m=0$, middle left: $\ell=1$, $m=1$, middle right: $\ell=2$, $m=0$, bottom left: $\ell=2$, $m=1$, bottom right: $\ell=2$, $m=2$}
\label{amplphasevelinfinite}
\end{figure*}

As the discriminant is not useful in the present analysis, the spectral lines are investigated by comparing the
amplitude and phase across the line profile with the ones obtained from simulations. This is the same procedure as used by \citet{telting1997} for \object{$\beta$ Cephei} and by \citet{decat2005} for the complex line profile behaviour due to the multi periodic
gravity mode oscillations in slowly pulsating B stars. The amplitude and phase 
for each velocity in the cross correlation profile are determined by fitting a
harmonic function, with the dominant frequencies of $\langle \mathrm{v} \rangle$, to the flux values at each velocity pixel of the time series of spectra \citep{schrijvers1997}. The amplitudes and phases as function of velocity determined from simulated line profiles with $\ell=0,1,2$, and positive $m$ values are shown in Fig.~\ref{amplphasevelinfinite}, assuming infinite lifetimes. The simulations are shown for $i=35\degr$, but the shape of the amplitude and phase distribution does not change significantly with inclination angle.

\begin{figure*}
\begin{minipage}{5.6cm}
\centering
\includegraphics[width=5.5cm]{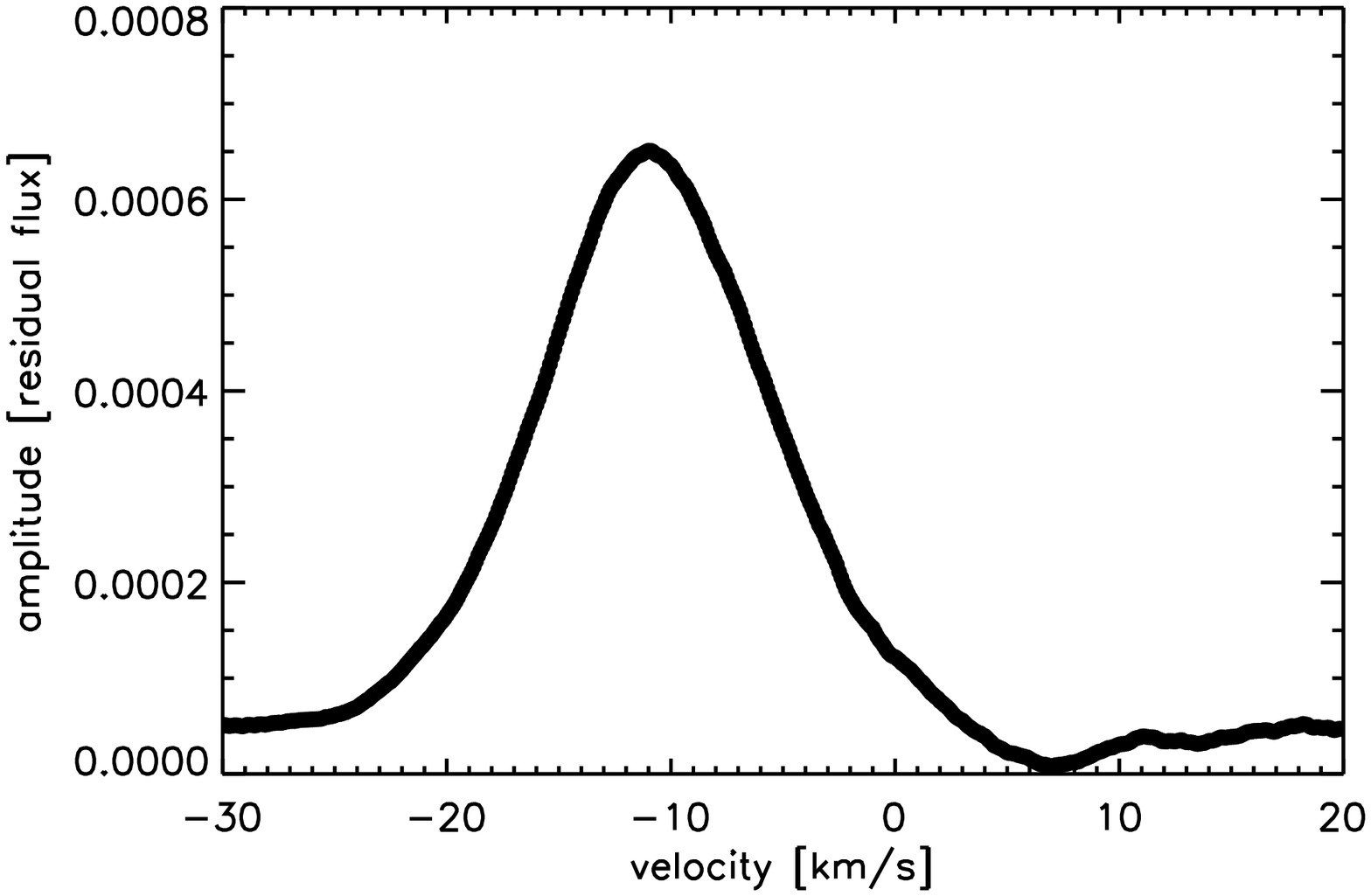}
\end{minipage}
\hfill
\begin{minipage}{5.6cm}
\centering
\includegraphics[width=5.5cm]{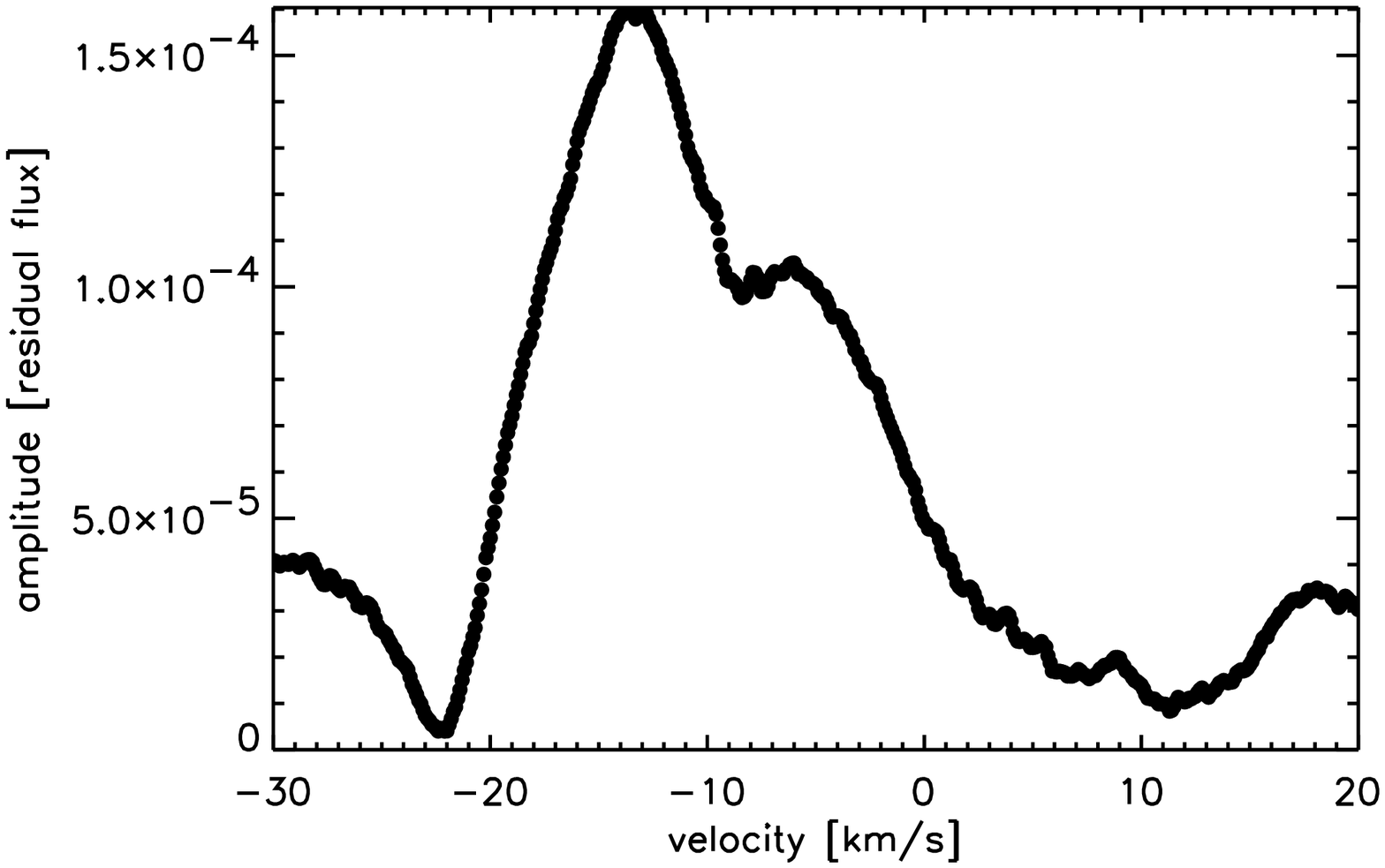}
\end{minipage}
\hfill
\begin{minipage}{5.6cm}
\centering
\includegraphics[width=5.5cm]{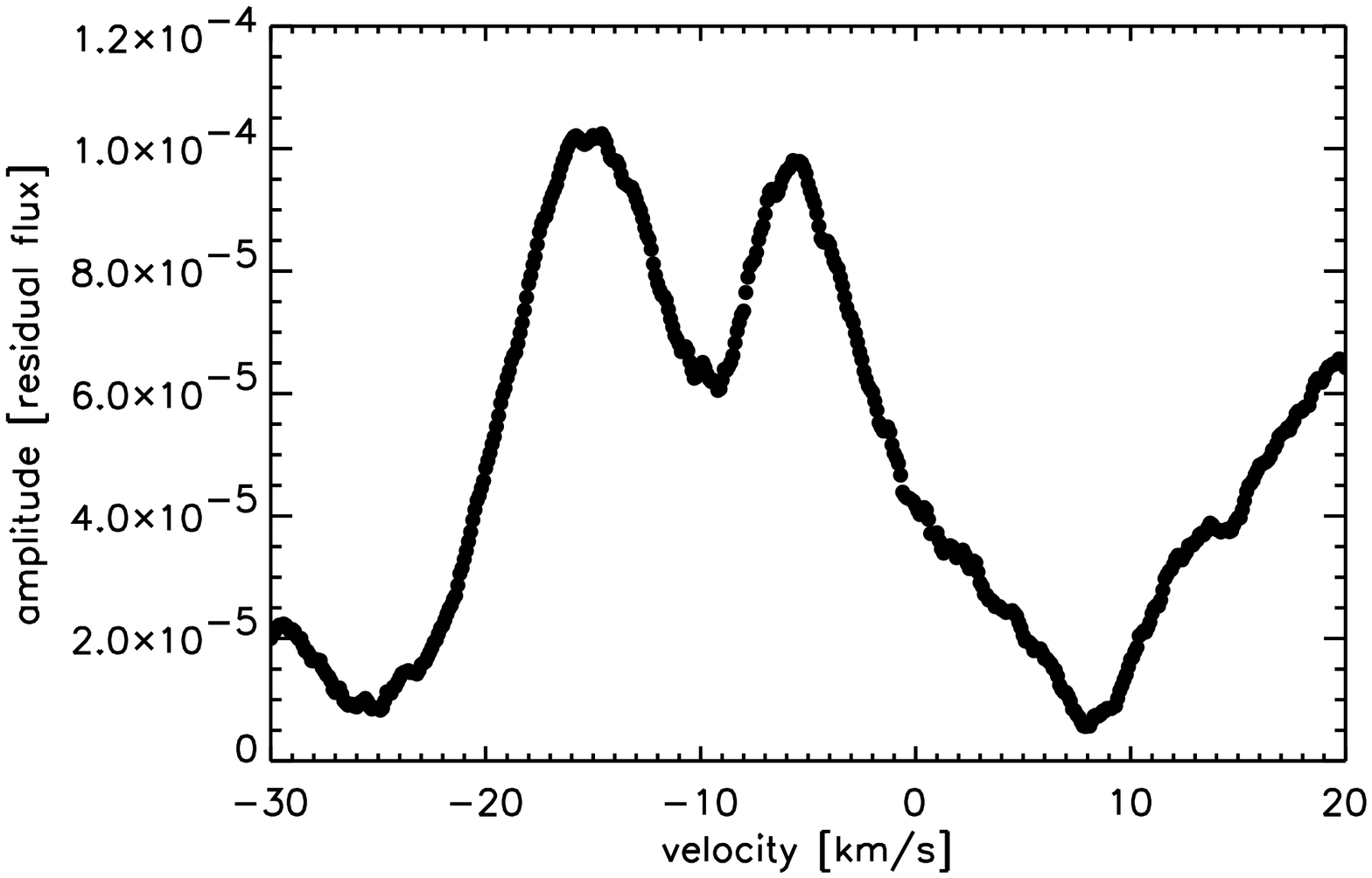}
\end{minipage}
\hfill
\begin{minipage}{5.6cm}
\centering
\includegraphics[width=5.5cm]{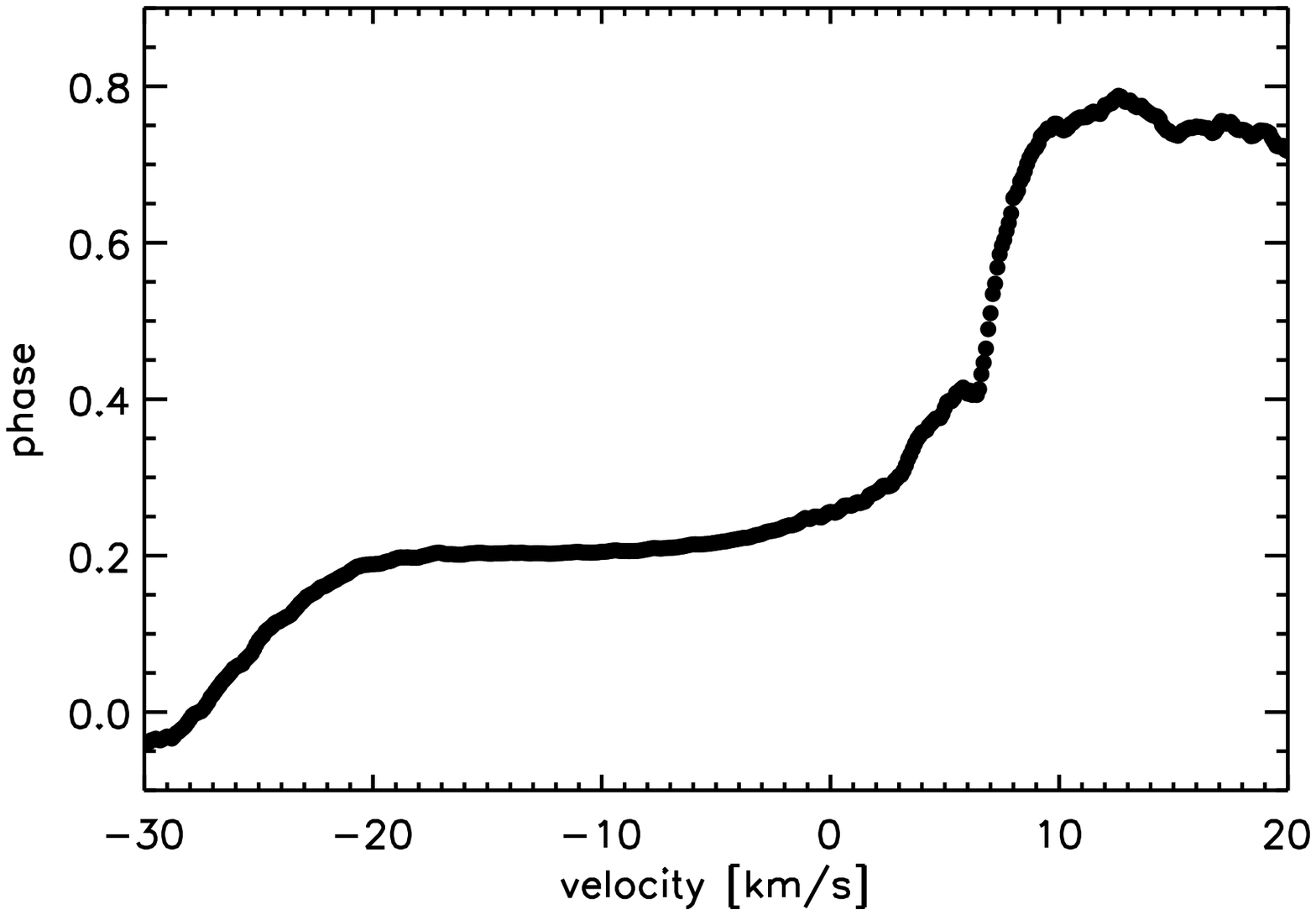}
\end{minipage}
\hfill
\begin{minipage}{5.6cm}
\centering
\includegraphics[width=5.5cm]{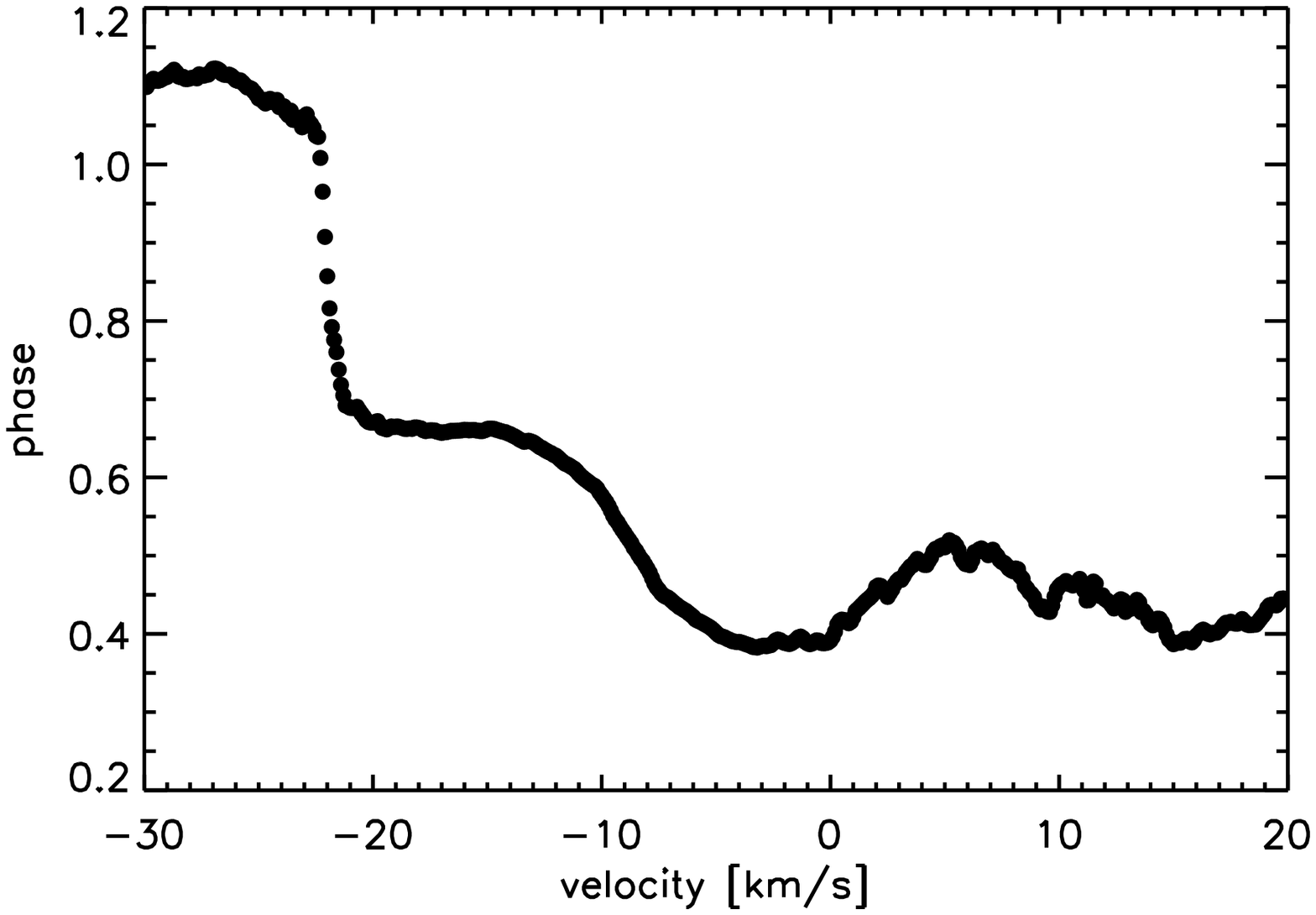}
\end{minipage}
\hfill
\begin{minipage}{5.6cm}
\centering
\includegraphics[width=5.5cm]{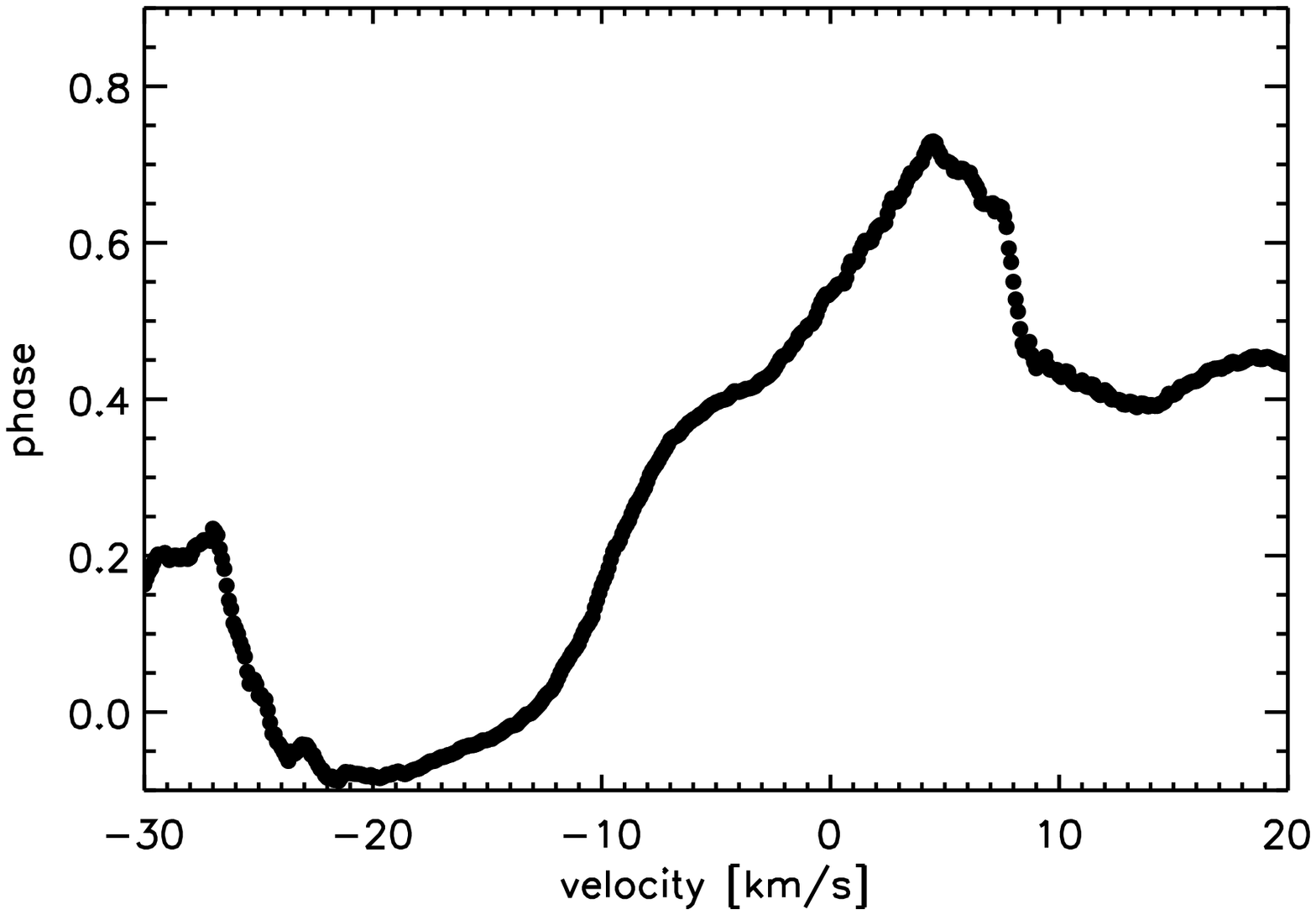}
\end{minipage}
\caption{The amplitude (top) and phase (bottom) distributions as a function of
velocity across the line profile of \object{$\epsilon$ Ophiuchi} for the three significant frequencies obtained from $\langle \mathrm{v} \rangle$: $\nu_{\langle \mathrm{v} \rangle}=5.03$ ~c\,d$^{-1}$ ($58.2 \mu$Hz) (left) $\nu_{\langle \mathrm{v} \rangle}=5.46$ ~c\,d$^{-1}$ ($63.2 \mu$Hz) (middle) and $\nu_{\langle \mathrm{v} \rangle}=5.83$ ~c\,d$^{-1}$ ($67.5 \mu$Hz) (right). The mean radial velocity of the star, is found to be approximately $-9.4$~km\,s$^{-1}$.}
\label{amplphaseHD146791}
\end{figure*}

\begin{figure*}
\begin{minipage}{5.6cm}
\centering
\includegraphics[width=5.5cm]{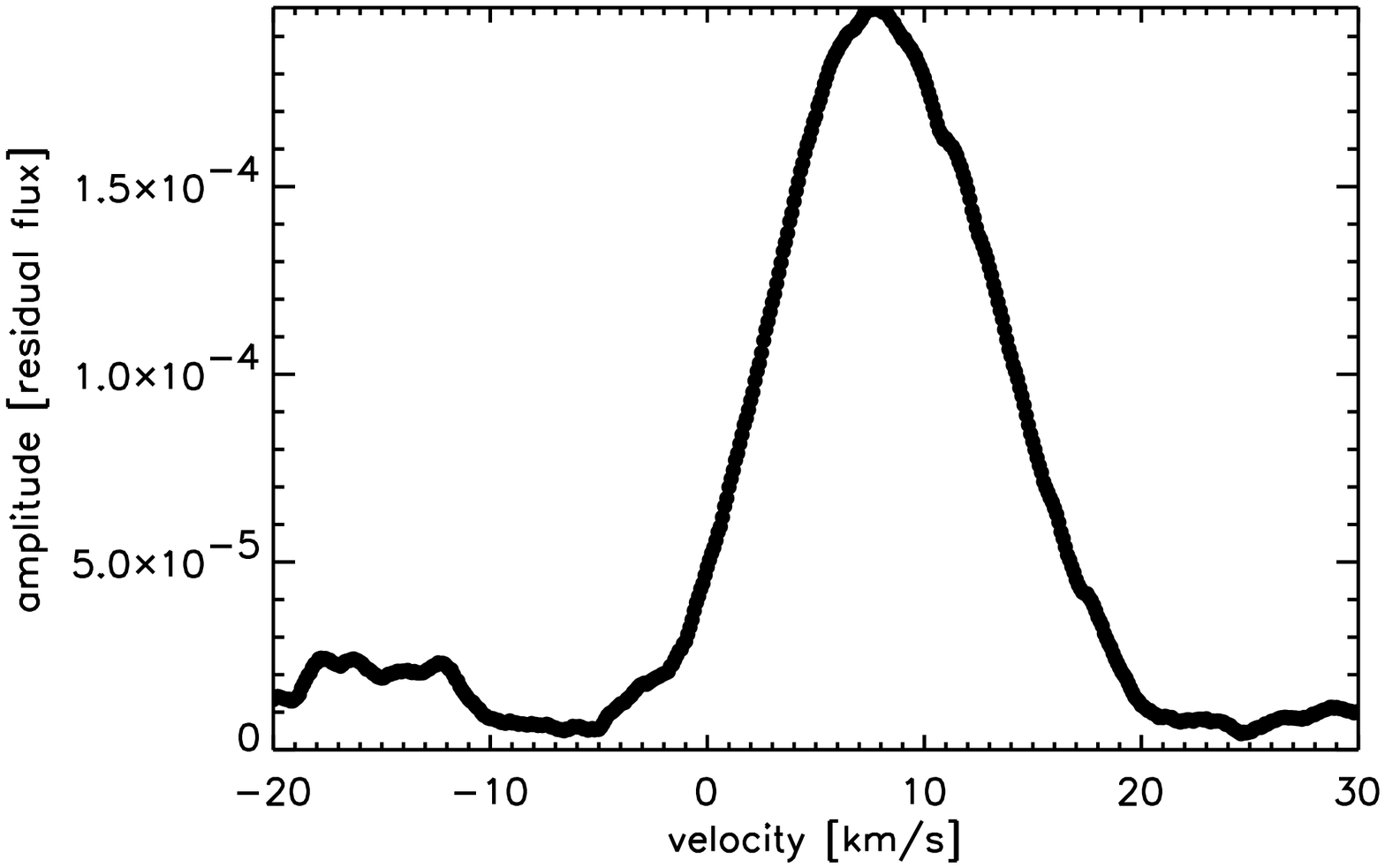}
\end{minipage}
\hfill
\begin{minipage}{5.6cm}
\centering
\includegraphics[width=5.5cm]{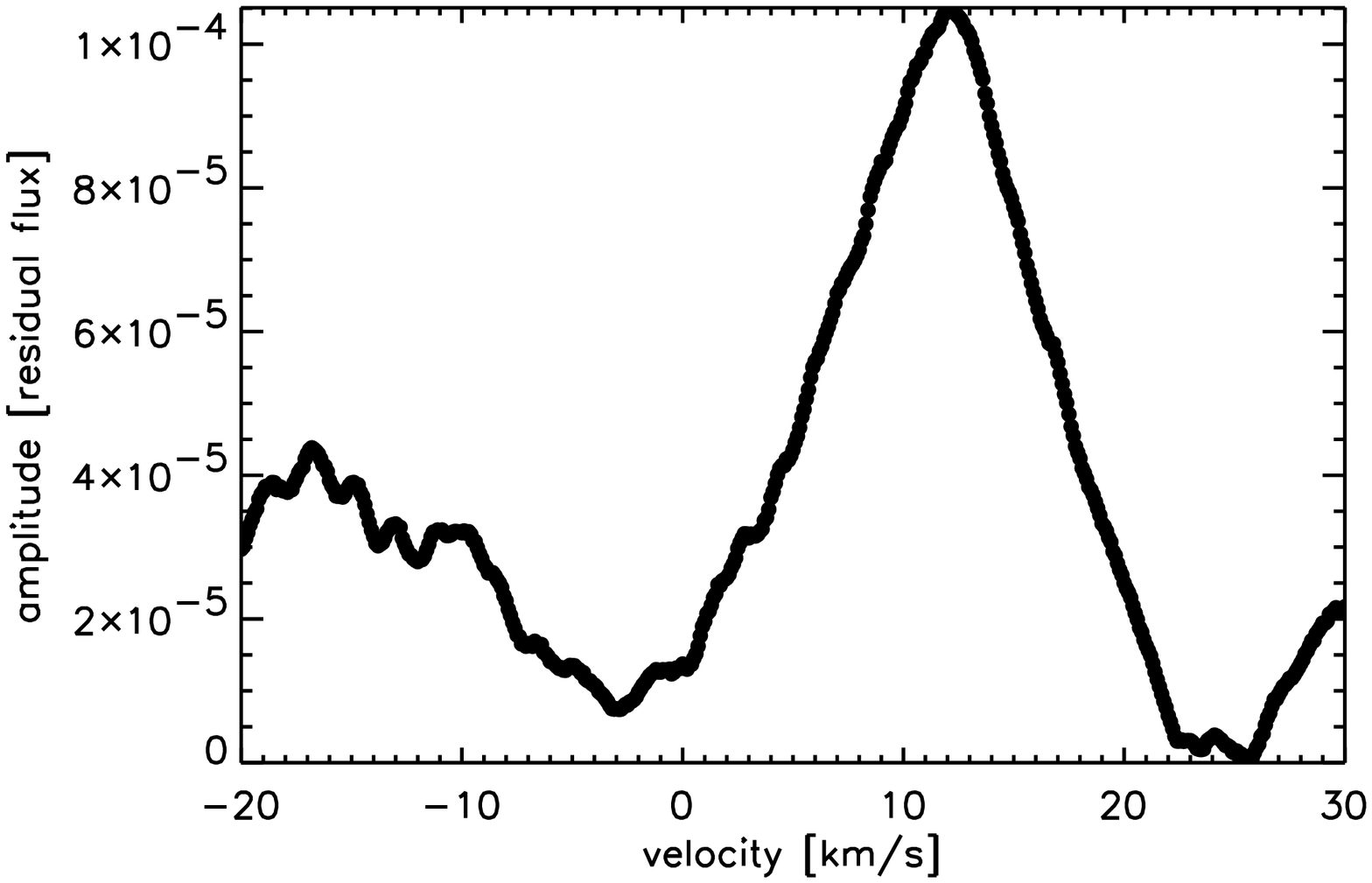}
\end{minipage}
\hfill
\begin{minipage}{5.6cm}
\centering
\includegraphics[width=5.5cm]{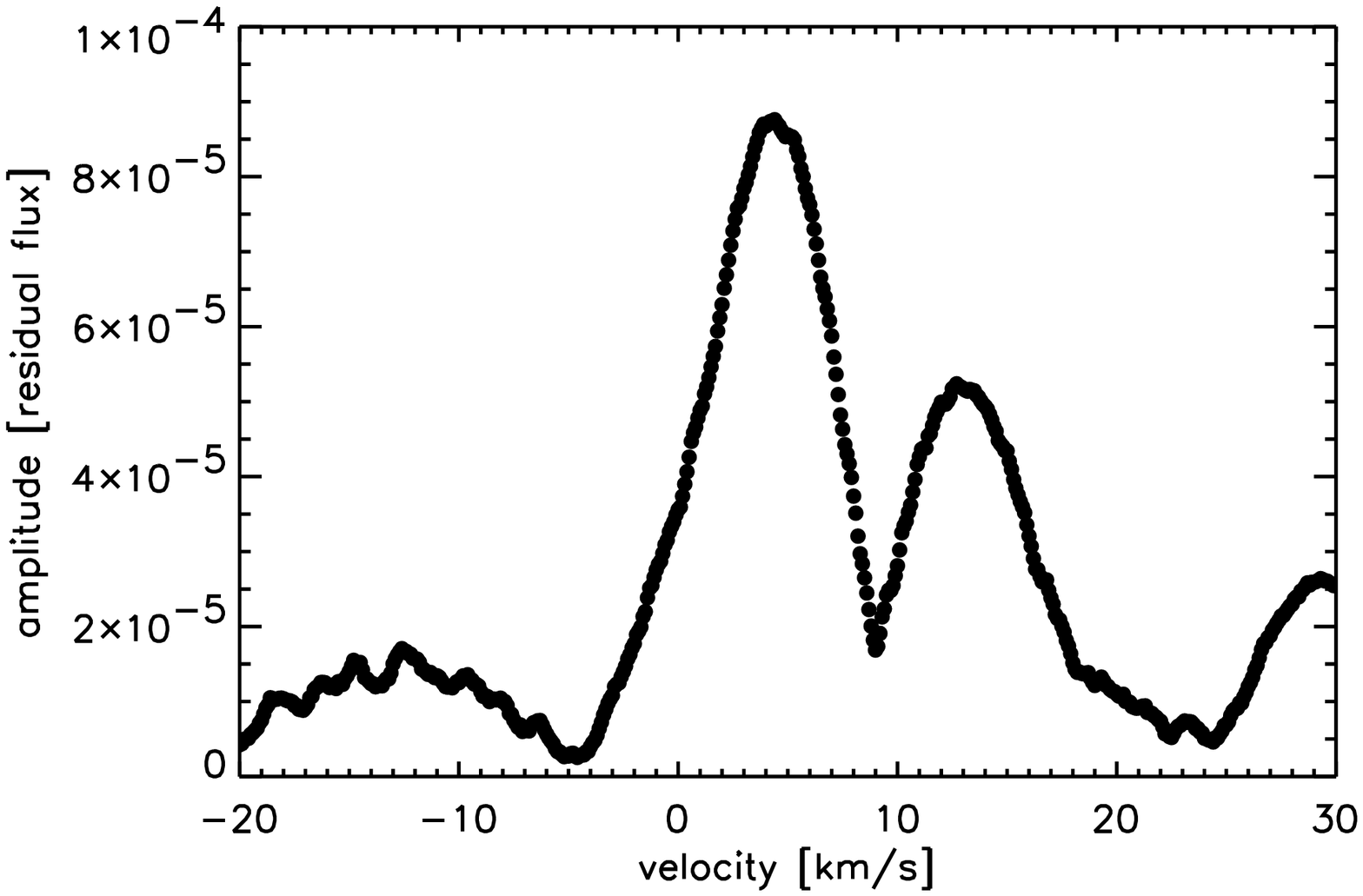}
\end{minipage}
\hfill
\begin{minipage}{5.6cm}
\centering
\includegraphics[width=5.5cm]{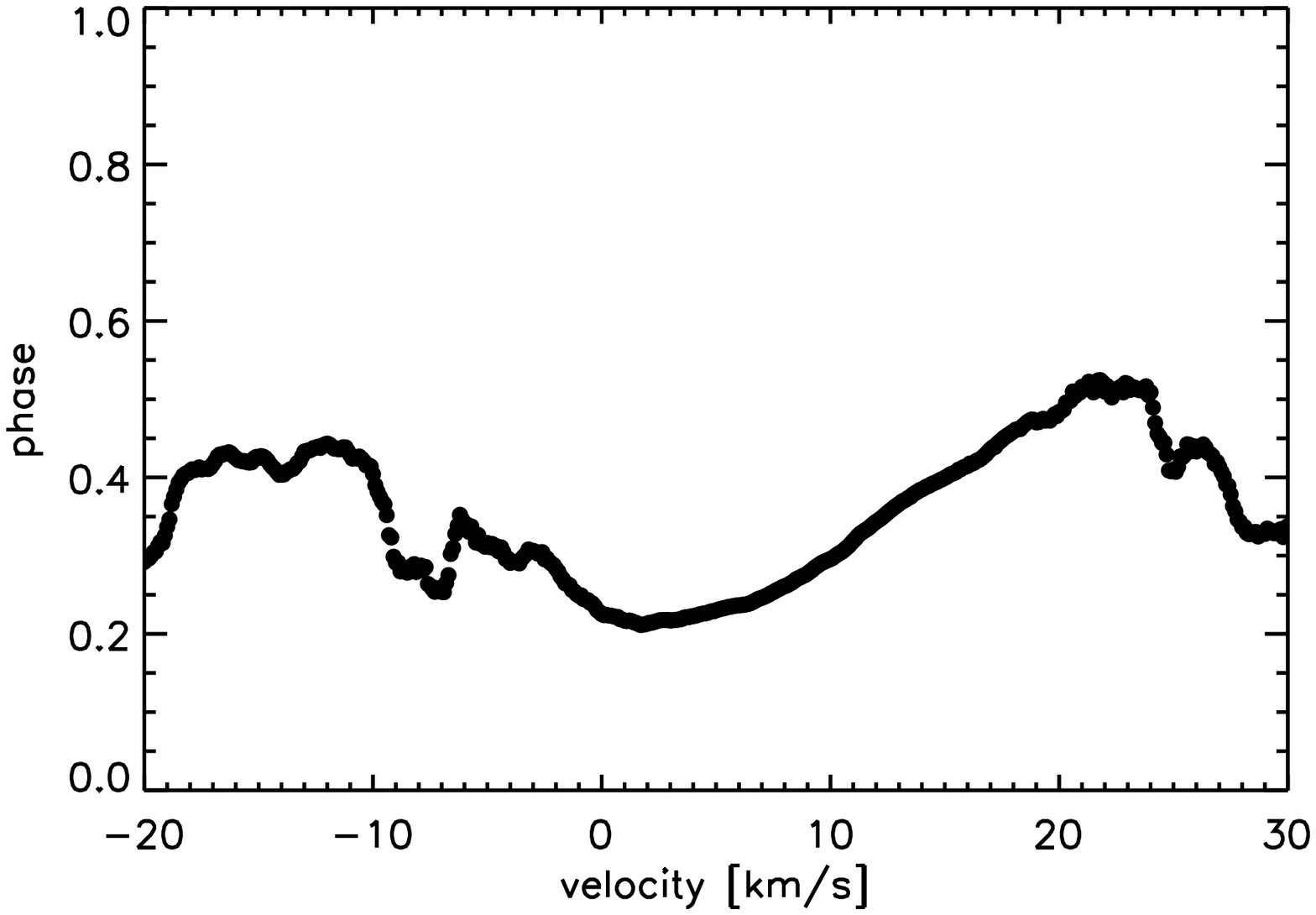}
\end{minipage}
\hfill
\begin{minipage}{5.6cm}
\centering
\includegraphics[width=5.5cm]{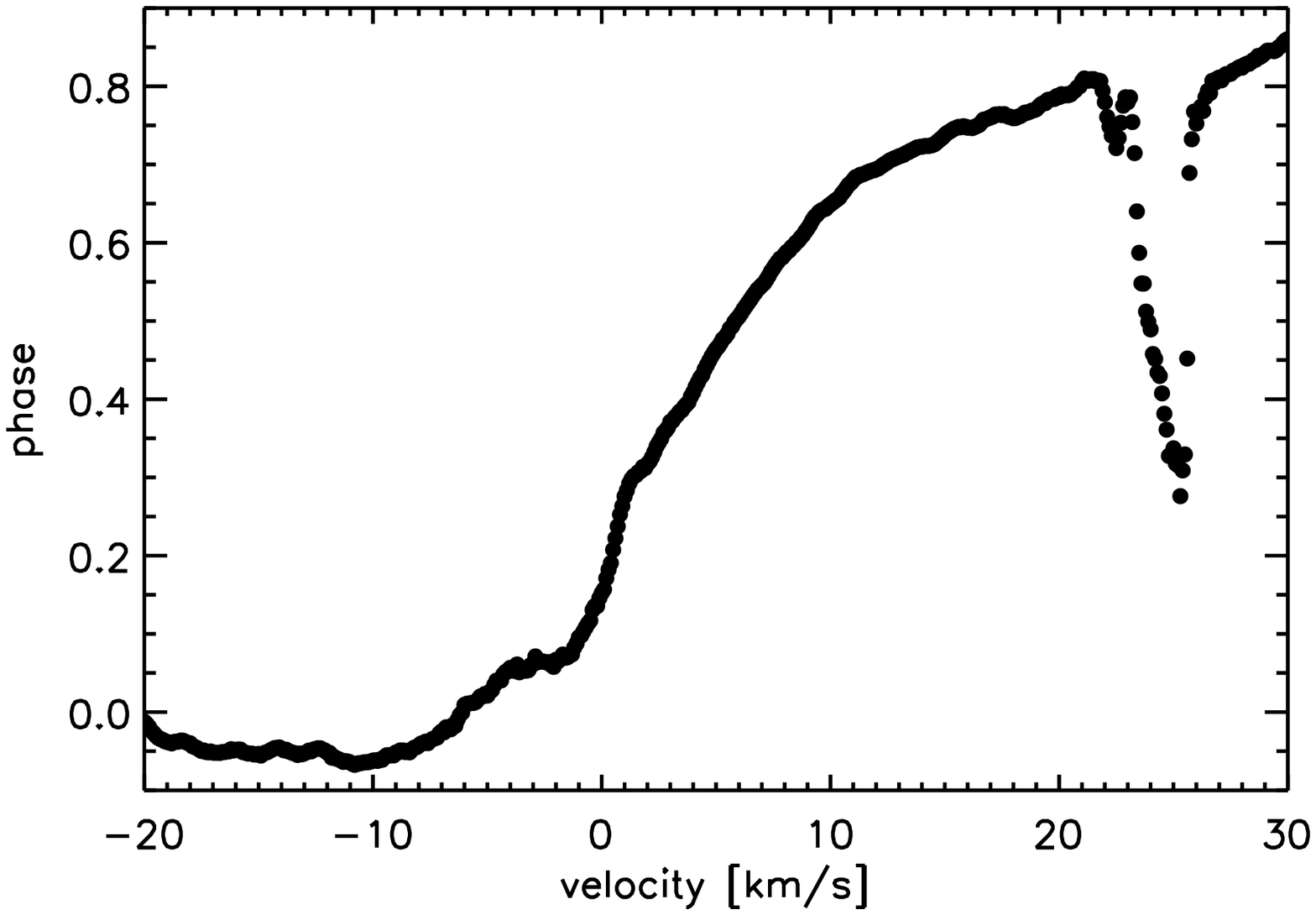}
\end{minipage}
\hfill
\begin{minipage}{5.6cm}
\centering
\includegraphics[width=5.5cm]{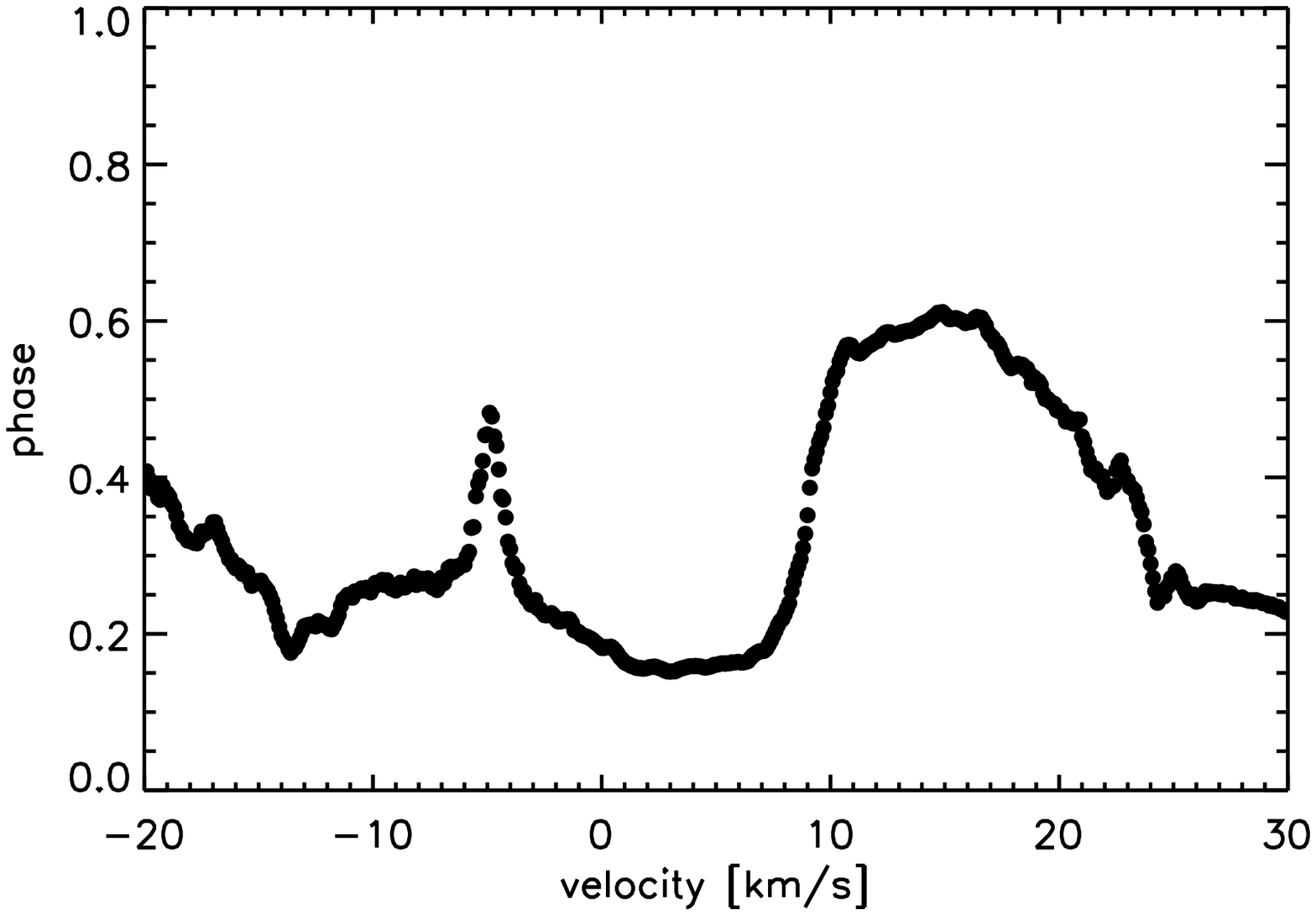}
\end{minipage}
\caption{The amplitude (top) and phase (bottom) distributions as a function of
velocity across the line profile of \object{$\eta$ Serpentis} for the significant frequencies obtained from $\langle \mathrm{v} \rangle$: $\nu_{\langle \mathrm{v} \rangle}=11.17$~c\,d$^{-1}$ ($129.3 \mu$Hz) (left), $\nu_{\langle \mathrm{v} \rangle}=11.71$~c\,d$^{-1}$ ($135.5 \mu$Hz) (middle) and $\nu_{\langle \mathrm{v} \rangle}=10.38$~c\,d$^{-1}$ ($120.2 \mu$Hz) (right). The mean radial velocity of the star, is found to be approximately $9.4$~km\,s$^{-1}$.}
\label{amplphaseHD168723}
\end{figure*}

\begin{figure}
\begin{minipage}{\columnwidth}
\centering
\includegraphics[width=5.5cm]{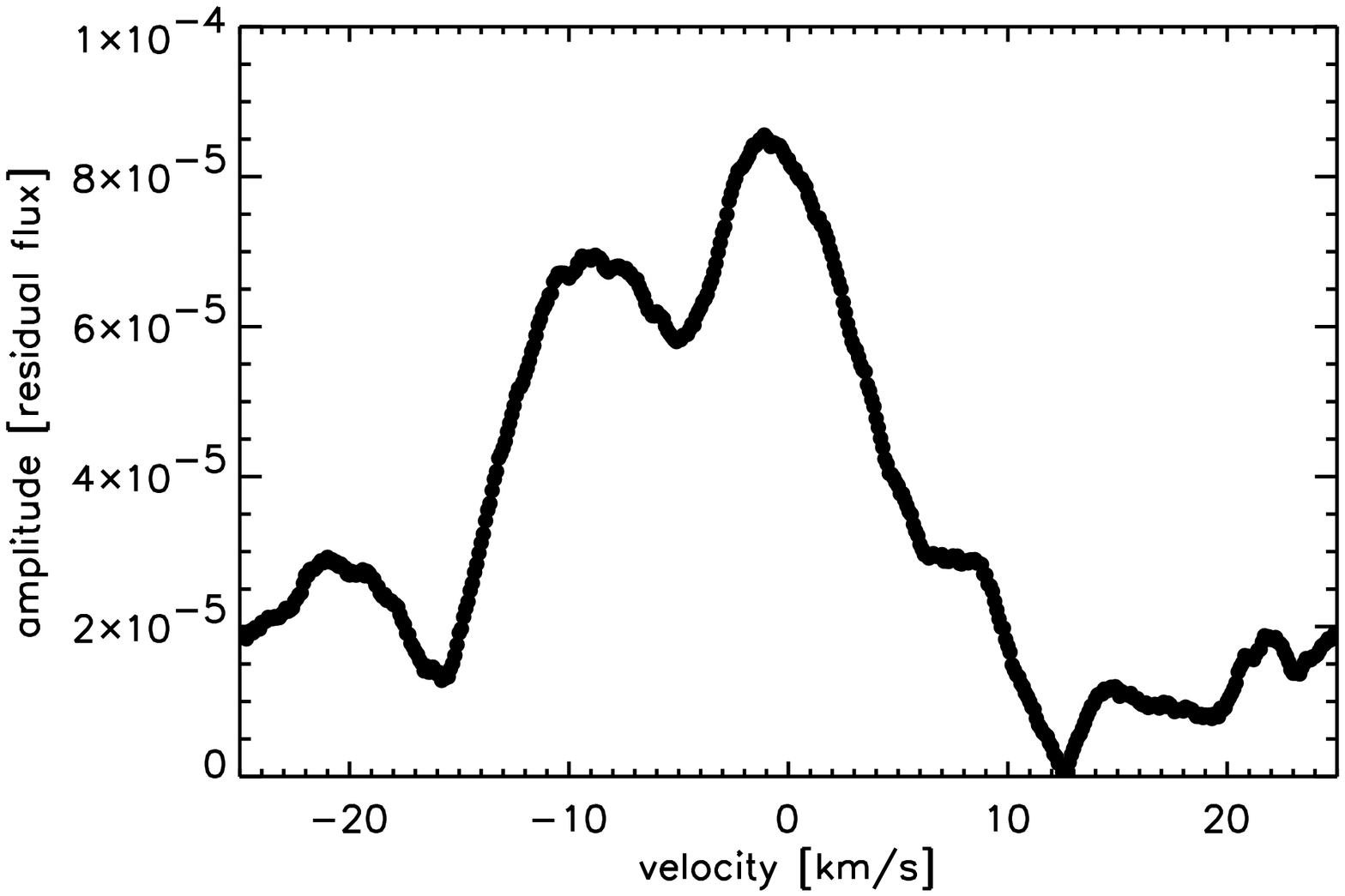}
\end{minipage}
\hfill
%\begin{minipage}{\columnwidth}
%\flushleft
%\includegraphics[width=5.6cm]{amplnoerr7.9838.eps}
%\end{minipage}
%\hfill
\begin{minipage}{\columnwidth}
\centering
\includegraphics[width=5.5cm]{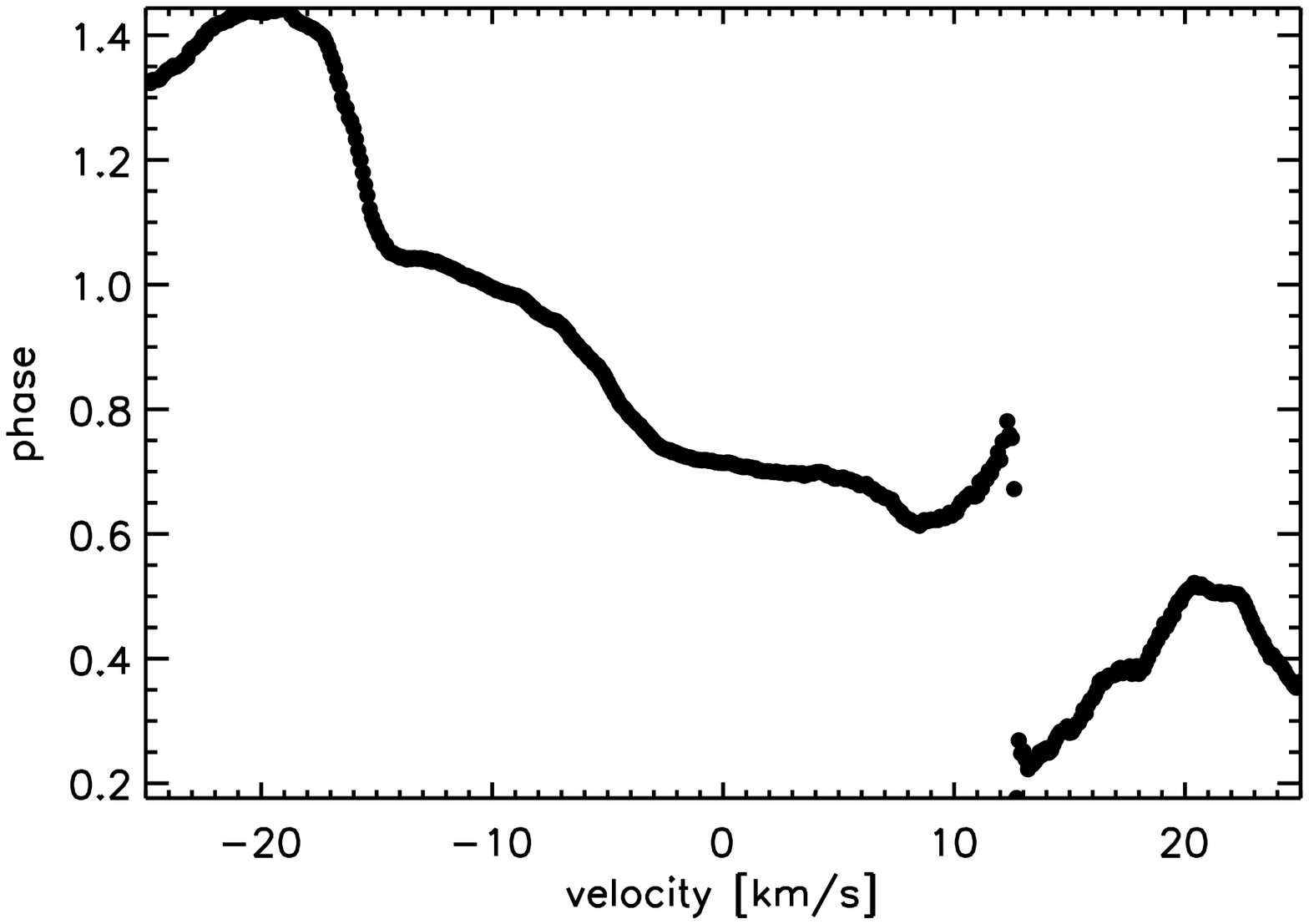}
\end{minipage}
\hfill
%\begin{minipage}{\columnwidth}
%\flushleft
%\includegraphics[width=5.6cm]{phasenoerr7.9838.eps}
%\end{minipage}
\caption{The amplitude (top) and phase (bottom) distributions as a function of
velocity across the line profile of \object{$\xi$ Hydrae} for the most dominant
frequency obtained from $\langle \mathrm{v} \rangle$: $\nu_{\langle \mathrm{v} \rangle}=8.42$~c\,d$^{-1}$ ($97.5 \mu$Hz). The mean radial velocity of the star, is found to be approximately $-4.6$~km\,s$^{-1}$.}
\label{amplphaseHD100407}
\end{figure}

\begin{figure*}
\begin{minipage}{5.6cm}
\centering
\includegraphics[width=5.5cm]{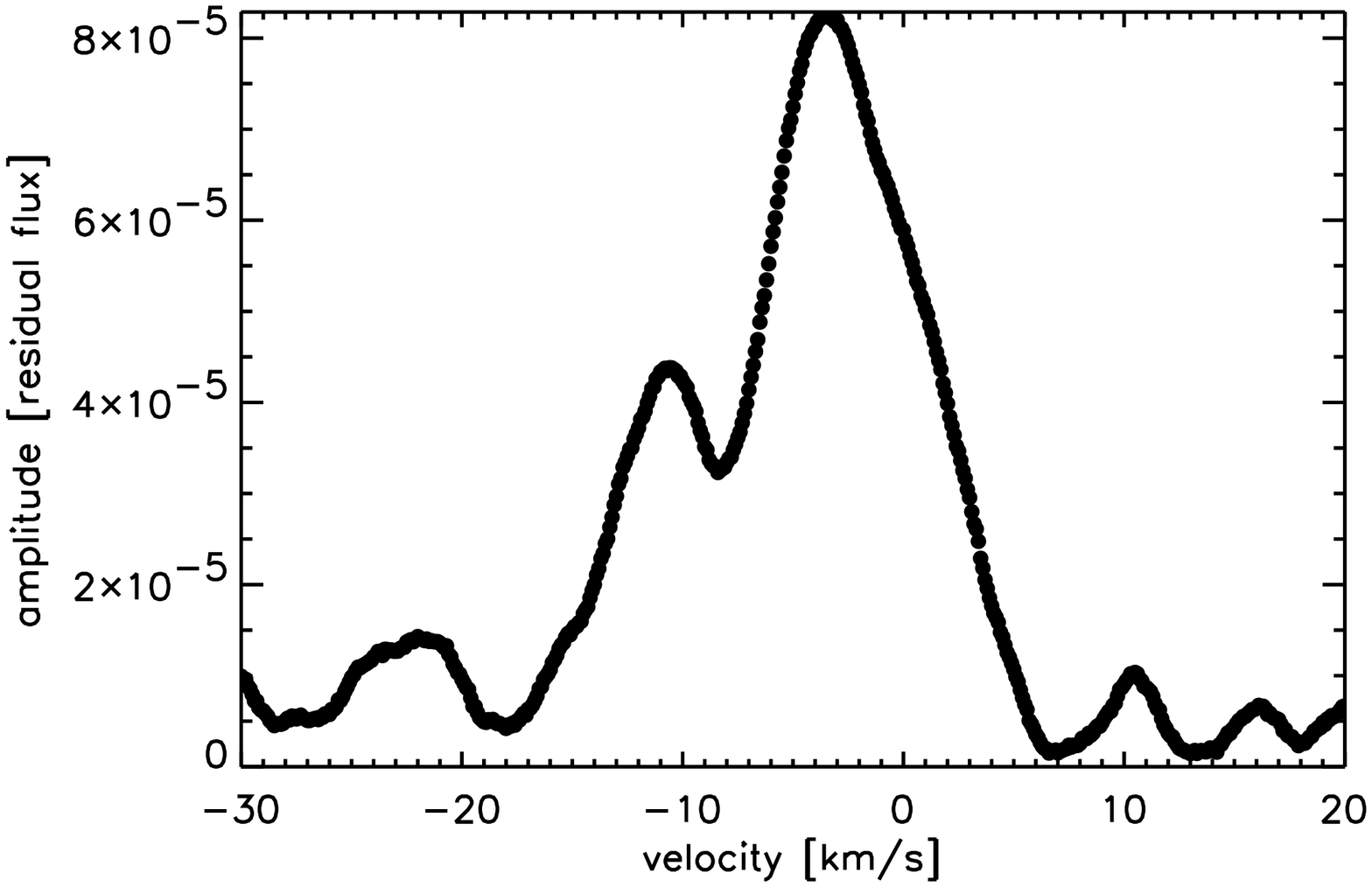}
\end{minipage}
\hfill
\begin{minipage}{5.6cm}
\centering
\includegraphics[width=5.5cm]{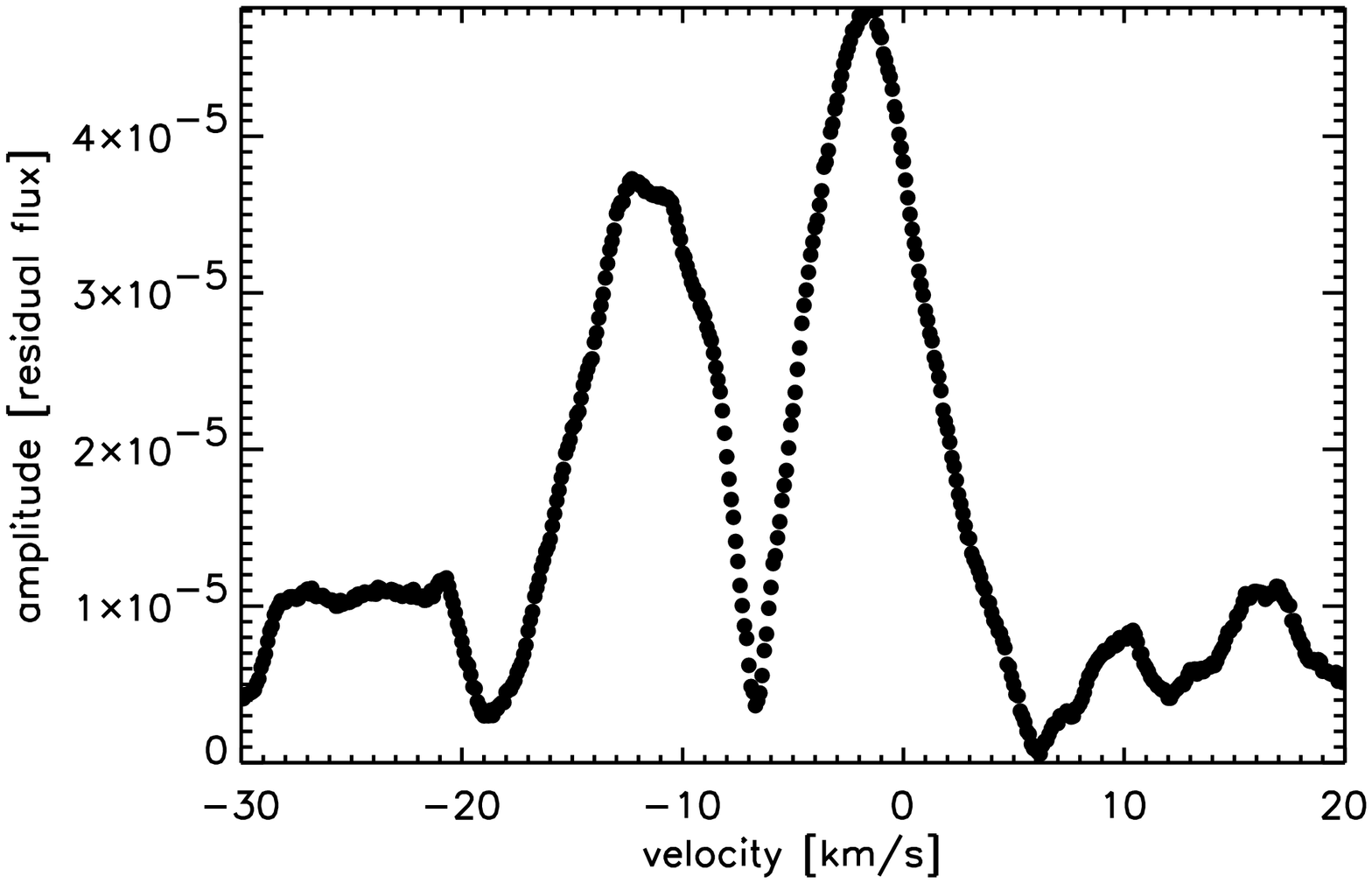}
\end{minipage}
\hfill
\begin{minipage}{5.6cm}
\centering
\includegraphics[width=5.5cm]{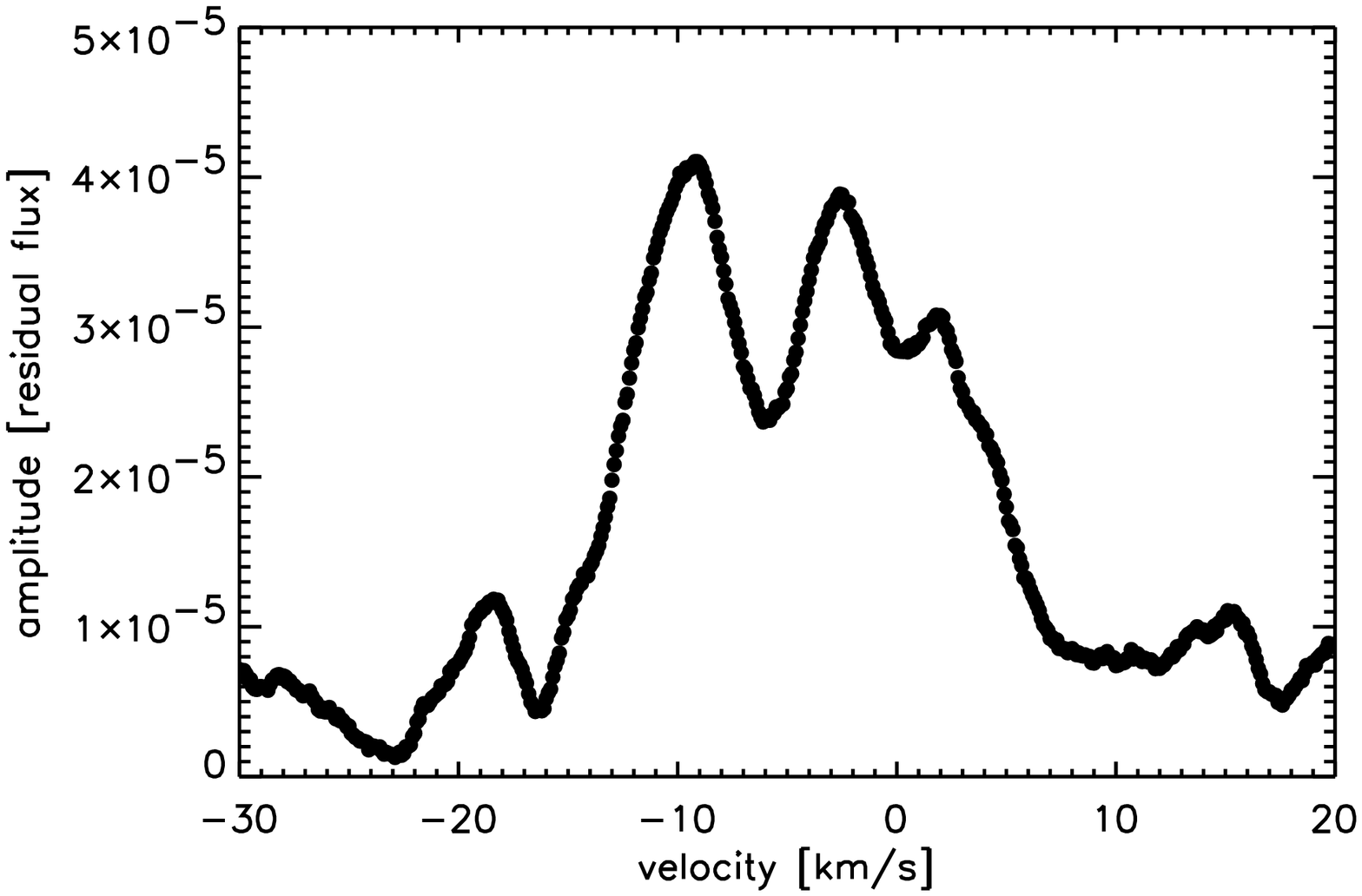}
\end{minipage}
\hfill
\begin{minipage}{5.6cm}
\centering
\includegraphics[width=5.5cm]{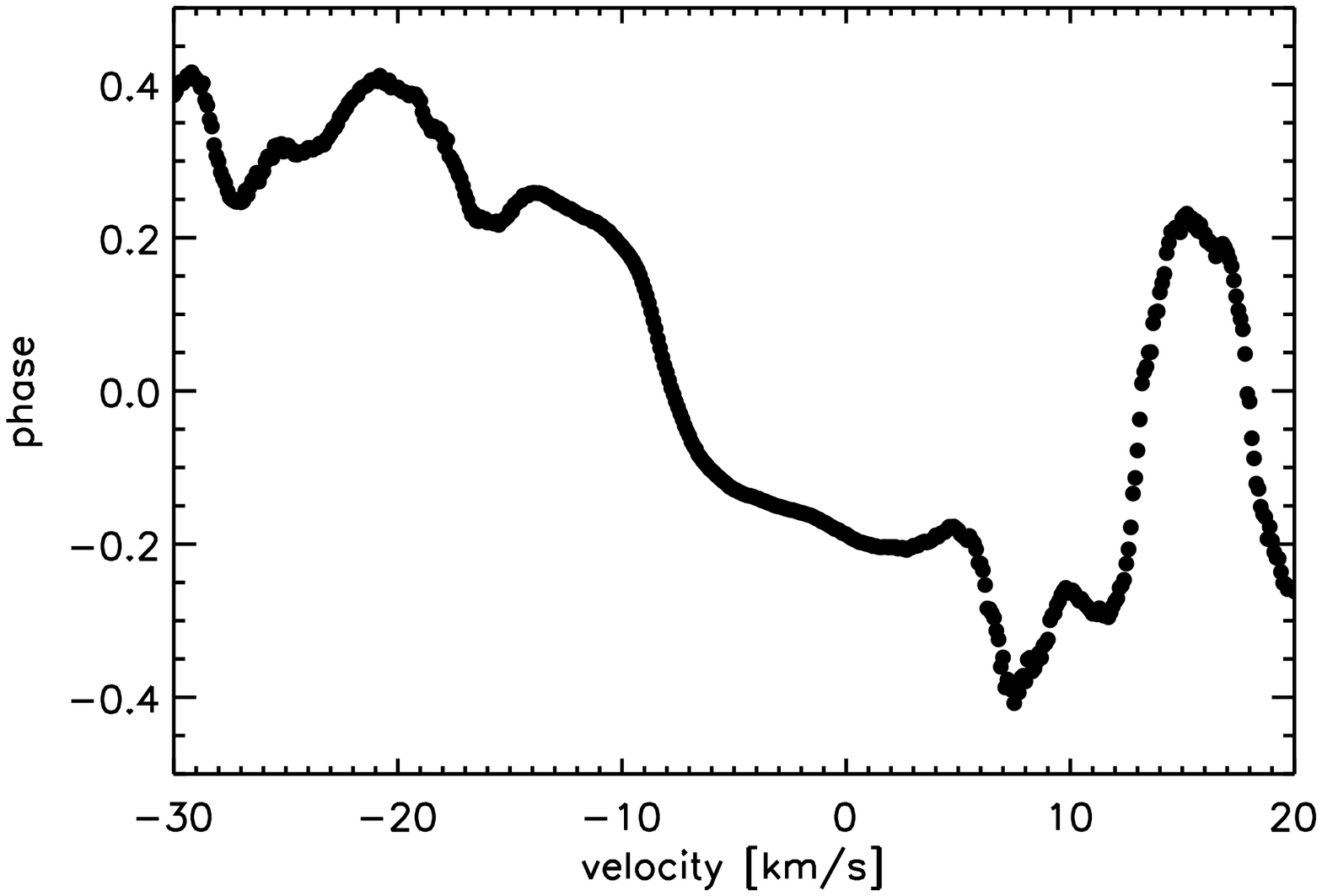}
\end{minipage}
\hfill
\begin{minipage}{5.6cm}
\centering
\includegraphics[width=5.5cm]{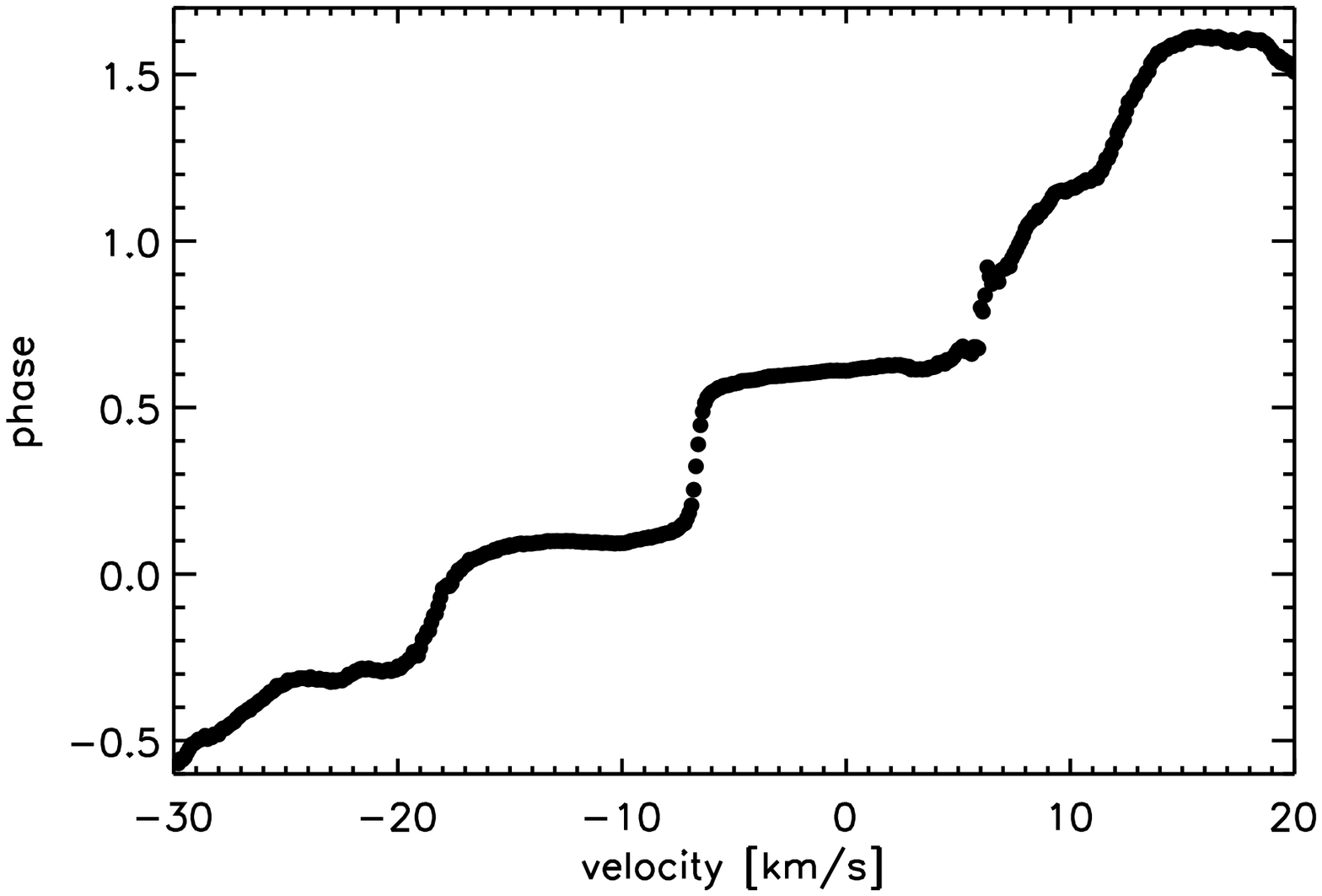}
\end{minipage}
\hfill
\begin{minipage}{5.6cm}
\centering
\includegraphics[width=5.5cm]{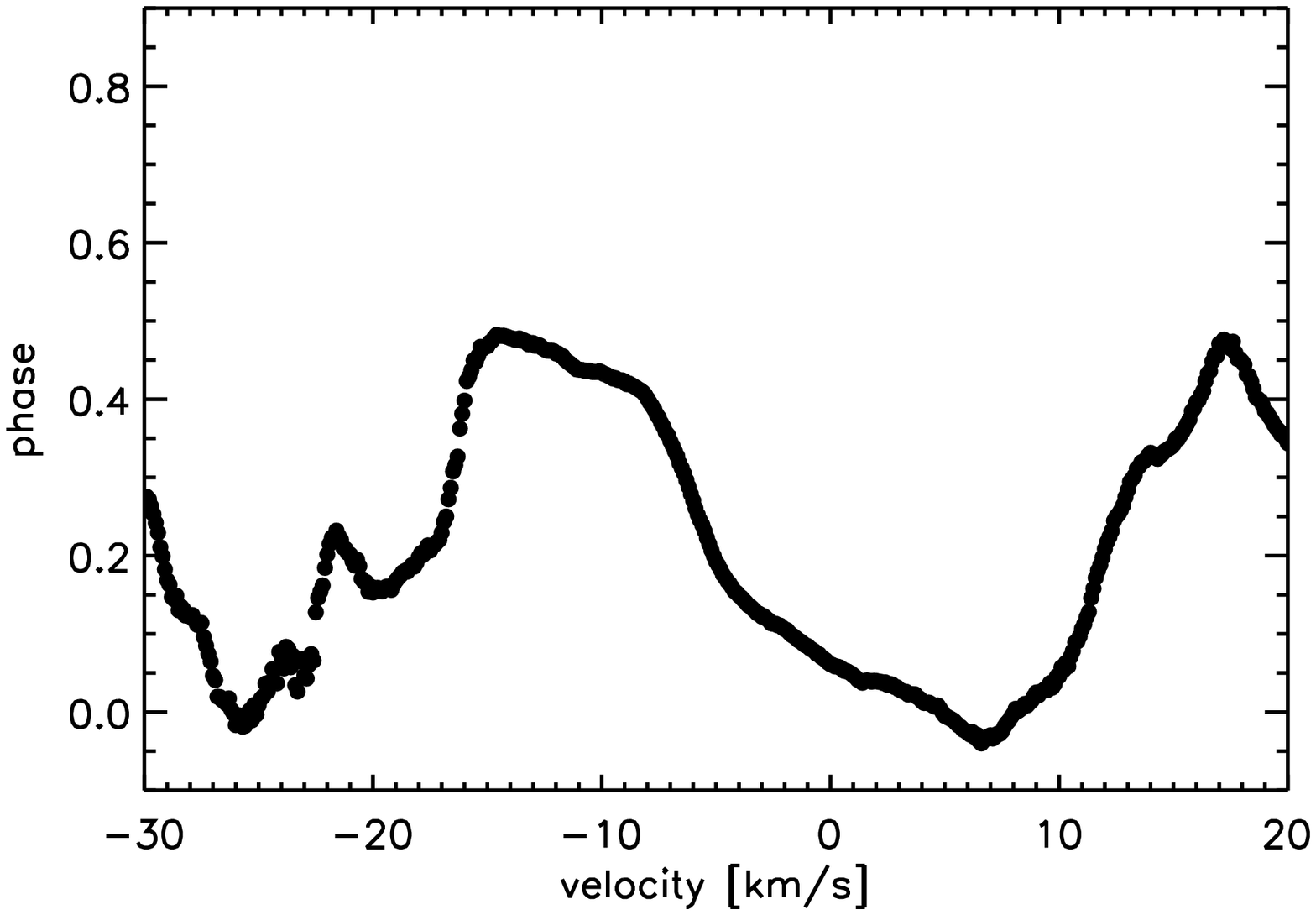}
\end{minipage}
\caption{The amplitude (top) and phase (bottom) distributions as a function of
velocity across the line profile of \object{$\delta$ Eridani} for the significant frequencies obtained from $\langle \mathrm{v} \rangle$: $\nu_{\langle \mathrm{v} \rangle}=59.40$~c\,d$^{-1}$ ($687.5 \mu$Hz) (left), $\nu_{\langle \mathrm{v} \rangle}=61.08$~c\,d$^{-1}$ ($706.9 \mu$Hz) (middle) and $\nu_{\langle \mathrm{v} \rangle}=52.68$~c\,d$^{-1}$ ($609.7 \mu$Hz) (right). The mean radial velocity of the star, is found to be approximately $-6.3$~km\,s$^{-1}$.}
\label{amplphaseHD23249}
\end{figure*}

The amplitude and phase distribution across the line profile of \object{$\epsilon$ Ophiuchi}, \object{$\eta$ Serpentis}, \object{$\xi$ Hydrae} and \object{$\delta$ Eridani} are shown in Figs~\ref{amplphaseHD146791},~\ref{amplphaseHD168723},~\ref{amplphaseHD100407} and \ref{amplphaseHD23249} respectively. The dominant frequencies obtained from $\langle \mathrm{v} \rangle$ are used for the harmonic fits at each velocity in the line profile. The frequencies obtained for each star are listed in Tables~\ref{fmomepsoph},~\ref{fmometaser},~\ref{fmomxihy} and \ref{fmomderi}. The windowfunction for \object{$\eta$ Serpentis} changes slightly in case only CORALIE data is used compared to the one obtained from both CORALIE and ELODIE data \citep{carrier2006a}. For the CORALIE data we find an extra aliasfrequency of 0.17 c\,d$^{-1}$. The frequency 11.17 c\,d$^{-1}$, only obtained for the CORALIE data, is therefore an alias of the 10.33 c\,d$^{-1}$ frequency obtained by \citet{carrier2006a}. The frequency $\nu = 3.00$ c\,d$^{-1}$ for \object{$\delta$ Eridani} is indicated in bold because this is an alias of the diurnal cycle of the observations and not due to solar like oscillations.

\begin{table}
\begin{minipage}{\columnwidth}
\caption{Comparison between frequencies found in $\langle \mathrm{v} \rangle$  ($\nu_{\langle \mathrm{v} \rangle}$) and the ones found by
\citet{deridder2006b} in the radial velocity of \object{$\epsilon$ Ophiuchi}.
The significance of $\nu_{\langle \mathrm{v} \rangle}$ is calculated with respect to the
average amplitude of the periodogram after prewhitening as described by
\citet{kuschnig1997}: $4\sigma \propto 99.9\%$ confidence interval, $3.6\sigma
\propto 99\%$ confidence interval and $3.25\sigma \propto 95\%$ confidence
interval.}
\label{fmomepsoph}
\centering
\renewcommand{\footnoterule}{}
\begin{tabular}{ccccc}
\hline\hline
$\nu_{\langle \mathrm{v} \rangle}$ & $\nu_{\langle \mathrm{v} \rangle}$ & significance & $\nu$\footnote{\citet{deridder2006b}}
& $\nu^{a}$ \\
c\,d$^{-1}$ & $\mu$Hz & & c\,d$^{-1}$ & $\mu$Hz\\
\hline
5.03 & 58.2 & 4.69$\sigma$ & 5.03 & 58.2\\
5.46 & 63.2 & 3.99$\sigma$ & 5.44 & 63.0\\
5.83 & 67.5 & 3.98$\sigma$ & 5.82 & 67.4\\
 & & & 4.59 & 53.1\\
3.17 & 36.7 & 3.43$\sigma$ & 4.17 & 48.3\\
 & & & 4.46 & 51.6\\
 & & & 5.59 & 64.7\\
 & & & 6.19 & 71.7\\
 & & & 5.18 & 59.9\\
 & & & 6.00 & 69.5\\
 & & & 6.43 & 74.4\\
\hline
\end{tabular}
\end{minipage}
\end{table}

\begin{table}
\begin{minipage}{\columnwidth}
\caption{Comparison between frequencies found in $\langle \mathrm{v} \rangle$ and the ones found by
\citet{carrier2006a} in the radial velocity of \object{$\eta$ Serpentis}. The
significance (sign) of $\nu_{\langle \mathrm{v} \rangle}$ is calculated with respect to the
average amplitude of the periodogram after prewhitening as described by
\citet{kuschnig1997}: $4\sigma \propto 99.9\%$ confidence interval, $3.6\sigma
\propto 99\%$ confidence interval and $3.25\sigma \propto 95\%$ confidence
interval.}
\label{fmometaser}
\centering
\renewcommand{\footnoterule}{}
\begin{tabular}{ccccccc}
\hline\hline
$\nu_{\langle \mathrm{v} \rangle}$ & $\nu_{\langle \mathrm{v} \rangle}$ & signi-  & $\nu$\footnote{Only CORALIE data from \citet{carrier2006a} taken into
account.} & $\nu^{a}$ & $\nu$\footnote{\citet{carrier2006a}} &
$\nu^{b}$\\
c\,d$^{-1}$ & $\mu$Hz & ficance & c\,d$^{-1}$ & $\mu$Hz & c\,d$^{-1}$ & $\mu$Hz\\
\hline
11.71 & 135.5 & 3.80$\sigma$ & 10.74 & 124.3 & 11.74 & 135.9\\
10.38 & 120.2 & 3.61$\sigma$ & 10.38 & 120.2 & 10.40 & 120.4 \\
& & & & & 11.49 & 133.0 \\
& & & & & 13.27 & 153.6\\
& & & & & 10.96 & 126.8\\
& & & & & 10.90 & 126.2\\
11.17 & 129.3 & 4.23$\sigma$ & 11.17 & 129.3 & 10.33 & 119.6\\
& & & & & 11.69 & 135.3\\
& & & & & 7.48 & 86.6\\
& & & & & 7.28 & 84.3\\
& & & & & 6.20 & 71.8\\
\hline
\end{tabular}
\end{minipage}
\end{table}

\begin{table}
\begin{minipage}{\columnwidth}
\caption{Comparison between frequencies found in $\langle \mathrm{v} \rangle$ and the ones found by
\citet{frandsen2002} in the radial velocity of \object{$\xi$ Hydrae}. The
significance of $\nu_{\langle \mathrm{v} \rangle}$ is calculated with respect to the average
amplitude of the periodogram after prewhitening as described by
\citet{kuschnig1997}. A significance of $3\sigma \propto 85\%$ confidence
interval.}
\label{fmomxihy}
\centering
\renewcommand{\footnoterule}{}
\begin{tabular}{ccccc}
\hline\hline
$\nu_{\langle \mathrm{v} \rangle}$ & $\nu_{\langle \mathrm{v} \rangle}$ & significance & $\nu$\footnote{\citet{frandsen2002}}
& $\nu^{a}$ \\
c\,d$^{-1}$ & $\mu$Hz & & c\,d$^{-1}$ & $\mu$Hz\\
\hline
 & & & 5.1344(26) & 59.43\\
 & & & 6.8366(27) & 79.13\\
8.42 & 97.5 & 3.00$\sigma$ & 7.4265(29) & 85.96\\
 & & & 8.2318(32) & 95.28\\
 & & & 9.3507(33) & 108.22\\
 & & & 8.7399(36) & 101.16\\
 & & & 10.0287(43) & 116.07\\
 & & & 9.0831(44) & 105.13\\
 & & & 8.5339(40) & 98.77\\
\hline
\end{tabular}
\end{minipage}
\end{table}

\begin{table}
\begin{minipage}{\columnwidth}
\caption{Comparison between frequencies found in $\langle \mathrm{v} \rangle$ and the ones found by
\citet{carrier2003} in the radial velocity of \object{$\delta$ Eridani}. The
significance (sign) of $\nu_{\langle \mathrm{v} \rangle}$ is calculated with respect to the
average amplitude of the periodogram after prewhitening as described by
\citet{kuschnig1997}: $4\sigma \propto 99.9\%$ confidence interval, $3.6\sigma
\propto 99\%$ confidence interval and $3.25\sigma \propto 95\%$ confidence
interval.}
\label{fmomderi}
\centering
\renewcommand{\footnoterule}{}
\begin{tabular}{crccccc}
\hline\hline
$\nu_{\langle \mathrm{v} \rangle}$ & $\nu_{\langle \mathrm{v} \rangle}$ & significance & $\nu$\footnote{\citet{carrier2003}} &
$\nu^{a}$ \\
c\,d$^{-1}$ & $\mu$Hz & & c\,d$^{-1}$ & $\mu$Hz\\
\hline
& & & 43.55 & 504.0 \\
& & & 47.12 & 545.4 \\
& & & 49.59 & 573.9 \\
& & & 52.73 & 610.3\\
52.68 & 609.7 & 4.84$\sigma$ & 54.67 & 632.7\\
& & & 56.54 & 654.4\\
& & & 58.06 & 672.0\\
59.40 & 687.5 & 7.45$\sigma$ & 58.40 & 675.9\\
& & & 60.65 & 702.0\\
61.08 & 706.9 & 6.01$\sigma$ & 62.08 & 718.5\\
60.25 & 697.3 & 4.08$\sigma$ & 62.27 & 720.7\\
57.55 & 666.1 & 4.45$\sigma$ & 64.55 & 747.1\\
65.28 & 755.6 & 3.71$\sigma$ & 66.28 & 767.1\\
& & & 68.07 & 787.9\\
& & & 75.63 & 875.4\\
\textbf{3.00} & 34.7 & 5.89$\sigma$ & &\\

\hline
\end{tabular}
\end{minipage}
\end{table}

\section{Simulations}
In order to test the robustness of moment diagnostics and the amplitude and phase behaviour across the line profiles against finite lifetimes
of oscillation modes, spectral lines are generated with damped and re-excited
modes. The moments of these lines are obtained and their dominant frequencies derived.  Furthermore, the amplitude and phase distribution across the spectral line is investigated.

The simulations are done for a single lineforming region. Therefore, the simulated phase distributions are not necessarily comparable to the observed ones, which are averaged phase distributions over a whole range of line forming regions because they are based on a cross correlation function. Our interpretation will therefore mainly be based on the amplitude distribution which represents an average amplitude across the whole lineforming region considered in the cross correlation function.

\subsection{Damping and re-excitation equations}
A damped and re-excited oscillation mode is damped by a factor $e^{-\eta t}$, 
with $\eta$ the damping rate, and re-excited before it is able to damp out.
As a consequence both the amplitude and the phase of the oscillation are time
dependent:
\begin{equation}
f(t)=A(t)\sin(2 \pi \nu t+ \psi (t)).
\label{sinos}
\end{equation} 
To simulate such an oscillator we follow the description of \citet{deridder2006a},
and we compute
\begin{equation}
f(t)= B(t) \sin(2\pi\nu t)+ C(t) \cos(2\pi\nu t),
\label{dampreex}
\end{equation}
where we let the amplitudes $B$ and $C$ vary with a first order autoregressive
process in a discrete time domain with time step $\Delta t$:
\begin{equation}
B_{n}= e^{-\eta\Delta t} B_{n-1} + \varepsilon_{n+1}
\end{equation}
where $\varepsilon_{n+1}$ is a Gaussian distributed excitation kick. 
For more details we refer to \citet{deridder2006a}. The above is applied
to the equations for the pulsation velocity which are obtained by taking
the time derivative of the displacement components mentioned in Section 3.2,
Eq.(2) of \citet{deridder2002}. Whenever rotation is neglected, the three
spherical components of the pulsation velocity, i.e. $\upsilon_{r}$,
$\upsilon_{\theta}$ and $\upsilon_{\varphi}$, for the damped and re-exited case
become:

\begin{eqnarray}
\upsilon_{r} & = & -\upsilon_{p}N_{\ell}^{m}P_{\ell}^{m}(\cos\theta)*
\nonumber\\
 & & e^{-\eta(t-n\Delta t_{kick})}(B_{n}\sin(m\varphi+2 \pi \nu t) +
C_{n}\cos(m\varphi +2 \pi\nu t)),
\label{vrad}
\end{eqnarray}
\begin{eqnarray}
\upsilon_{\theta} & = &
-K\upsilon_{p}N_{\ell}^{m}\frac{\partial}{\partial\theta}(P_{\ell}^{m}
(\cos\theta))*
\nonumber\\
& & e^{-\eta(t-n\Delta t_{kick})}(B_{n}\sin(m\varphi+2 \pi \nu t) +
C_{n}\cos(m\varphi +2 \pi\nu t)),
\label{vtheta}
\end{eqnarray}
\begin{eqnarray}
\upsilon_{\varphi} & = &
-mK\upsilon_{p}N_{\ell}^{m}\frac{1}{\sin\theta}P_{\ell}^{m}(\cos\theta)*
\nonumber\\
& & e^{-\eta(t-n\Delta t_{kick})}(B_{n}\cos(m\varphi+2 \pi \nu t) -
C_{n}\sin(m\varphi +2 \pi\nu t)),
\label{vfi}
\end{eqnarray} 
with $\upsilon_{p}$ proportional to the pulsation amplitude, $N_{\ell}^{m}$ the
normalisation factor for the spherical harmonics $Y_{\ell}^{m}(\theta,\varphi)
\equiv P_{\ell}^{m}(\cos\theta)e^{im\varphi}$ and $K$ the ratio of the
horizontal to the vertical velocity amplitude.

Line profiles with the same parameters as described in section 3.1 are generated with a damping time ($\eta^{-1}$) of two
days (the estimate for \object{$\xi$ Hydrae} by \citet{stello2004}). The used pulsation amplitude is 0.04 km\,s$^{-1}$. The moments, with the dominant frequencies and the amplitude and phase distribution across the line profiles are investigated.

\subsection{Frequencies}
$\langle \mathrm{v} \rangle$, $\langle \mathrm{v}^{2} \rangle$ and $\langle \mathrm{v}^{3} \rangle$ are determined for the generated series of line profiles and the frequencies are obtained in the same way as described in
section~2.3.1. 
Rather than fitting a Lorenz profile to the power spectrum, we simply used the highest peak to obtain the oscillation frequency. This implies that the obtained frequencies differ slightly from the input frequency as predicted by the Lorenzian probability distribution.
The frequencies obtained from $\langle \mathrm{v}^{3}\rangle$ are the
same as the ones obtained for $\langle \mathrm{v} \rangle$, which is as expected from theory and also
seen in the observations of the four red (sub)giants. None of the modes reveal a relation between the frequencies obtained from  $\langle \mathrm{v} \rangle$ and $\langle \mathrm{v}^{2} \rangle$, which is due to the small amplitudes as described in section 2.3.1.

\subsection{Amplitude and phase distribution}

\begin{figure*}
\begin{minipage}{4.25cm}
\centering
\includegraphics[width=4.25cm]{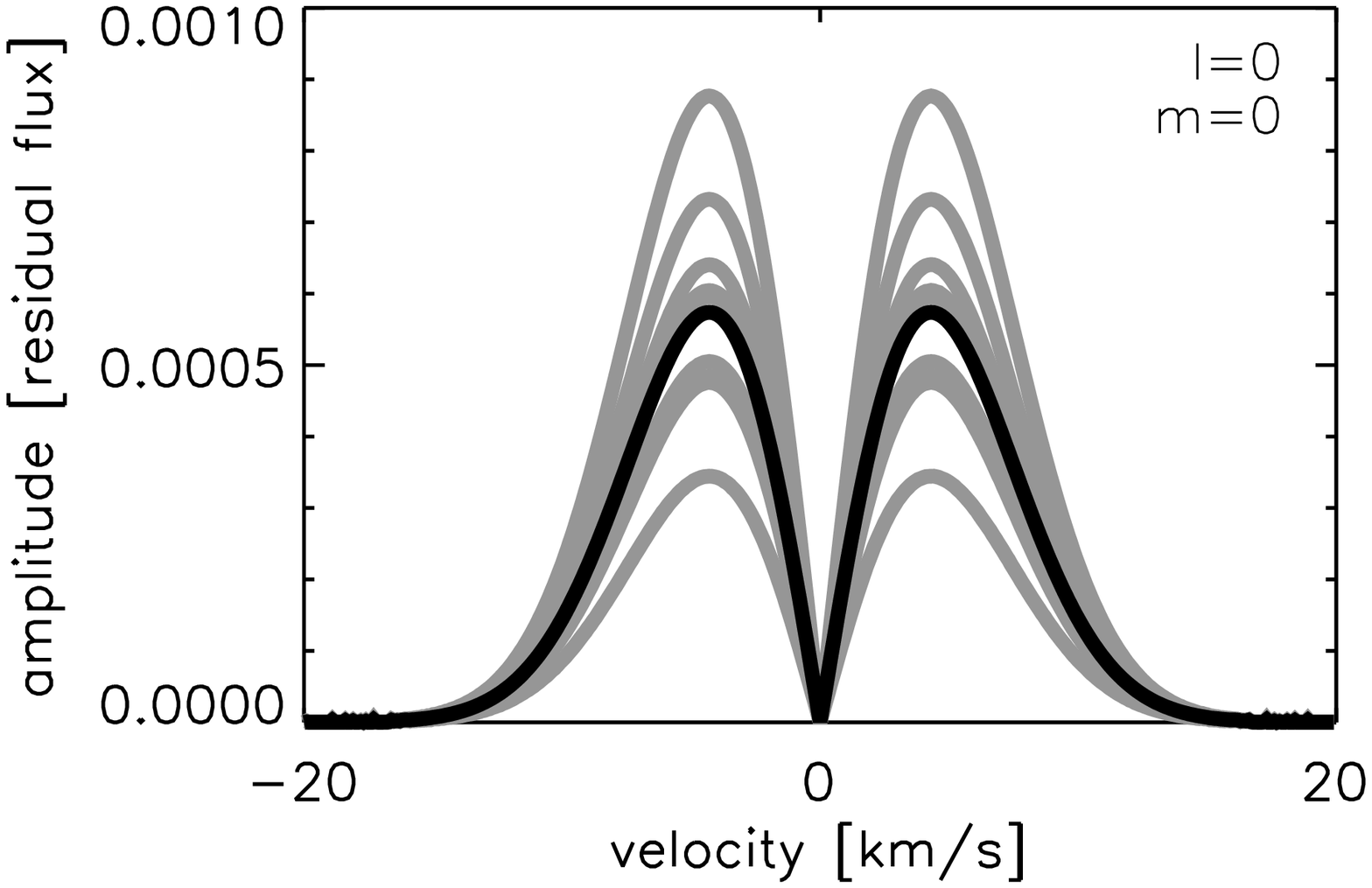}
\end{minipage}
\hfill
\begin{minipage}{4.25cm}
\centering
\includegraphics[width=4.25cm]{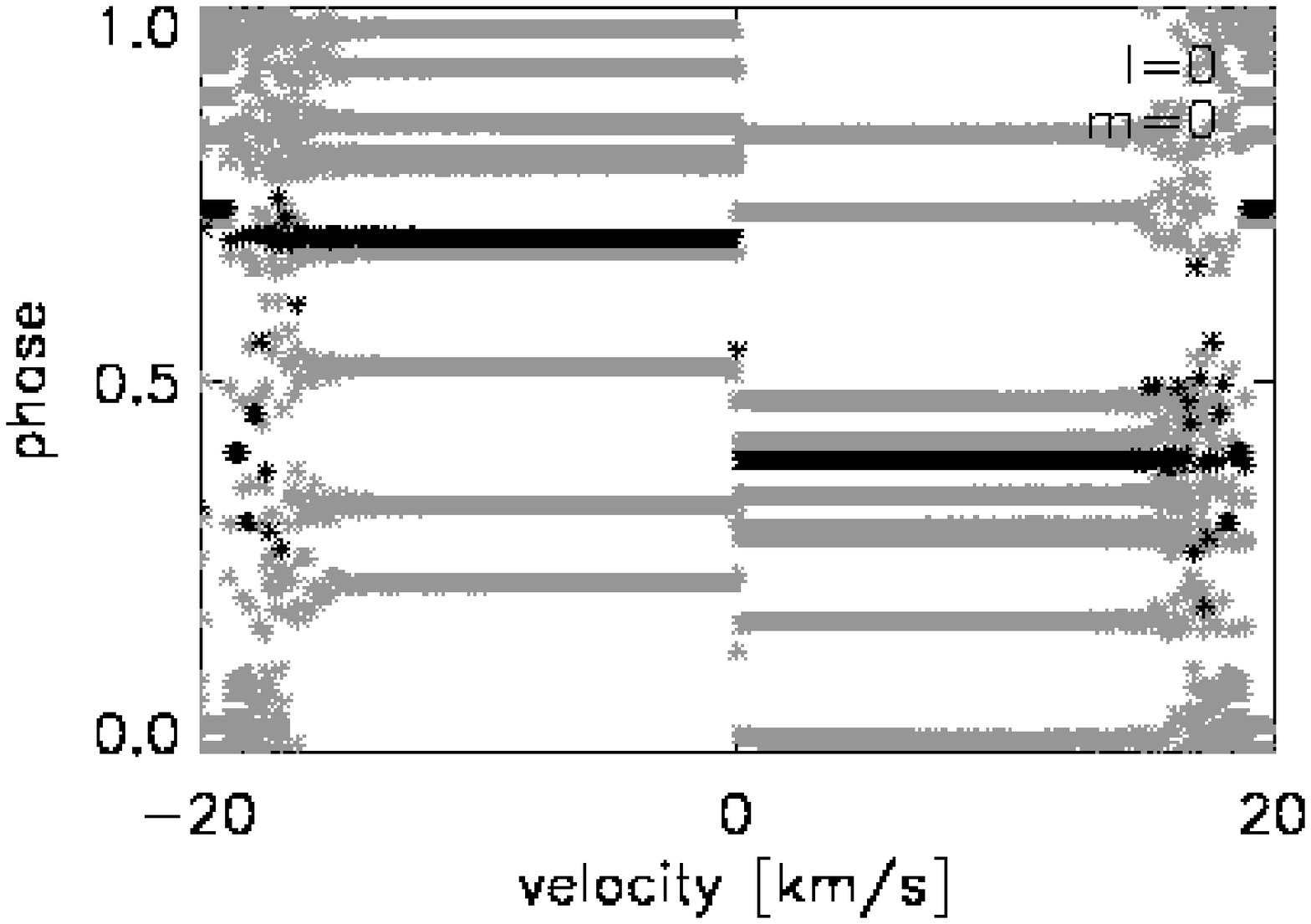}
\end{minipage}
\hfill
\begin{minipage}{4.25cm}
\centering
\includegraphics[width=4.25cm]{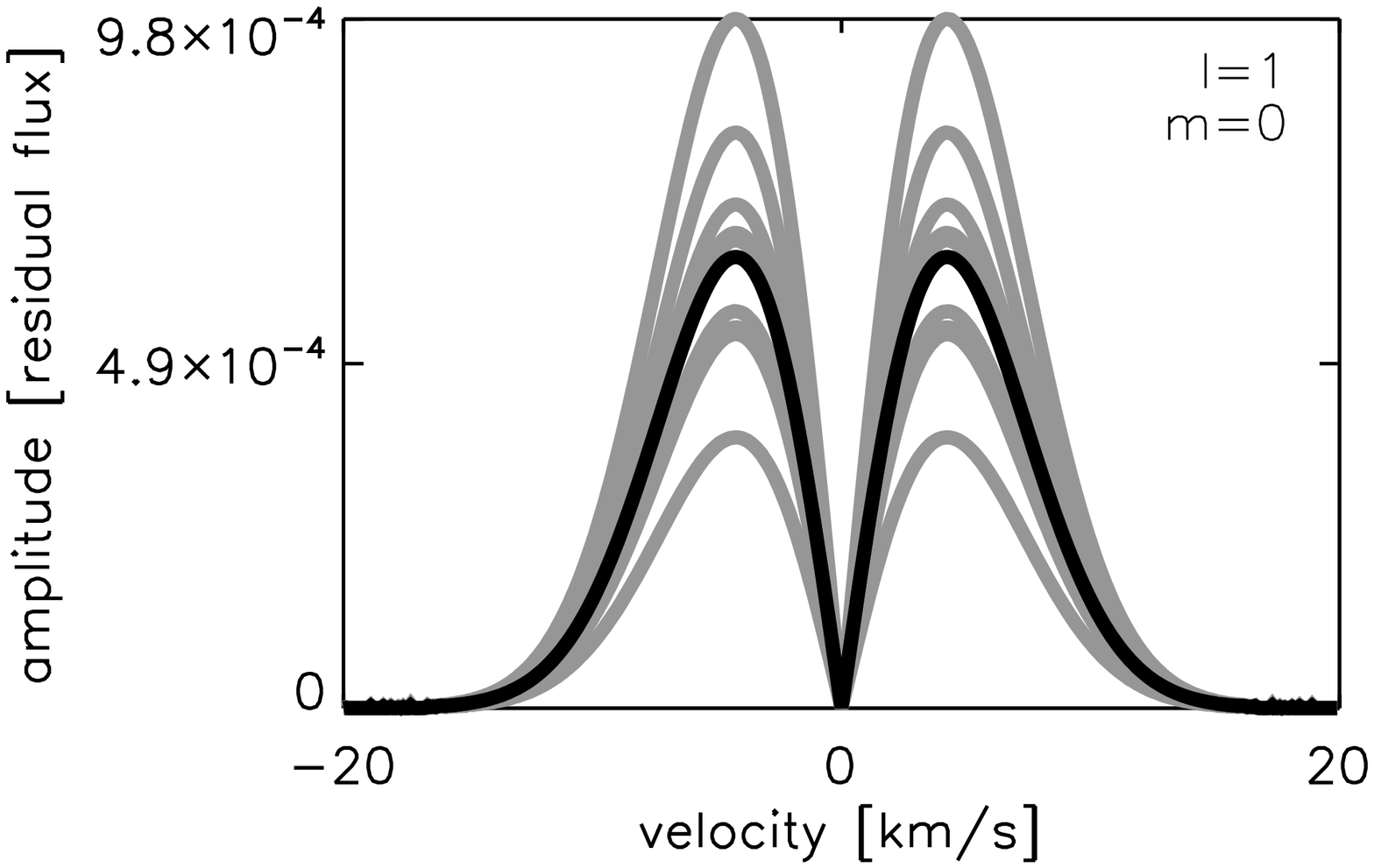}
\end{minipage}
\hfill
\begin{minipage}{4.25cm}
\centering
\includegraphics[width=4.25cm]{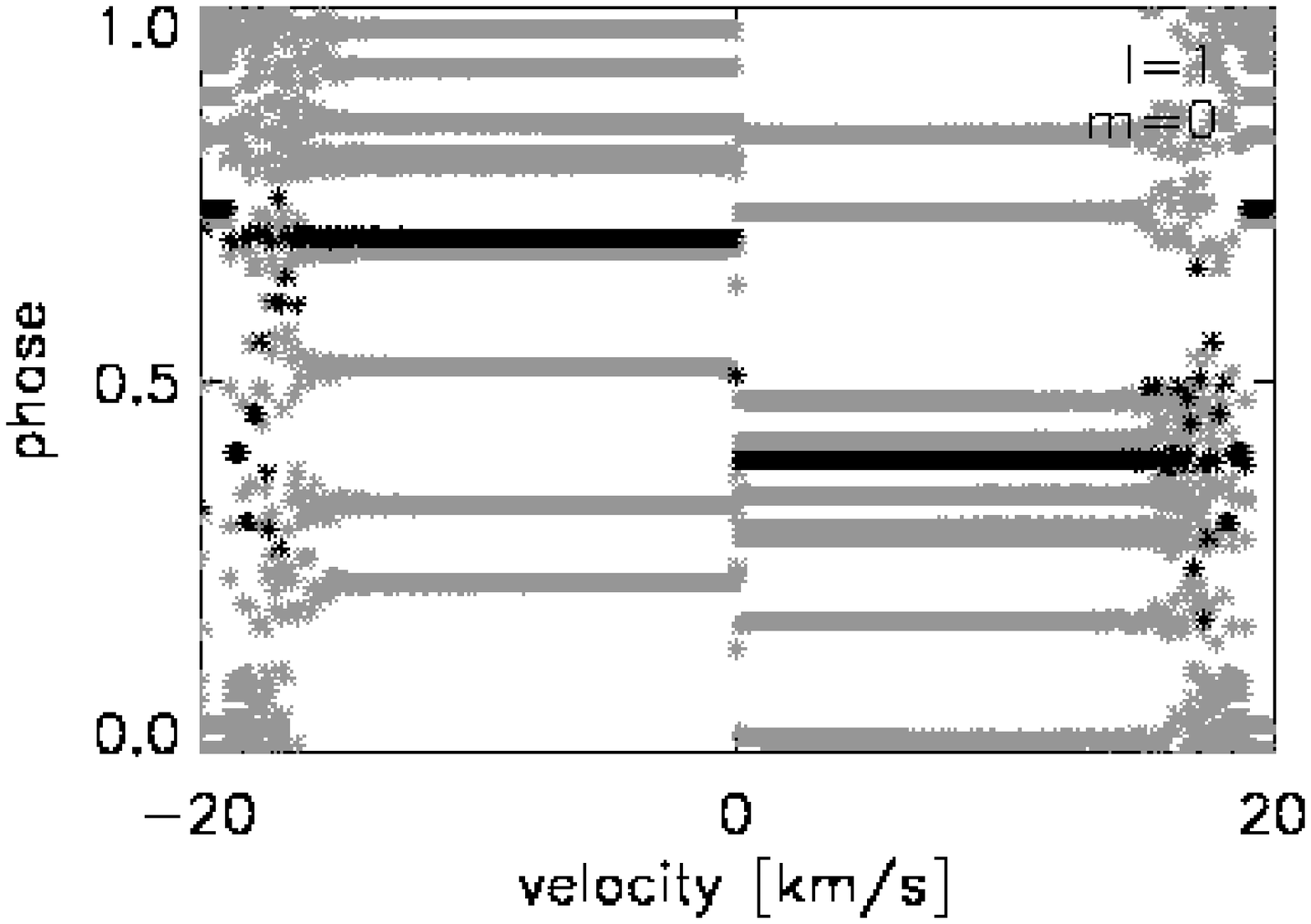}
\end{minipage}
\hfill
\begin{minipage}{4.25cm}
\centering
\includegraphics[width=4.25cm]{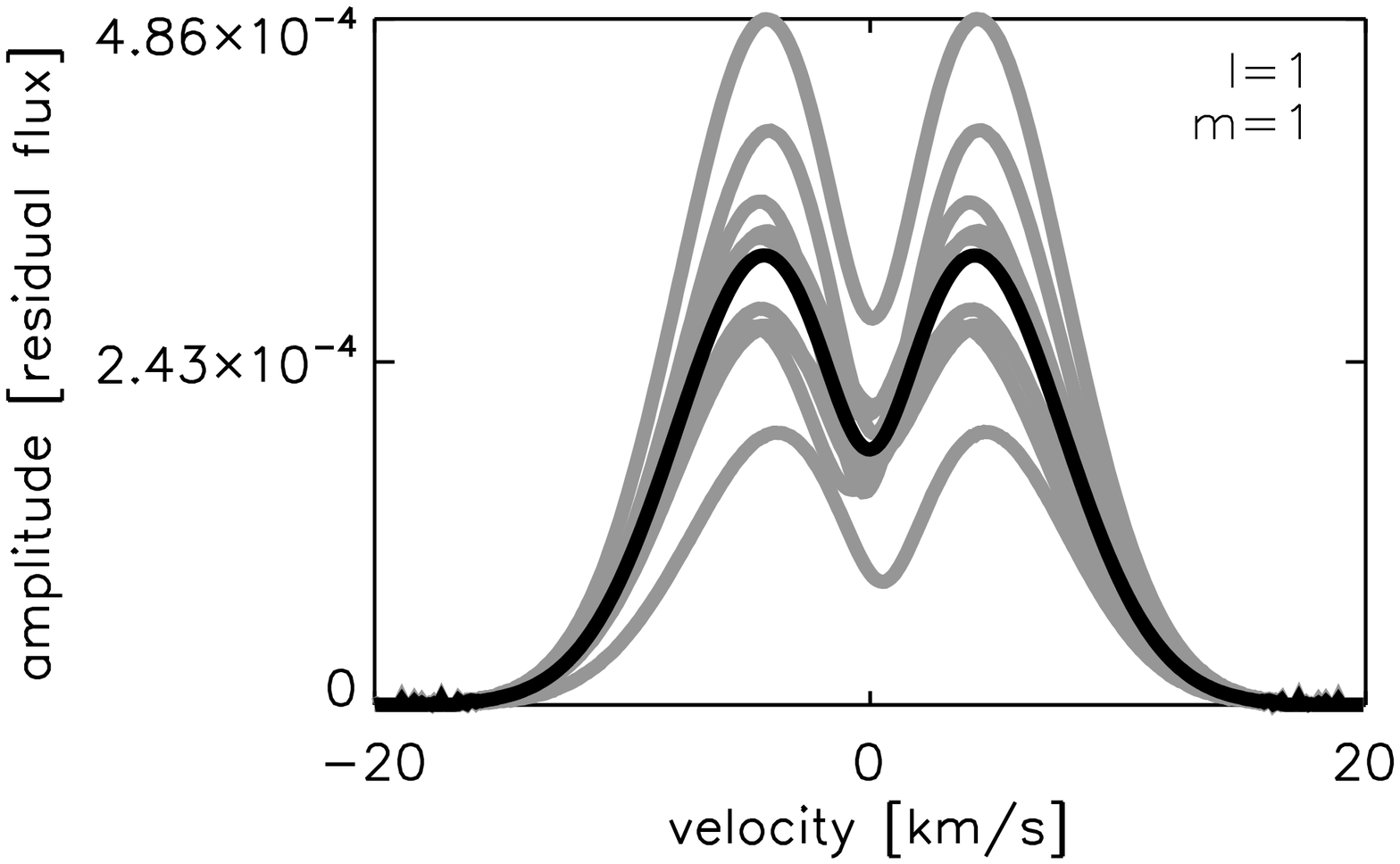}
\end{minipage}
\hfill
\begin{minipage}{4.25cm}
\centering
\includegraphics[width=4.25cm]{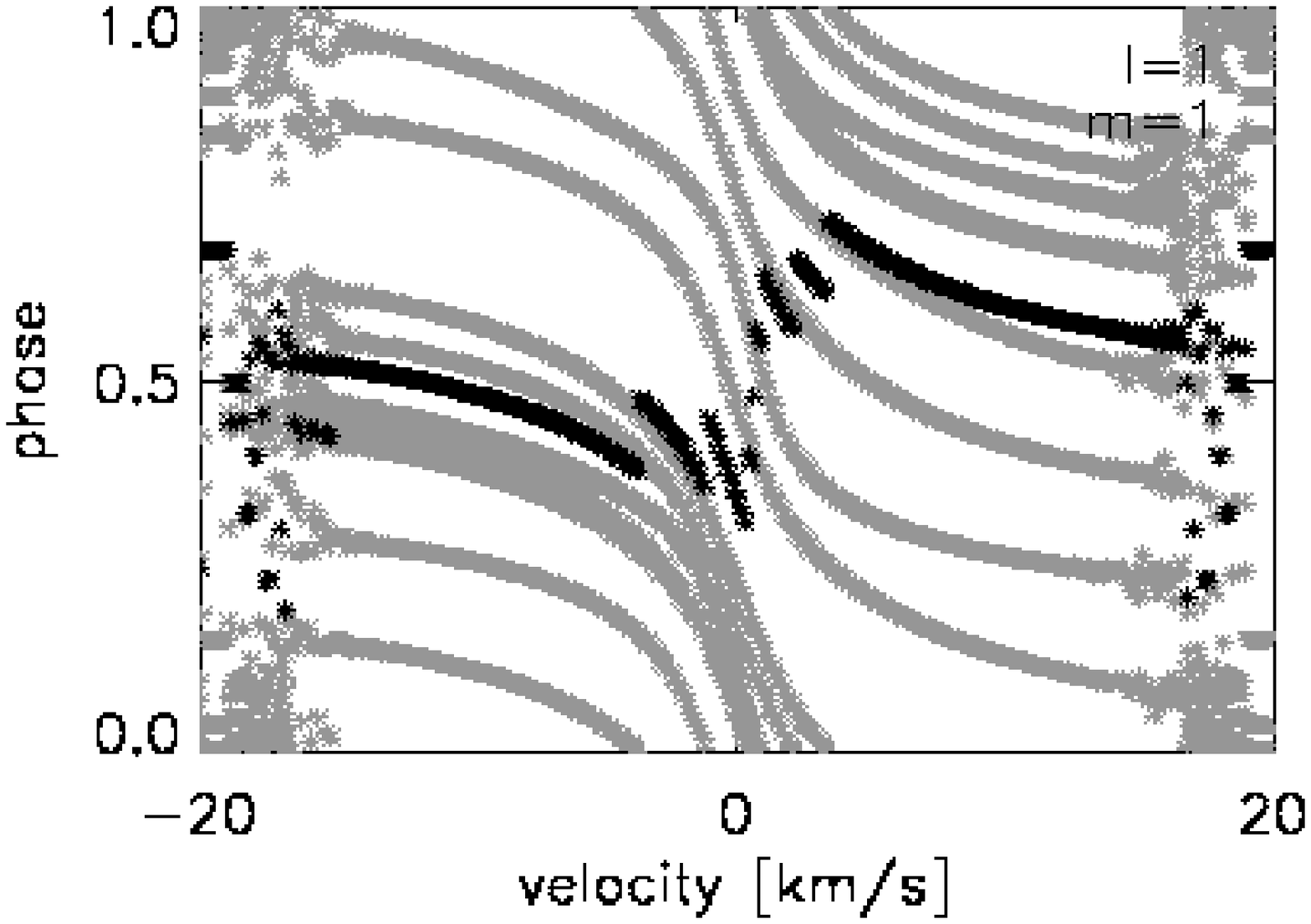}
\end{minipage}
\hfill
\begin{minipage}{4.25cm}
\centering
\includegraphics[width=4.25cm]{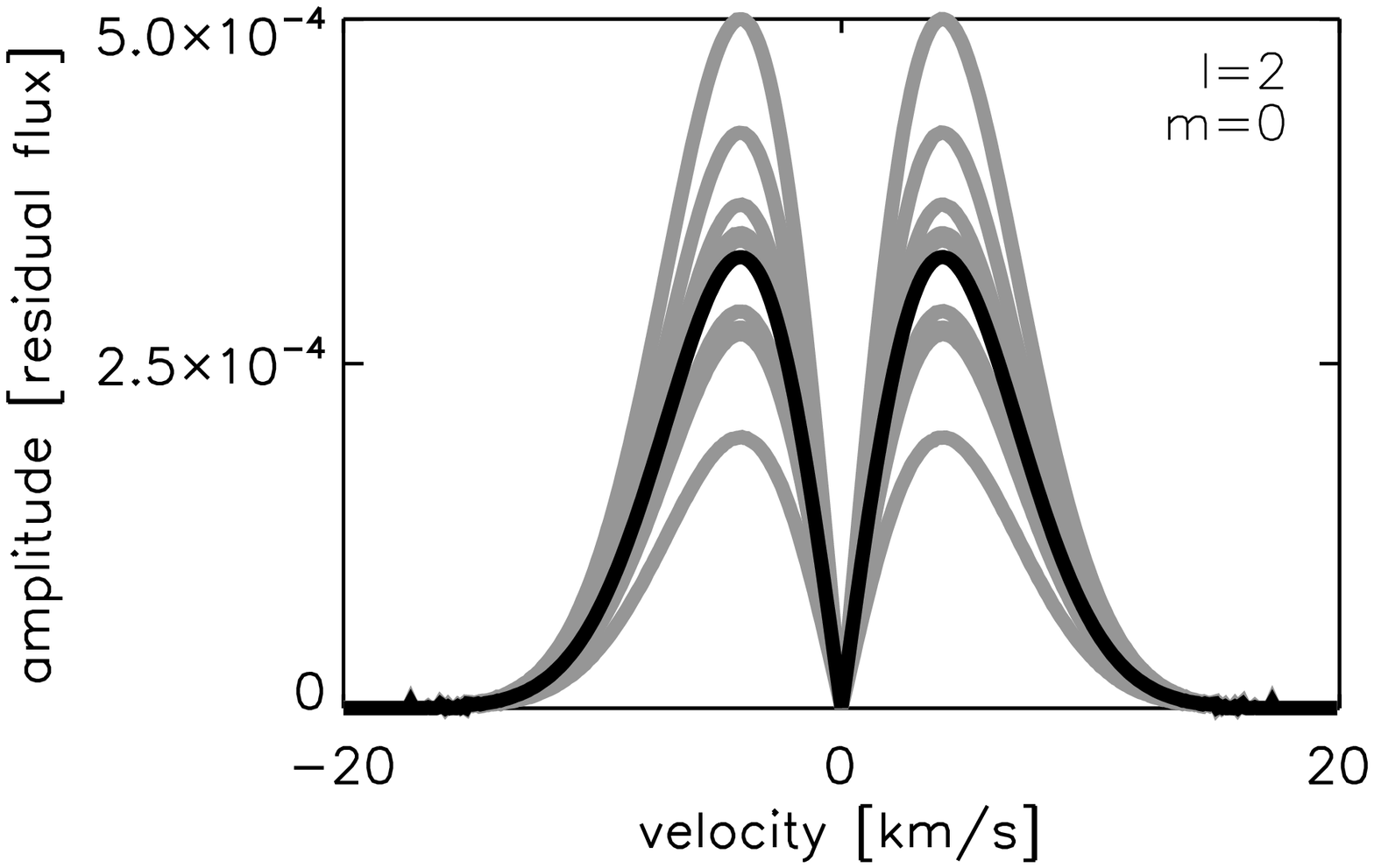}
\end{minipage}
\hfill
\begin{minipage}{4.25cm}
\centering
\includegraphics[width=4.25cm]{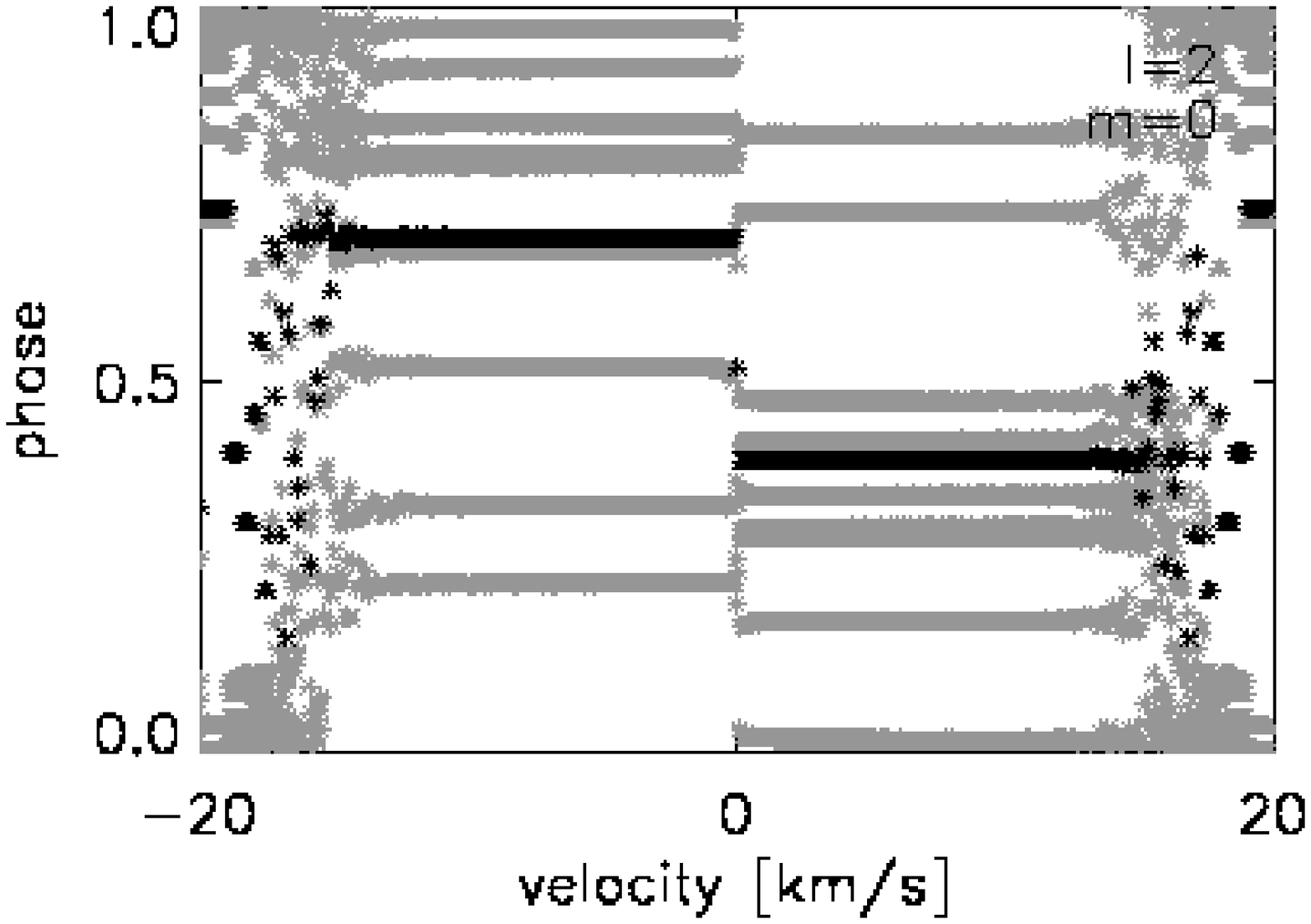}
\end{minipage}
\hfill
\begin{minipage}{4.25cm}
\centering
\includegraphics[width=4.25cm]{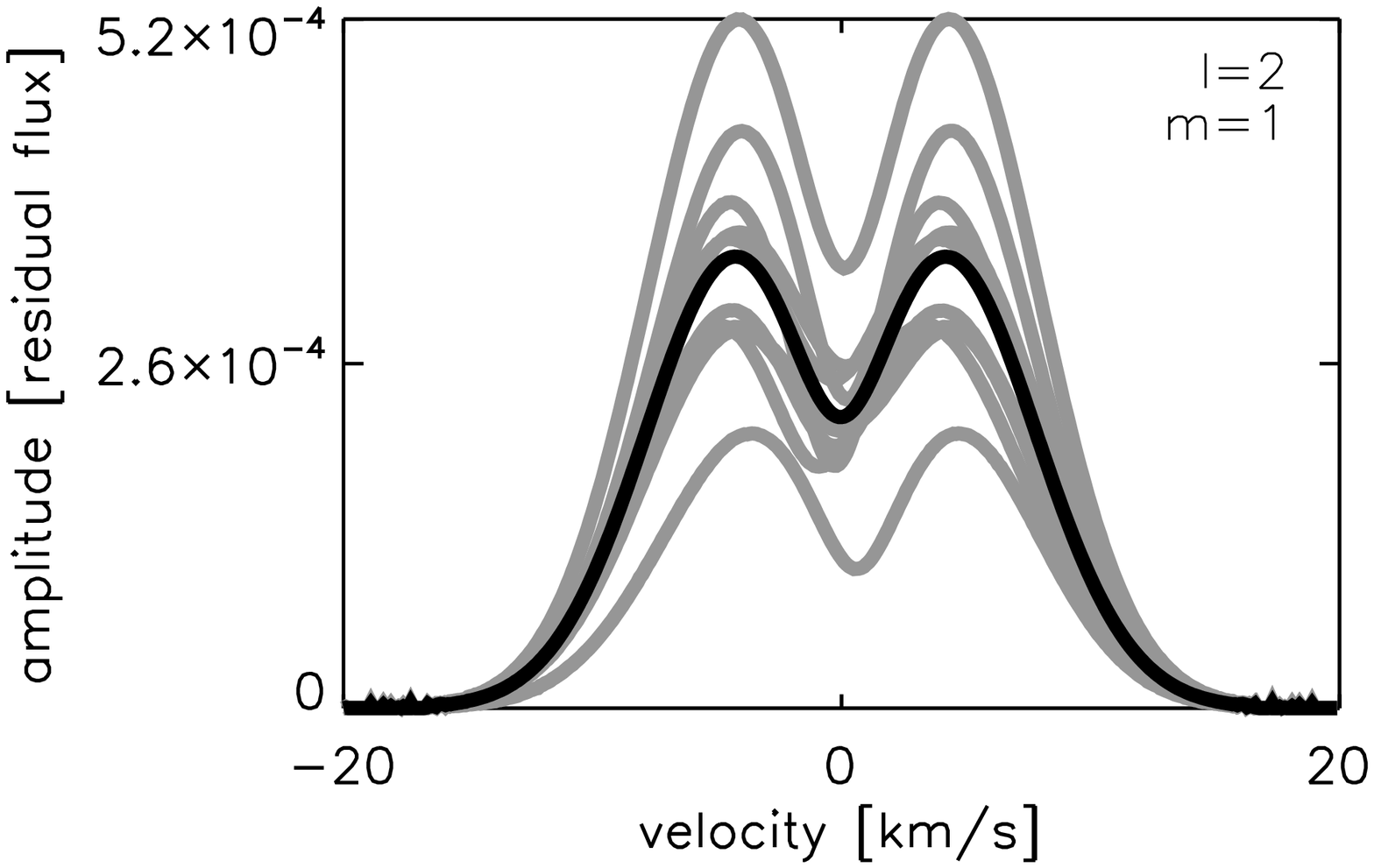}
\end{minipage}
\hfill
\begin{minipage}{4.25cm}
\centering
\includegraphics[width=4.25cm]{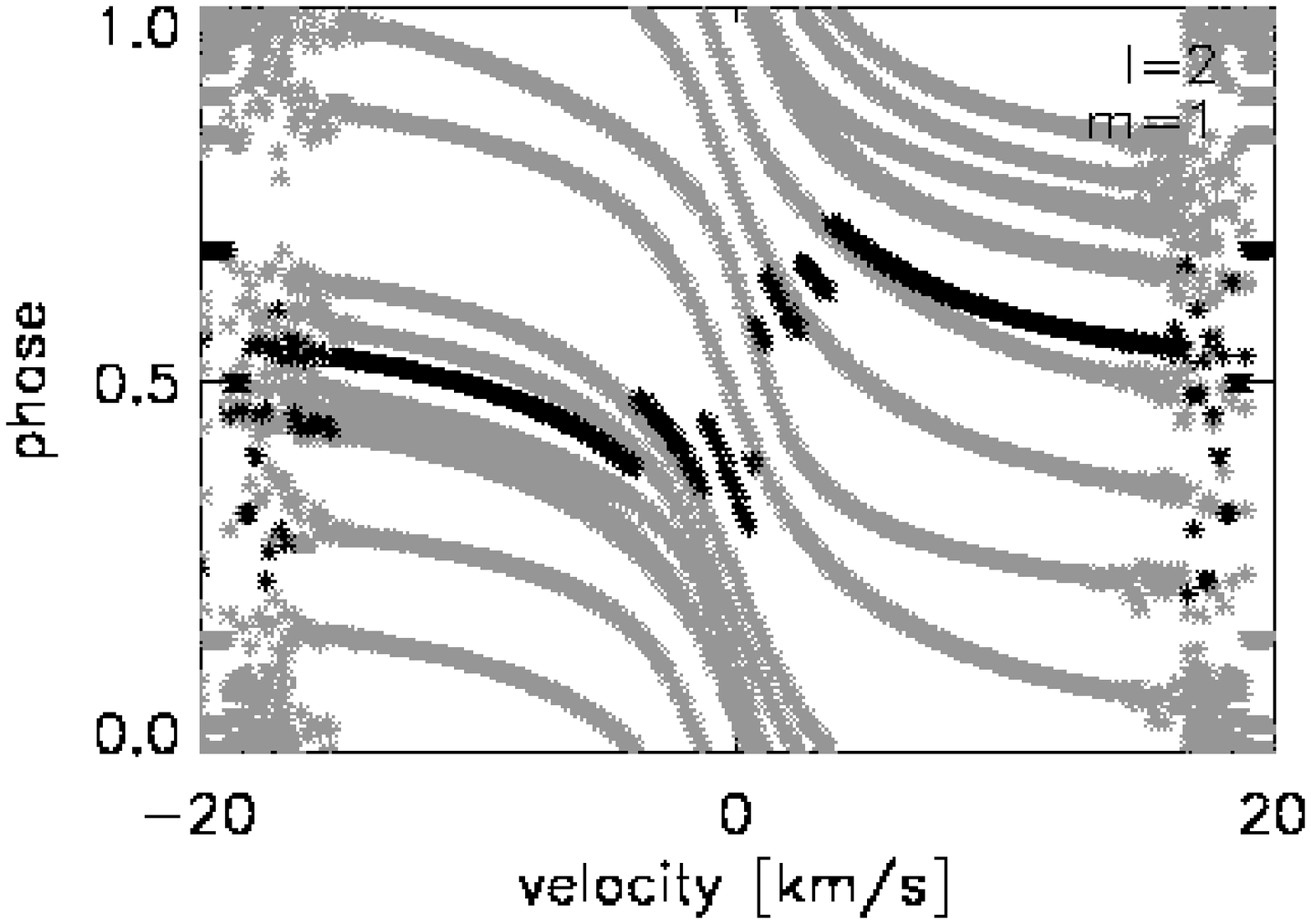}
\end{minipage}
\hfill
\begin{minipage}{4.25cm}
\centering
\includegraphics[width=4.25cm]{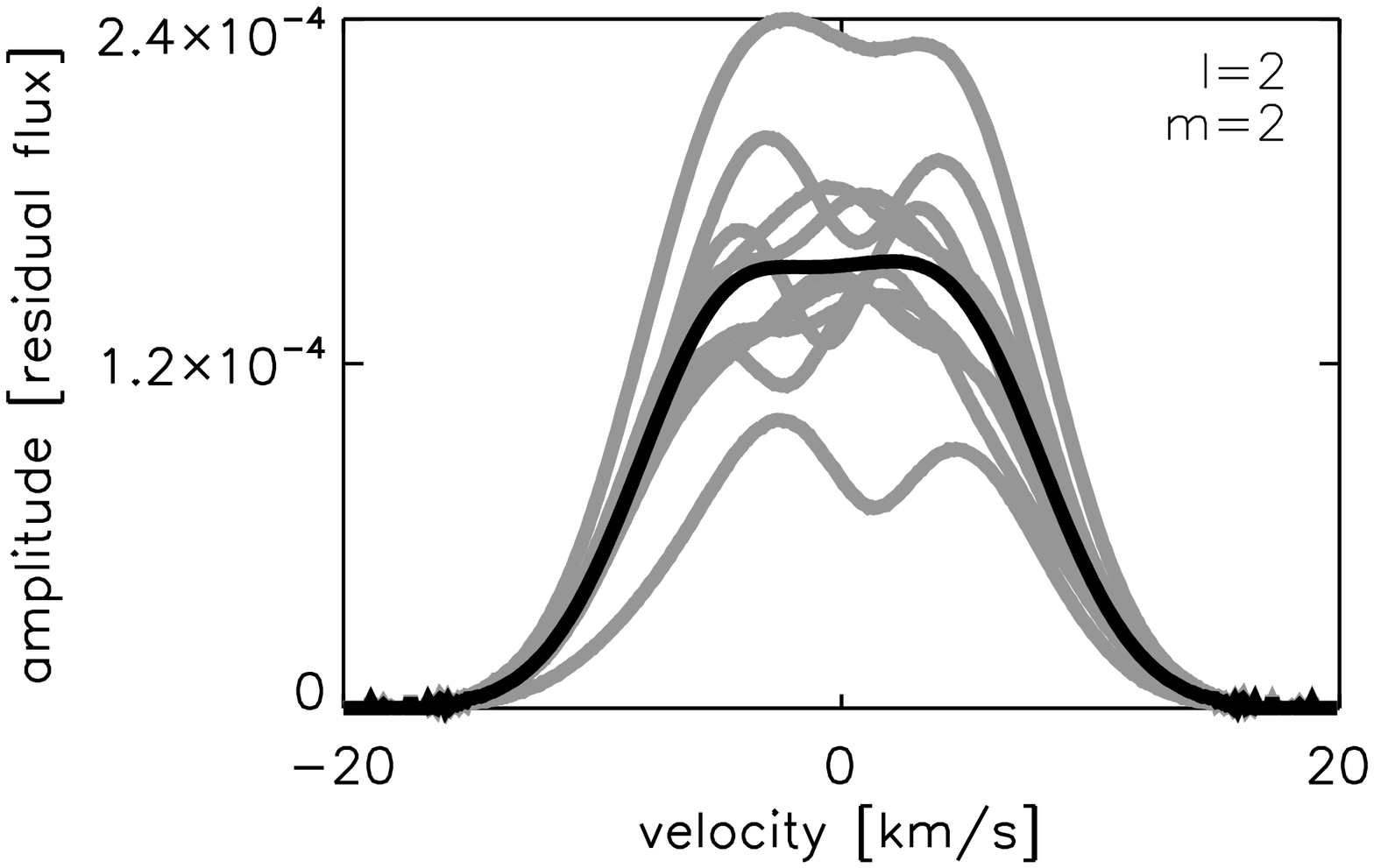}
\end{minipage}
\hfill
\begin{minipage}{4.25cm}
\centering
\includegraphics[width=4.25cm]{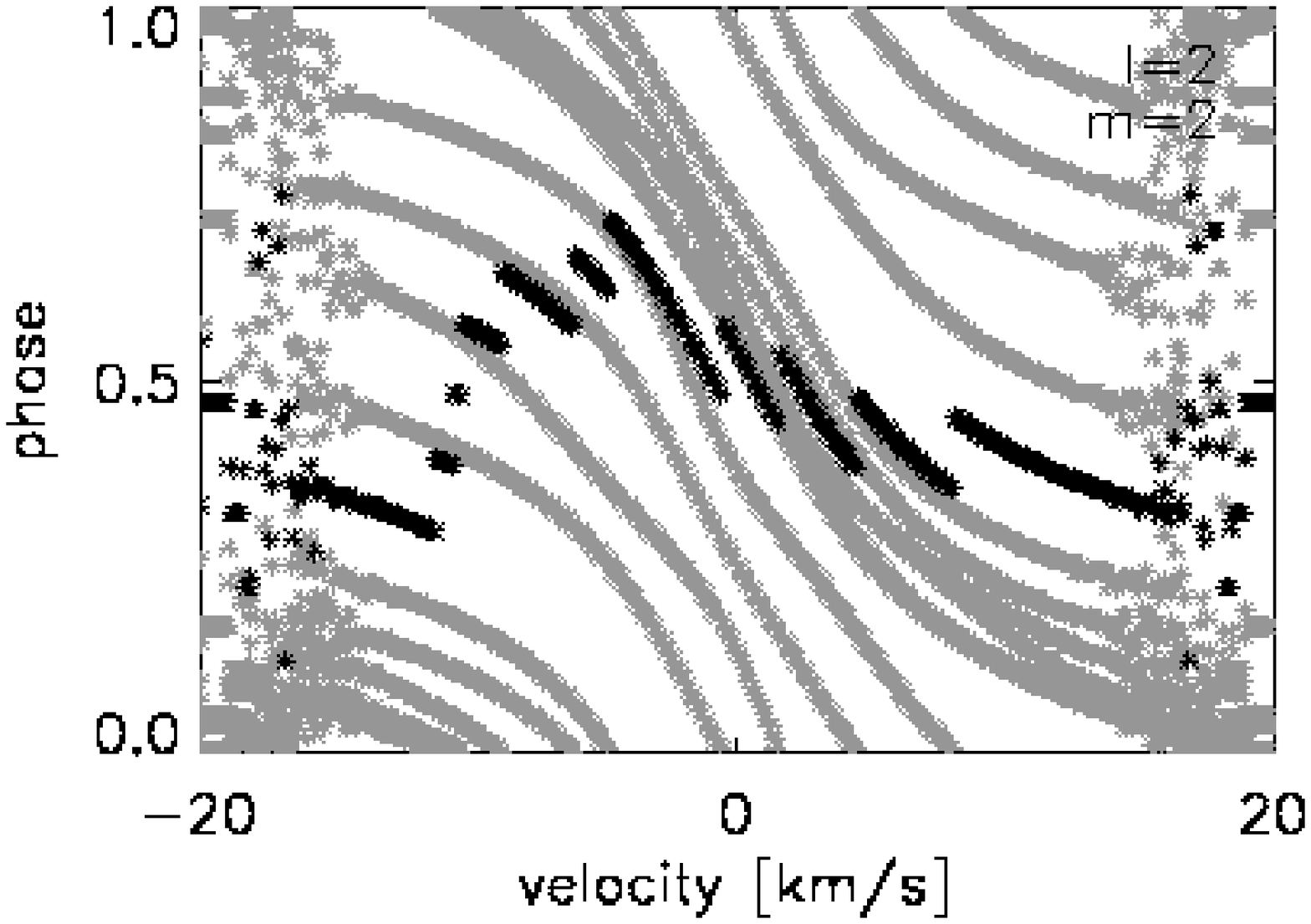}
\end{minipage}
\hfill
\caption{The amplitude (left) and phase (right) distributions for simulated line profiles with a two day damping time at an inclination angle of $i=35\degr$. For each mode ten different realisations are shown in grey, with the average shown in black. Top left: $\ell=0$, $m=0$, top right: $\ell=1$, $m=0$, middle left: $\ell=1$, $m=1$, middle right: $\ell=2$, $m=0$, bottom left: $\ell=2$, $m=1$, bottom right: $\ell=2$, $m=2$}
\label{amplphaseveleta2}
\end{figure*}

The amplitude and phase distribution as a function of the velocity across the
line profiles are investigated for the simulated spectral lines. The profiles for the cases with $\ell=0,1,2$ and positive $m$ values, with a damping time of $\eta=2$~days, are shown in Fig.~\ref{amplphaseveleta2}. The dominant frequency obtained from $\langle \mathrm{v} \rangle$ is used for the harmonic fits. This frequency is different for each realisation due to the stochastic nature of the forcing. For each mode ten realisations are shown in grey, with the average shown in black. The shape of the amplitude of axisymmetric and tesseral modes do not change, although the amplitudes per velocity pixel can have different values for different realisations. The drop to zero amplitude in the centre of the line is very characteristic for all the amplitude distributions. However, for the sectoral modes, the shape and value of the amplitudes may be different for different realisations and the amplitude distribution does not necessarily drop in the centre of the line for such modes. The behaviour does not change significantly with inclination.

\section{Interpretation}

The amplitude distribution across the line profile is used for mode identification, as in \citet{telting1997} and \citet{decat2005}. For each star we find different distributions for different dominant mode frequencies. This clearly indicates that modes with different wavenumbers $(\ell,m)$ must be present in the data. Thus, we detect non-radial modes in the observed cross correlation profiles. In order to identify the wavenumbers of the individual modes, observed amplitude distributions are compared with simulated ones, as is commonly done for mode identification. Only very recently, \citet{zima2006} implemented a statistical measure to quantify the identification of $(\ell,m)$ as a function of amplitude across the profile for modes with infinite lifetime. A similar quantitative measure in the case of damped oscillations is not yet available, but visual inspection of amplitude diagrams allows a clear discrimination between axisymmetric and non-axisymmetric modes which is what we do here, based on simulations. The averaged amplitudes over different realisations (indicated in black in Fig.~\ref{amplphaseveleta2}) are used for this comparison, keeping in mind that the observations are single realisations. Due to the fact that single realisations of different modes are in some cases comparable, different sets of wavenumbers are occasionally likely for observed modes.

For \object{$\epsilon$ Ophiuchi} the amplitude of the first frequency $\nu =5.03$ c\,d$^{-1}$ ($58.2 \mu$Hz) resembles the one of modes with $\ell=2$, $m=2$ mode. The mode with frequency $\nu=5.46$ c\,d$^{-1}$ ($63.2 \mu$Hz) resembles an $m=1$ mode, although it does not exclude an $m=0$ mode. The mode with $\nu=5.83$ c\,d$^{-1}$ ($67.5 \mu$Hz) resembles an $\ell=2$, $m=2$ mode.

The modes with frequency $\nu=11.17$ c\,d$^{-1}$ ($129.3 \mu$Hz) and $\nu=11.71$ c\,d$^{-1}$ ($135.5 \mu$Hz) obtained for \object{$\eta$ Serpentis} resembles a realisation of the $\ell=2$, $m=2$ mode, while the mode with the third frequency ($\nu=10.38$ c\,d$^{-1}$ ($102.2 \mu$Hz)) is likely to have $m=0$.

For \object{$\xi$ Hydrae} the most dominant frequency $\nu=8.42$ c\,d$^{-1}$ ($97.5 \mu$Hz) reveals a mode with either $m=0$ or $m=1$.

For \object{$\delta$ Eridani} the dominant mode ($\nu=59.40$ c\,d$^{-1}$ ($687.5 \mu$Hz)) resembles an $m=1$, while the second one ($\nu=61.08$ c\,d$^{-1}$ ($706.9 \mu$Hz)) is characteristic for an $m=0$ mode. The third mode ($\nu=52.68$ c\,d$^{-1}$ ($609.7 \mu$Hz)) shows an additional dip in the second peak, but $m=1$ as well as $m=0$ seem to be possible.

From the comparison of the observations with the simulations we can conclude that we see modes with $m\neq0$ and thus non-radial modes among the dominant modes in the (sub) giants. This result is robust against a change in the value of the inclination angle for the simulations.

\section{Discussion and conclusions}

We have implemented a line profile generation code for damped and re-excited solar like oscillations. We found that the quantities $\langle \mathrm{v} \rangle$, $\langle \mathrm{v}^{3} \rangle$ and amplitude across the profile are good line diagnostics to characterise these oscillations. Our simulations were made for a single lineforming region while we compared them with cross correlation profiles based on a mask. For this reason, the phase variations across the profile are difficult to interpret, since they are an average over the depth in the atmosphere, and this average depends on the shape of the eigenfunctions. Moreover, we learned from our simulations that the phase behaviour can be quite different for realisations, as illustrated in Fig.~\ref{amplphaseveleta2}. More extensive simulations across the depth would be needed to interpret the observed phase behaviour in detail. In order to do so, we would need to know the shape of the excited eigenmodes as a function of depth in the outer regions of the star. 

From the frequency analysis performed on the bisector velocity span, it becomes clear that it is not a good diagnostic to unravel solar like oscillations.

Individual mode identification is not yet possible with the method
used in the present work. This is not surprising because we did a first exploration of how well-working methods for modes with infinite lifetimes could be adapted for damped oscillations. The discriminant of the moments is not useful at the low amplitudes present in the sub(giants) investigated in the present work. Nevertheless, from the resemblance between the observed amplitude distribution with the simulated ones, and comparison between different modes of the same star, we presented clear observational evidence for modes with different m-values in the same star for three of our four examples, i.e. non-radial modes are present in red (sub)giants. The amplitudes for \object{$\xi$ Hydrae} are too low to make firm conclusions, and since we see only one frequency in the line profile variations, a comparison of the amplitude shapes for different modes is not possible. Therefore we can not exclude that this star has only radial modes as proposed by \citet{frandsen2002}.

\citet{dziembowski2001} and \citet{houdek2002} show theoretically that stochastically excited radial pulsation modes
are most likely to be observed in red giants. They predict
that, for non-radial modes, damping effects will have influence in the p-mode
cavity as well as in the g-mode cavity, while for radial modes only damping in the
p-mode cavity is present. Therefore, according to this theory, radial modes are less
severely influenced by damping than non-radial modes. Our result pointing towards the presence of non-radial modes, calls for a re-evaluation of the theoretical predictions.

\begin{acknowledgement}
SH wants to thank the staff at the Instituut voor Sterrenkunde at the Katholieke
Universiteit Leuven for their hospitality during her three month visit. JDR is a postdoctoral fellow of the fund for Scientific Research, Flanders. FC was supported financially by the Swiss National Science Foundation. The authors are supported by the Fund for Scientific Research of Flanders (FWO) under grant G.0332.06 and by the Research Council of the University of Leuven under grant GOA/2003/04.
\end{acknowledgement}

\bibliographystyle{aa}
\bibliography{bibpuls}
\listofobjects
\end{document}